\documentclass[useAMS,usenatbib]{mn2e}

\usepackage{graphicx,amssymb,amsmath}
\usepackage{natbib}
\usepackage{color}
\usepackage{longtable,lscape}
\usepackage{enumitem}

\def\kms{$\mbox{km s}^{-1}$}

\def\rh{r_{\rm h}}
\def\msun{{\rm M}_\odot}
\def\limepy{{\sc limepy}}

\def\spes{{\sc spes}}

\def\rj{r_{\rm J}}

\usepackage[normalem]{ulem}

\usepackage{txfonts}
\usepackage[table]{xcolor}

\usepackage{hyperref}
\usepackage[all]{hypcap}

\hypersetup{
    breaklinks=true,
    colorlinks,
    citecolor=blue,
    linkcolor=blue,
    urlcolor=blue,
    pdftitle={Mass modelling globular clusters: a method comparison},
    pdfauthor={V. H\'{e}nault-Brunet et al.}
}

\def\equationautorefname~#1\null{%
  equation~(#1)\null
}
\title[Mass modelling globular clusters in the Gaia era]{Mass modelling globular clusters in the Gaia era: a method comparison using mock data from an $N$-body simulation of M\,4}
\author[V. H\'{e}nault-Brunet et al.]{V. H\'{e}nault-Brunet\thanks{Contact e-mail: vincent.henault-brunet@nrc-cnrc.gc.ca}$^{1,2}$, M. Gieles$^{3,4,5}$, A. Sollima$^6$, L. L. Watkins$^7$, A. Zocchi$^{8,9}$, \newauthor I. Claydon$^3$, E. Pancino$^{10,11}$, H. Baumgardt$^{10}$\\
$^1$ National Research Council, Herzberg Astronomy \& Astrophysics, 5071 West Saanich Road, Victoria, BC, V9E 2E7, Canada\\
$^2$ Department of Astrophysics/IMAPP, Radboud University, PO Box 9010, 6500 GL, Nijmegen, The Netherlands\\
$^3$ Department of Physics, Faculty of Engineering and Physical Sciences, University of Surrey, Guildford, GU2 7XH, UK\\
$^4$ Institut de Ci\`{e}ncies del Cosmos (ICCUB), Universitat de Barcelona, Mart\'{i} i Franqu\`{e}s 1, E08028 Barcelona, Spain\\
$^5$ ICREA, Pg. Lluis Companys 23, 08010 Barcelona, Spain.\\
$^6$ INAF, Osservatorio di Astrofisica e Scienza dello Spazio di Bologna, via Gobetti 93/3, Bologna, 40129, Italy\\
$^7$ Space Telescope Science Institute, 3700 San Martin Drive, Baltimore, MD 21218, USA \\
$^{8}$ European Space Research and Technology Centre (ESA/ESTEC), Keplerlaan 1, 2201 AZ Noordwijk, Netherlands\\
$^9$ Dipartimento di Fisica e Astronomia, Universit{\`a} degli Studi di Bologna, viale Berti Pichat 6/2, I40127, Bologna, Italy\\
$^{10}$ INAF-Osservatorio Astrofisico di Arcetri, Largo Enrico Fermi 5, 50125, Firenze, Italy\\
$^{11}$ Space Science Data Center - Agenzia Spaziale Italiana, via del Politecnico, s.n.c., I-00133, Roma, Italy\\
$^{12}$School of Mathematics and Physics, University of Queensland, St. Lucia, QLD 4072, Australia
}

\begin{document}

\date{Accepted 2018 November 22. Received 2018 November 20 ; in original form 2018 October 10}

\pagerange{\pageref{firstpage}--\pageref{lastpage}} \pubyear{2018}

\maketitle
\label{firstpage}

\begin{abstract}
As we enter a golden age for studies of internal kinematics and dynamics of Galactic globular clusters (GCs), it is timely to assess the performance of modelling techniques in recovering the mass, mass profile, and other dynamical properties of GCs. Here, we compare different mass-modelling techniques (distribution-function (DF)-based models, Jeans models, and a grid of $N$-body models) by applying them to mock observations from a star-by-star $N$-body simulation of the GC M 4 by Heggie. The mocks mimic existing and anticipated data for GCs: surface brightness or number density profiles, local stellar mass functions, line-of-sight velocities, 
and {\it Hubble Space Telescope}- and {\it Gaia}-like proper motions. 
We discuss the successes and limitations of the methods. We find that multimass DF-based models, Jeans, and $N$-body models provide more accurate mass profiles compared to single-mass DF-based models. We highlight complications in fitting the kinematics in the outskirts due to energetically unbound stars associated with the cluster (``potential escapers", not captured by truncated DF models nor by $N$-body models of clusters in isolation), which can be avoided with DF-based models including potential escapers, or with Jeans models. We discuss ways to account for mass segregation. For example, three-component DF-based models with freedom in their mass function are a simple alternative to avoid the biases of single-mass models (which systematically underestimate the total mass, half-mass radius, and central density), while more realistic multimass DF-based models with freedom in the remnant content represent a promising avenue to infer the total mass and the mass function of remnants.

\end{abstract}


\section{Introduction}

Since the half-mass relaxation time of globular clusters (GCs) is typically shorter than their age, two-body relaxation has modified the phase-space distribution of their stars during their lifetime, resulting in an approximately Maxwellian velocity distribution and a tendency toward kinetic energy equipartition and mass segregation.  Dynamical models capturing the internal kinematics of GCs come in two main flavours: (1) static (equilibrium) models, and (2) evolutionary models. The first category includes approaches like Jeans modelling and 
distribution function (DF) based models (see below) as well as Schwarzschild's orbit superposition method \citep[e.g.][respectively]{1992AJ....104.2104L, 1979AJ.....84..752G, 2006A&A...445..513V}, while the second category includes $N$-body, Monte Carlo, Fokker-Planck and gas models \citep*[for reviews see e.g.][]{MeylanHeggie1997,2010ARA&A..48..431P}.

Being generally much faster to compute and allowing to efficiently explore phase-space configurations of GCs without needing to specify their initial conditions, equilibrium models (in particular DF-based and Jeans models) are well suited to tackle mass-modelling problems, i.e. to infer the present-day total mass and mass distribution within a cluster from observables such as surface brightness or number density profiles \citep[e.g.][]{1995AJ....109..218T}, line-of-sight (LOS) velocities \citep[e.g.][]{1986A&A...166..122M}, proper motions \citep[PMs, e.g.][]{2006A&A...445..513V}, deep stellar luminosity/mass functions \citep[e.g.][]{2004A&A...428..469P}, and pulsar accelerations \citep[e.g.][]{1993ASPC...50..141P}. They can be used, for example, to estimate mass-to-light ratios, to infer global mass functions from local mass functions, to estimate the total mass in faint/dark populations that cannot be directly observed (low-mass stars, white dwarfs, neutron stars, black holes), and to measure cluster distances by comparing LOS velocities and PMs. Knowledge of the present-day mass profiles of GCs can also tell us about their dynamical evolution since GC structure and evolution are closely related \citep{1961AnAp...24..369H}, but note that clusters with different initial conditions may at the present day look very similar. Finally, equilibrium mass-modelling analyses can provide boundary conditions and inform more detailed evolutionary models for which initial conditions must be specified, for example to infer the initial mass function (IMF) from the present-day mass function and correct for preferential escape of low-mass stars, which is particularly important for GCs with low mass and near the Galactic centre but would also affect massive tidally filling clusters \citep[e.g.][]{2003MNRAS.340..227B}. 
With improvements in the speed of computers and the sophistication of codes, detailed evolutionary models have however become a possible option on their own for mass-modelling applications of individual GCs. Simulations of GCs with up to $10^6$ stars are now possible either through direct $N$-body \citep{Heggie2014, 2016MNRAS.458.1450W} or Monte Carlo simulations \citep{2011MNRAS.410.2698G, 2017MNRAS.464L..36A}. Star-by-star $N$-body simulations with a million particles are still computationally too intensive to run a large number of models of massive GCs, but computing an extensive grid of (scaled) models with $\sim10^5$ particles is now feasible \citep[e.g.][]{2017MNRAS.464.2174B}.

With the amount and quality of internal kinematic data (both LOS velocities and PMs) of Milky Way GCs steadily increasing in recent years, and with the advent of \textit{Gaia}, we are entering a golden age for studies of the kinematics and dynamics of these systems \citep[e.g.][]{2015AJ....149...53K, 2015ApJ...803...29W, 2017MNRAS.464.2174B, 2018MNRAS.473.5591K, Libralato2018, 2018MNRAS.478.1520B}. \textit{Gaia} Data Release 2 \citep[DR2,][]{Gaia2018} has provided parallaxes and PMs for more than a billion stars over a $\sim$2-year baseline. These PMs have proven good enough to analyse mean motions -- i.e. bulk motions of the clusters \citep{Helmi2018, Watkins2018, Posti2018, 2018arXiv180709775V} and their internal rotations \citep{Bianchini2018,2018MNRAS.479.5005M} -- but are not yet good enough for studies of GC internal velocity dispersions, which limits the analyses of the internal kinematics that can be done. However, future data releases will increase the PM baseline to $\sim$5 years and include a sophisticated crowding treatment (particularly necessary in 
GC fields), which will enable kinematic studies with internal PMs for nearby GCs. For more distant objects, membership information will greatly facilitate efficient target selection for follow-up spectroscopic observations \citep{Pancino2017}, especially in the low-density outskirts where genuine cluster members are usually the needle in a haystack due to contamination by Milky Way stars.

Given these recent and upcoming observational advances, it is timely to assess the successes, strengths, limitations, and biases of various mass-modelling techniques in recovering the mass and other structural and dynamical properties of GCs. This is particularly important as these methods have not yet been confronted to data (or even mock data) that extends out to a significant fraction of the Jacobi radius of GCs, nor have they been systematically compared with each other. The kinematics in the outskirts of GCs hide important information about the interplay between internal cluster dynamics, Galactic tides, and potentially a small dark matter halo, but these outer parts have been so far poorly explored, and velocity measurements are typically confined within central regions that represent a small fraction of the Jacobi volume \citep[see][Fig. 16]{2017MNRAS.466.3937C}. An additional motivation for testing mass-modelling methods is to examine how they perform at inferring the mass-to-light ratio profiles and dark stellar remnant content of GCs. With the recent detections of gravitational waves \citep[e.g.][]{2016PhRvL.116f1102A} from merging stellar-mass black holes (BHs), there has been a renewed interest in the dark content of GCs because dynamical formation of BH-BH binaries in the dense cores of GCs and subsequent mergers has been proposed as one of the formation channels for these events \citep{2016ApJ...818L..22A, 2016PhRvD.93h4029R, 2016ApJ...831..187A}. 

The first class of models that we consider in this work are DF-based models, which are equilibrium models satisfying the collisionless Boltzmann equation and described by a parametric form for the phase-space distribution (mass, position and velocity) of stars capturing a few key physically-motivated ingredients \citep[e.g.][]{1963MNRAS.125..127M, 1966AJ.....71...64K, 1979AJ.....84..752G}. From the DF, the potential is obtained from solving Poisson's equation self-consistently, and spatial density and velocity distributions can be computed, projected on the plane of the sky, and compared to data to constrain the free parameters of the assumed DF. Despite their relative simplicity, such DF-based models have been extensively and successfully applied to GC observations for decades. 
This is due to the fact that {\it i)} they are typically fast to compute, {\it ii)} they do not require knowledge of poorly constrained input like orbits and tidal effects which are complex to account for, {\it iii)} they still 
capture (approximately) basic physical ingredients such as two-body relaxation effects and tidal truncation, and {\it iv)} they can be easily extended to include additional ingredients such as a mass spectrum \citep{1976ApJ...206..128D,1979AJ.....84..752G} or anisotropy in the distribution of velocities \citep{1963MNRAS.125..127M}. 
Here we consider more specifically several flavours DF-based models from the \limepy\footnote{The \limepy\ (Lowered Isothermal Model Explorer in PYthon) code is available from \url{https://github.com/mgieles/limepy}.} family \citep{2015MNRAS.454..576G} and from the Spherical Potential Escapers Stitched models family \citep[\spes;][]{claydon18}, as well as multimass Michie-King models \citep{1963MNRAS.125..127M, 1966AJ.....71...64K, 1979AJ.....84..752G}.

The second class of models that we consider here are Jeans models, which also provide a static description of the cluster and are based on solving the Jeans equations to derive the global mass profile from the density and velocity dispersion profiles of a tracer population. These models can include both anisotropy and rotation \citep[e.g.][]{Cappellari2008}, and allow for radially varying mass-to-light ratios, which encompass underlying differences in the spatial distributions of different mass populations, although they do not (yet) include an explicit mass function. They are typically fast to run and have proved to be useful tools for understanding a variety of stellar systems including the Milky Way disk \citep{Buedenbender2015}, globular clusters \citep{Watkins2013}, dwarf galaxies \citep{Zhu2016a}, and giant elliptical galaxies \citep{Zhu2016b}. They are also commonly used to provide initial insights for more sophisticated, albeit slower, alternatives such as Schwarzschild models. They have the advantage that no functional form needs to be assumed for the underlying DF, leaving significant freedom in determining the mass, anisotropy, and rotation profiles. This freedom however comes at the cost that such a method can lead to unphysical and/or unstable results, especially when the data is noisy (see \autoref{Sect_Jeans} and references therein).

The third class of models that we consider are $N$-body models, specifically a mass-modelling approach based on comparing a grid of $N$-body models to kinematic, structural, and stellar mass function data as recently applied to real data of Milky Way GCs \citep{2017MNRAS.464.2174B, 2017MNRAS.472..744B, 2018MNRAS.478.1520B}. While these are not as flexible as the equilibrium models introduced above, they have the advantage of being self-consistent, of including the effects of collisional dynamics from first principles and of offering a way to constrain the initial conditions of GCs.

We apply these different mass-modelling techniques to realistic mock data from a star-by-star $N$-body simulation of the GC M\,4 by \citet{Heggie2014} and compare their performance using a number of metrics. We include PMs in the mock data similar to those that would be provided by the \textit{Hubble Space Telescope} (\textit{HST}) in the central regions of GCs and, most importantly, the expected (end-of-mission) performance of the data from {\it Gaia} in the outskirts (out to twice the Jacobi radius). We restrict ourselves to non-rotating models and do not consider putative intermediate-mass black holes (IMBH) since the mock data that we use refers to a cluster with no significant rotation and no IMBH. That said, extending this kind of work to consider these additional ingredients in the future would be a worthwhile exercise.

We describe the mock data\footnote{Data available to download on the {\it Gaia Challenge} workshops wiki: \url{http://astrowiki.ph.surrey.ac.uk/dokuwiki/doku.php?id=tests:collision:
mock\_data:challenge\_4}. } in \autoref{section:mock_data} and then present the different methods (labelled as models A to H) and modelling results in the following sections: single-mass models without and with potential escapers (\autoref{section:single-mass}), multimass models (\autoref{Sect_multimass}), Jeans models (\autoref{Sect_Jeans}), and a grid of $N$-body models (\autoref{Nbody_Holger}). A comparison of the enclosed mass and mass-to-light ratio profiles from the various methods is presented in \autoref{comparison}, a discussion of their pros and cons can be found in \autoref{Sect_discussion}, and concluding remarks summarizing our results are presented in \autoref{summary}.

\section{Mock data}
\label{section:mock_data}

Our mock data is extracted from a snapshot of the star-by-star $N$-body model of the globular cluster M\,4 by \citet{Heggie2014}, which included a full mass spectrum, primordial binaries, the effects of stellar evolution, and the tidal field of a point-mass galaxy. Data for all particles (mass, 3D position, 3D velocity, stellar type, stellar radius, luminosity, $V-$band magnitude, $B-V$ colour) are available at more than 300 instants throughout the lifetime of the cluster\footnote{The raw snapshots are available here: \url{https://datashare.is.ed.ac.uk/handle/10283/618}}.

We selected the snapshot at $t = 12023.9$ Myr (comparable to the current age of GCs), at which point the cluster is dynamically evolved and has lost $\sim80\%$ of its initial mass due to stellar mass loss and dynamical evolution. The main cluster properties for the selected snapshot are: total cluster mass $M =69144.5$~M$_{\odot}$, total $V$-band luminosity $L_{V} = 35547$ L$_{V,\odot}$, mass-to-light ratio $\Upsilon_V = 1.95$~M$_{\odot}/L_{V,\odot}$, and half-mass radius $r_{\rm h}=3.14$~pc. The Jacobi radius ($\rj$) of the model at the selected time is 20.3\,pc. The distance to the real M\,4 is $D=1862.1$~pc \citep{2015ApJ...799..165B}, which corresponds to a distance modulus of 11.35~mag. We adopt this distance here to transform projected positions into angular distances and transverse velocities into PMs, and conversely it is this distance we will seek to recover from our modelling when LOS velocities and PMs are used. In the $N$-body model the Galactic centre is along the $x$-axis and the orbit is in the $xy$-plane. We assume the observer views the cluster along the $z$-axis, which is not the best representation of the way we view the real M\,4. Because the tidal potential is triaxial at large distances from the GC centre, certain properties of the mock data depend on the viewing angle. However, we compared the dispersion profiles at large radii for different viewing angles and found that they are similar.

From the $N$-body model snapshot we extract photometric and kinematic information that generally is (or will be) available to observers: surface brightness and number density profiles, LOS velocities, \textit{HST}-like PMs, and \textit{Gaia}-like PMs. In each case, we mimic as closely as possible the size, quality, and spatial distribution of real observations, as we describe below.

Different mass-modelling methods either use the number density profile or the surface brightness profile as the main observational constraint on the structural properties of a GC, so we construct both from the mock data. We build a projected number density profile for all the stars brighter than an apparent magnitude of $V=17$ (at the adopted distance of M\,4). We also extract a $V-$band surface brightness profile from the positions and apparent $V$ magnitudes of all the stars in the snapshot (with no magnitude cut).

We assume that LOS velocities are available for the subset of red giant branch stars brighter than $V=15$ (the main sequence turn-off is around $V=16.5$, corresponding to a turnoff mass of $\sim0.85 \ \msun$), for a total of $N=635$ stars. 
 An uncertainty of 1 \kms \ is assumed on the LOS velocities of all these stars. This mimics the type of measurements available or achievable with spectrographs on ground-based telescopes. 
For methods that fit on the dispersion profile (as opposed to discrete velocities), we build a LOS velocity dispersion profile with 70 stars per radial bin (and any additional leftover stars included in the last bin).

We extract two PM catalogues, the first designed to mimic the kind of kinematic data enabled by {\it HST} \citep[e.g.][]{Bellini2014}. We restrict this sample to stars in the inner 100$^{\prime\prime}$ of the cluster centre in projection (comparable to the field of view of the Advanced Camera for Surveys (ACS) on-board {\it HST}) and magnitudes in the range $16<V<17.5$, for a total of 2567 stars with a mean mass of 0.76~$\msun$. We assume a typical uncertainty of 0.1~mas~yr$^{-1}$ on the PM of all these stars. We build PM dispersion profiles for the radial and tangential components in the plane of the sky with 120 stars per radial bin (and any additional leftover stars included in the last bin).

The second PM catalog is obtained by using the method by \citet{Pancino2017} to transform the $N$-body model snapshot into {\it Gaia}-like observables and extract internal PMs for cluster members that would be recovered by {\it Gaia}. 
The simulated cluster was projected, at the adopted distance of M\,4, on a 2$^{\circ}\times$2$^{\circ}$ field population simulated with the Galaxy models by \citet{Robin2003} and representing the actual
position of M\,4 ($\alpha=245.89675^{\circ}$, $\delta=-26.52575^{\circ}$, with a field population of approximately 100\,000 stars per square degree).

We adopted the bulk proper motion for M\,4 from \citet{2003AJ....126..247B}. The simulations were complemented with {\it Gaia} integrated magnitudes using the prescriptions by \citet{Jordi2010}. Errors were simulated using the post-launch {\it Gaia} science performances \citep{Gaia2016}, worsened by an appropriate amount to take into account stellar crowding effects, based on the current deblending pipeline simulations and assuming end-of-mission (i.e. 5 years nominal duration) performances, as described in detail by \citet{Pancino2017}. From these simulated {\it Gaia} end-of-mission data, we retain only recovered cluster members that are not significantly affected by blending or contamination (we assume perfect membership selection due to the very different proper motion of the cluster and background). Specifically, we discard classic blends (stars closer than {\it Gaia}'s point spread function - PSF - of 0.177$^{\prime\prime}$) that will not always be deblended by {\it Gaia} and we also discard binaries. With real {\it Gaia} data, these contaminants could be identified from the various quality flags and binary modeling provided with the final {\it Gaia} catalog. We also discard {\it Gaia} blends and contaminants \citep[for definitions see][]{Pancino2017} blended or contaminated by at least 1 per cent of their flux. We finally only retain stars above 0.7~$\msun$ ($V\lesssim 17$) since fainter lower-mass stars are not expected to have precise enough {\it Gaia} PMs for the purpose of the present study. These stars have a mean mass of 0.8~$\msun$. From this final sample we build PM dispersion profiles for the radial and tangential components in the plane of the sky with 400 stars per radial bin (and any additional leftover stars included in the last bin).

Note that, in order to test the effects of the background on the membership selection, we also considered other field populations, namely a nearly empty halo field at $\alpha=186.5^{\circ}$ and $\delta=36.0^{\circ}$, containing approximately 1000 stars per square degree, as well as an extremely crowded bulge field at $\alpha=264.5^{\circ}$ and $\delta=-22^{\circ}$, containing approximately 1 million stars per square degree. We found that the number of selected members with no or limited contamination depends only weakly on the adopted field population, unlike in \citet{Pancino2017}, because M\,4 is closer -- thus more spread out on the sky -- and the simulated field populations are also less crowded than the cases considered by \citet{Pancino2017}. Thus, in the remainder of this work we present results based on the mock {\it Gaia} data with the M\,4 background (7083 `clean' recovered member stars with {\it Gaia} PMs). One should however bear in mind that this should be seen as a best-case scenario and we caution that field contamination would be more critical for more distant/crowded clusters in real data.

Finally, we also extract a mock local stellar mass function of main-sequence stars as an additional constraint for methods that aim to infer the global stellar mass function of the cluster. This mock stellar mass function is built for stars between about 0.2 and 0.75~$\msun$ \citep[mimicking the depth typically achieved by {\it HST} observations of nearby GCs, e.g.][]{2017MNRAS.471.3668S} in an annular region between 250 and 350$^{\prime\prime}$ in projection from the cluster centre (around the half-mass radius). Actual measurements of the stellar luminosity function (and thus mass function) are limited to a restricted portion of GCs, usually close to the cluster core \citep[e.g.][]{2007AJ....133.1658S, 2017MNRAS.471.3668S} or sometimes near the half-mass radius \citep[e.g.][]{2000ApJ...534..870P}. 
For the location of the reference region of our mock local stellar mass function, we chose a field close to the half-mass radius, as is often done because there the local mass function most resembles the global mass function \citep{1986AJ.....92.1358P, 1997MNRAS.289..898V, 2003MNRAS.340..227B}. 
\footnote{The effect on the inferred global mass function of a non-optimal location of the measured luminosity function has been explored by \citet{2015MNRAS.451.2185S} by comparing multimass Michie-King models to mock data.}

We compare the performance of the different mass-modelling methods applied to the mock data in three cases:
\begin{enumerate}[label=\arabic*., leftmargin=0.3\parindent]
\item fits to the surface brightness profile (or number density profile) and LOS velocities (mimicking widely available ground-based data);
\item fits to the data from step 1 and additionally including {\it HST}-like PMs (currently available for a subset of $\sim20$ Milky Way GCs);
\item fits to all the above and additionally {\it Gaia}-like internal PMs (available in the near future for nearby GCs).
\end{enumerate}

We fix the distance to its known value in step 1, but we leave it as a free parameter when PMs are used. Distance determination via dynamical modelling indeed requires both PMs and LOS velocities \citep[e.g.][]{Watkins2015b} and is not possible with LOS velocities only. It is usual for studies of real clusters with only LOS velocities available to fix the distance to the value estimated via an alternative method. Note that {\it Gaia} end-of-mission data will allow to secure distances to better than 1 per cent for the majority of Milky Way GCs \citep{Pancino2017}.

By fitting these data, we compare how well the different methods recover the following quantities and observables: total cluster mass, half-mass radius, global mass-to-light ratio, mass-to-light ratio profile, global stellar mass function and mass in dark remnants (when applicable). We also assess the general quality of the fit to the LOS velocity and PM dispersion profiles, surface brightness or number density profile. In the next sections, we summarise each method in turn and we present its results and performance when fitting the three subsets of mock data introduced above.

\label{Sect_single-mass}

\section{Single-mass models}
\label{section:single-mass}

\subsection{(A) Single-mass {\sc limepy} DF}
\label{section:single-mass_limepy}

\begin{figure}
    \centering
        \includegraphics[width=\linewidth]{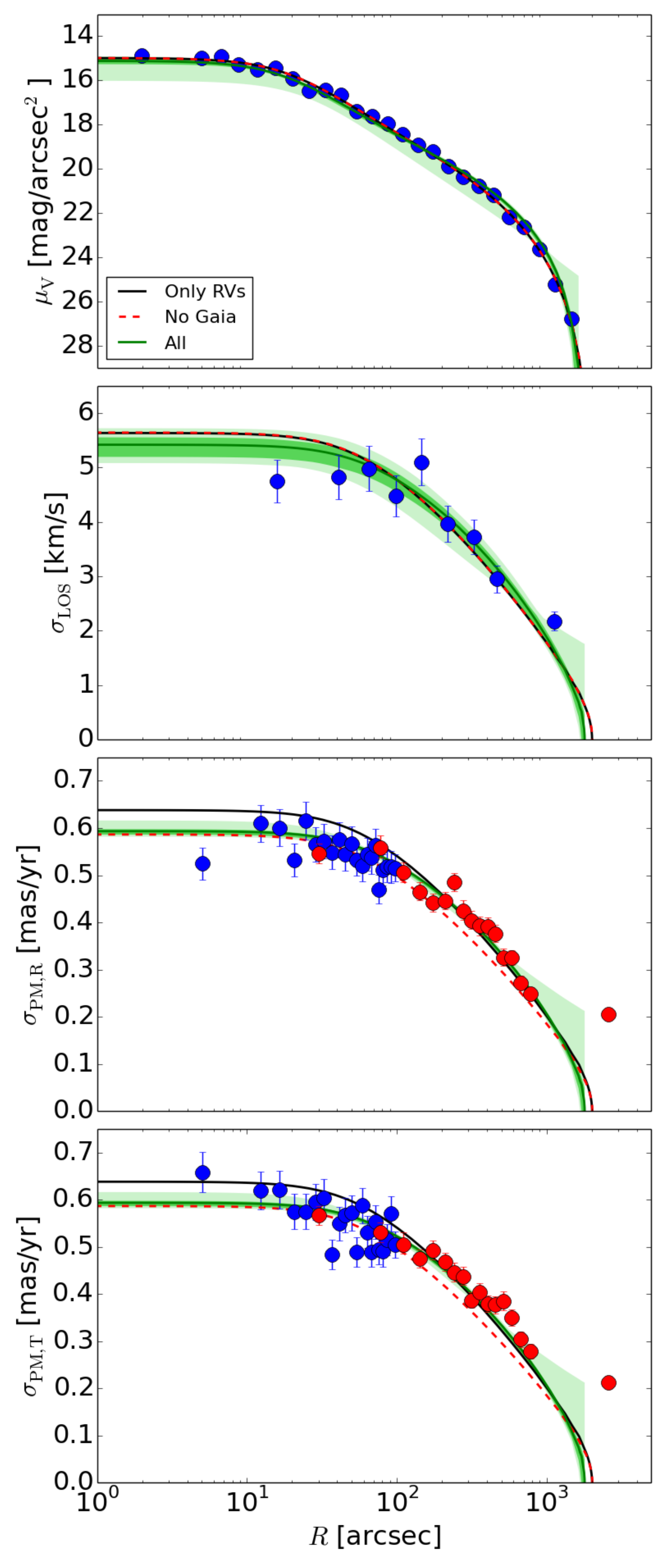}
    \caption{Fit of the single-mass {\sc limepy} models from \autoref{section:single-mass_limepy} to the datasets of case 1, 2 and 3. From top to bottom: surface brightness profile, LOS velocity dispersion profile and radial and tangential PM dispersion profiles. In each panel, the black solid line represents the result of the fit carried out only on the surface brightness profile and LOS velocity dispersion (case 1), the dashed red line the fit when also using {\it HST} PMs (case 2), and the green line the profile obtained when fitting also on {\it Gaia}-like PMs (case 3). The $1\sigma$ and $2\sigma$ contours for the model obtained when fitting on all the data are indicated as dark and light green shaded areas. The mock data is represented by blue filled circles, with the exception of {\it Gaia}-like PMs shown with red filled circles. The most distant {\it Gaia} PM data point at $\sim3000$ arcsec was not included in the fit.}
    \label{fig:fit_SM}
\end{figure}

We first consider the family of \limepy\ dynamical models, defined from a lowered-isothermal 
distribution function as described in \citet{2015MNRAS.454..576G}. 
In the single-mass variant of the \limepy\ models, stars of different masses are implicitly assumed to have the same dynamics. This implies that the mass-to-light ratio of the cluster is assumed to be independent of the distance from the centre and that the effect of mass segregation is ignored.

To compute a model in the \limepy\ family, it is necessary to specify the values of three parameters. The concentration parameter $W_0$ represents the central dimensionless potential, and is used as a boundary condition to solve the Poisson equation. The truncation parameter $g$ sets the sharpness of the truncation: models with larger values of $g$ are more extended and have a less abrupt truncation. For $g=1$ the well-known \citet{1966AJ.....71...64K} models are recovered.
The anisotropy radius $r_{\rm a}$ sets the amount of velocity anisotropy in the system: models with a small $r_{\rm a}$ are more radially anisotropic, and when $r_{\rm a}$ is large with respect to the truncation radius ($r_{\rm t}$), the velocity distribution is everywhere isotropic. Moreover, to match a particular cluster, the models need to be scaled, by using the value of the desired total mass, $M$, and half-mass radius, $r_{\rm h}$. 

We compute the projected mass density profile and the projected velocity dispersion components (along the LOS and on the plane of the sky) for the models, and we compare them with the 
mock binned profiles described in \autoref{section:mock_data} to determine the best-fit parameters. The fitting procedure we follow is similar to the one presented by \citet{2017MNRAS.468.4429Z} in their section~4.2. We adopt a
 log-likelihood function of the form $\sum_{i=1}^n (\mathcal{O}_i - \mathcal{M}_i)^2/\mathcal{E}_i^2$ 
 where $n$ is the number of bins, $\mathcal{O}_i$ are the mock data values and $\mathcal{E}_i$ are the errors on the data, and $\mathcal{M}_i$ are the model values at the same radial position as the data. The fitting parameters we consider are the three model parameters $W_0$, $g$, $r_{\rm a}$, the two scales $M$ and $r_{\rm h}$, and the mass-to-light ratio $\Upsilon_V =M/L_V$ needed to convert the model projected mass density profile to a surface brightness profile for comparison with the observations. 
A nuisance parameter is included as a fitting parameter to capture the unknown uncertainty in the surface brightness profile values (e.g. due to stochastic sampling of the stellar luminosity function). For the surface brightness data points, the log-likelihood function above thus takes the form $\sum_{i=1}^n (\mathcal{O}_i - \mathcal{M}_i)^2/\sigma_\mu^2$, where $\sigma_\mu^2$ is now a fitting parameter. The fits are performed by 
using \textsc{emcee} \citep{ForemanMackey2013} and adopting uniform priors on all parameters apart from $r_{\rm a}$ for which the prior is uniform in $\log(r_{\rm a})$. 

Figure~\ref{fig:fit_SM} 
shows the best-fit models together with the profiles from the snapshot of the M\,4 simulation for the three combinations of mock data considered. The figure shows that the best-fit models for the three cases similarly reproduce the surface brightness profile, but some differences are found when comparing their kinematic profiles. When considering the models that include the PM data and the distance as a fitting parameter, we see that the central part of the PM dispersion profiles are very similar, while some differences are seen beyond the extent of the {\it HST}-like sample: only when including the {\it Gaia}-like PM data do the best-fit single-mass models satisfyingly reproduce the PM dispersion profiles at large radii.

\begin{table} \renewcommand{\arraystretch}{1.2}\addtolength{\tabcolsep}{4pt}
\caption{Results of {\sc limepy} single-mass model fits to different dataset combinations.  The fitting parameters $M$, $\rh$, $\Upsilon_V$ and $D$ are given in \autoref{results_table}. 
}

\label{tab:singlemass}
\centering
\begin{tabular}{lccc}
\hline
Data & $W_0$ & $g$ &  $r_{\rm a}/r_{\rm t}$ \\
\hline
(1) Only LOS & $8.70^{+0.17}_{-0.16}$ & $0.93^{+0.07}_{-0.08}$ & $12.30^{+45.24}_{-10.78}$ \\
(2) No Gaia  & $8.70^{+0.17}_{-0.16}$ & $0.94^{+0.07}_{-0.08}$ & $7.64^{+36.71}_{-6.28}$ \\
(3) All      & $9.18^{+0.18}_{-0.21}$ & $0.56^{+0.08}_{-0.07}$ & $14.01^{+43.15}_{-10.90}$ \\
\hline
 \end{tabular}
\end{table}

The best-fit values and $\pm1\sigma$ uncertainties on the parameters of the {\sc limepy} single-mass models (obtained from their marginalized posterior probability distributions) are listed in \autoref{tab:singlemass}. We present the value of the anisotropy radius in relation to the truncation radius to make it clear that the anisotropy radius obtained is very large, and the models are therefore isotropic. In \autoref{results_table} we present the best-fit values obtained for the mass, half-mass radius, mass-to-light ratio, and distance of the cluster for the three cases considered. PM data sampling the entire extent of the cluster clearly improves the performance of single-mass models. Only when considering all the PM data it is possible to accurately recover the true value of the mass and a value of the half-mass radius that is consistent with the real one within $2\sigma$. 

It is important to point out that when considering {\it Gaia}-like data in this subsection, we did not include the outermost point of each of the PM dispersion profiles in the fit, where these profiles flatten in a way that cannot be reproduced by truncated models. We did this because, due to their small associated errors, these points would drive the fit towards models with an unrealistically large mass and half-mass radius, with the resulting best-fit surface brightness profile heavily underestimating the mock data in the centre and overestimating it in the outer parts, with a very large discrepancy. This illustrates a limitation of these models in the outskirts of stellar clusters significantly affected by Galactic tides, which can be overcome with the \spes\ models described in \autoref{Sect:PEs}.

\begin{table*}

\centering                 
\caption{Recovered quantities for different mass-modelling methods and combinations of mock datasets. For the ``Case" column: 1 = surface brightness profile (or number density) + LOS velocities; 2 = surface brightness profile (or number density) + LOS velocities + {\it HST} PMs; 3 = surface brightness profile (or number density) + LOS velocities + {\it HST} PMs + {\it Gaia} PMs. The local stellar mass function around the half-mass radius was used as an additional observational constraint when the dataset is marked by *.}

\footnotesize{            
\label{results_table}
\renewcommand{\arraystretch}{1.2}\addtolength{\tabcolsep}{4.6pt}    
\begin{tabular*}{\linewidth}{c l c l l l l l}     
\hline\hline                 
& & & \multicolumn{5}{c}{Recovered quantities} \\  \cline{4-8}
\multicolumn{2}{c}{Modelling Method} & Case & $M$ [M$_{\odot}$] & $r_{\rm h}$ [pc] & $\Upsilon_V$ [M$_{\odot}$/$L_{V, \odot}$] & $M_{\rm dark}$/$M$ & $D$ [kpc] \\
\hline
\multicolumn{3}{r}{True Values $\rightarrow$} & $69144.5$ & $3.14$ & $1.95$ & 0.393 & 1.862 \\
\hline
A & Single-mass {\sc limepy} DF & 1 & $57424^{+3679}_{-3509}$ & $2.16^{+0.12}_{-0.11}$ & $1.63^{+0.12}_{-0.11}$ & \ldots & 1.862 (fixed)  \\  
& (\autoref{section:single-mass_limepy}) & 2 & $62618^{+5795}_{-5298}$ & $2.35^{+0.15}_{-0.14}$ & $1.50^{+0.08}_{-0.07}$ & \ldots & $2.028^{+0.063}_{-0.061}$  \\ 
& & 3 & $68998^{+7911}_{-6073}$ &  $2.93^{+0.18}_{-0.15}$ & $1.71^{+0.11}_{-0.09}$ & \ldots & $1.925^{+0.063}_{-0.066}$ \\ 
\hline 
B & Single-mass \textsc{SPES} DF (with PEs) & 1& $57707^{+3918}_{-3687}$ &$2.26_{-0.10}^{+0.13}$ & $1.63^{+0.13}_{-0.12}$ & \ldots & $1.862$ (fixed) \\
& (\autoref{Sect:PEs}) & 2 & $62240^{+5461}_{-5856}$ & $2.42_{-0.15}^{+0.16}$ & $1.53^{+0.08}_{-0.07}$ & \ldots &$2.007^{+0.062}_{-0.069}$  \\
& & 3 & $66444^{+6223}_{-5945}$ & $2.80_{-0.16}^{+0.16}$ & $1.62^{+0.10}_{-0.09}$ & \ldots &$1.917^{+0.057}_{-0.058}$ \\
\hline
C & 3-component {\sc limepy} DF & 1 &$84215_{-9645}^{+10230}$&$3.35_{-0.35}^{+0.35}$ & $2.36_{-0.28}^{+0.31}$ &$0.26_{-0.07}^{+0.09}$ & 1.862 (fixed) \\
& (\autoref{Sect_3-comp}) & 2 & $77583_{-7789}^{+7833}$& $3.06_{-0.21}^{+0.21}$& $2.02_{-0.20}^{+0.23}$ & $0.32_{-0.08}^{+0.10}$ & $1.927_{-0.070}^{+0.066}$  \\
& & 3 & $78336_{-6535}^{+6865}$& $3.12_{-0.13}^{+0.14}$& $2.08_{-0.10}^{+0.12}$ & $0.31_{-0.06}^{+0.08}$ & $1.906_{-0.053}^{+0.055}$ \\
\hline 
D & Multimass {\sc limepy} DF & 1* & $79195^{+7318}_{-6665}$ & $3.05^{+0.35}_{-0.34}$ & $2.27^{+0.35}_{-0.29}$ & $0.46^{+0.05}_{-0.07}$ & 1.862 (fixed) \\
& (\autoref{Sect_multi_vincent}) & 2*& $75662^{+7625}_{-6706}$ & $2.79^{+0.25}_{-0.23}$ & $2.09^{+0.30}_{-0.24}$ &$0.42^{+0.07}_{-0.07}$ &$1.904^{+0.070}_{-0.065}$  \\
& & 3* & $75643^{+6721}_{-6193}$ & $2.78^{+0.14}_{-0.13}$ & $2.08^{+0.23}_{-0.20}$ & $0.44^{+0.05}_{-0.06}$ & $1.904^{+0.053}_{-0.054}$  \\
\hline
E & Multimass King DF & 1* & 70560$\pm$5013  & 2.87$\pm$0.12  & 1.97$\pm$0.14 &0.36$\pm$0.01  & 1.862 (fixed) \\
& (\autoref{Sect_multi_antonio})& 2* & 70560$\pm$3305 & 2.87$\pm$0.12  & 1.97$\pm$0.10 & 0.36$\pm$0.01 & 1.861$\pm$0.056 \\ 
& & 3* & 68890$\pm$3130  & 2.87$\pm$0.12 & 1.92$\pm$0.10 & 0.35$\pm$0.01 & 1.861$\pm$0.056 \\
\hline
F & Variable M/L Jeans (Sollima) & 1 & 83345$\pm$23850 & 3.71$\pm$2.55 & 2.32$\pm$0.66 & \ldots & 1.862 (fixed) \\
& (\autoref{Jeans_Sollima})& 2 & 58757$\pm$17417 & 2.33$\pm$1.37 & 1.66$\pm0.49$ & \ldots & 1.861$\pm$0.056 \\
& & 3 & 65991$\pm$13676 & 2.59$\pm$0.71 & 1.84$\pm$0.38 & \ldots & 1.861$\pm$0.056 \\
\hline   
G & Variable M/L Jeans (JAM) & 1 & $64269^{+15635}_{-12389}$ & $2.31^{+1.17}_{-0.60}$ & $1.88^{+0.46}_{-0.35}$ & \ldots & 1.862 (fixed) \\
& (\autoref{Jeans_Watkins}) & 2 & $65924^{+16438}_{-12218}$ & $2.37^{+1.05}_{-0.47}$ & $1.89^{+0.50}_{-0.36}$ & \ldots & $1.878^{+0.090}_{-0.081}$ \\
& & 3 & $71856^{+8119}_{-7815}$ & $2.97^{+0.48}_{-0.36}$ & $2.08^{+0.15}_{-0.15}$ & \ldots & $1.869^{+0.059}_{-0.056}$ \\
\hline 
H & $N$-body grid & 1 & $68159\pm3820$ & $2.72\pm0.18$ & $1.91\pm0.11$ & $0.36\pm0.05$ & 1.862 (fixed) \\
& (\autoref{Nbody_Holger}) & 2 & $70000\pm1300$ & $2.84\pm0.11$ & $1.82\pm0.03$ & $0.38\pm0.05$ & $1.935\pm0.055$ \\
& & 3 & $70540\pm1020$ & $2.82\pm0.11$ & $1.87\pm0.03$ & $0.37\pm0.05$ & $1.914\pm0.055$ \\
\hline 
\end{tabular*}
}

\end{table*}

\subsection{(B) Single-mass \textsc{SPES} DF (with potential escapers)}
\label{Sect:PEs}

\begin{figure}
    \centering
        \includegraphics[width=\linewidth]{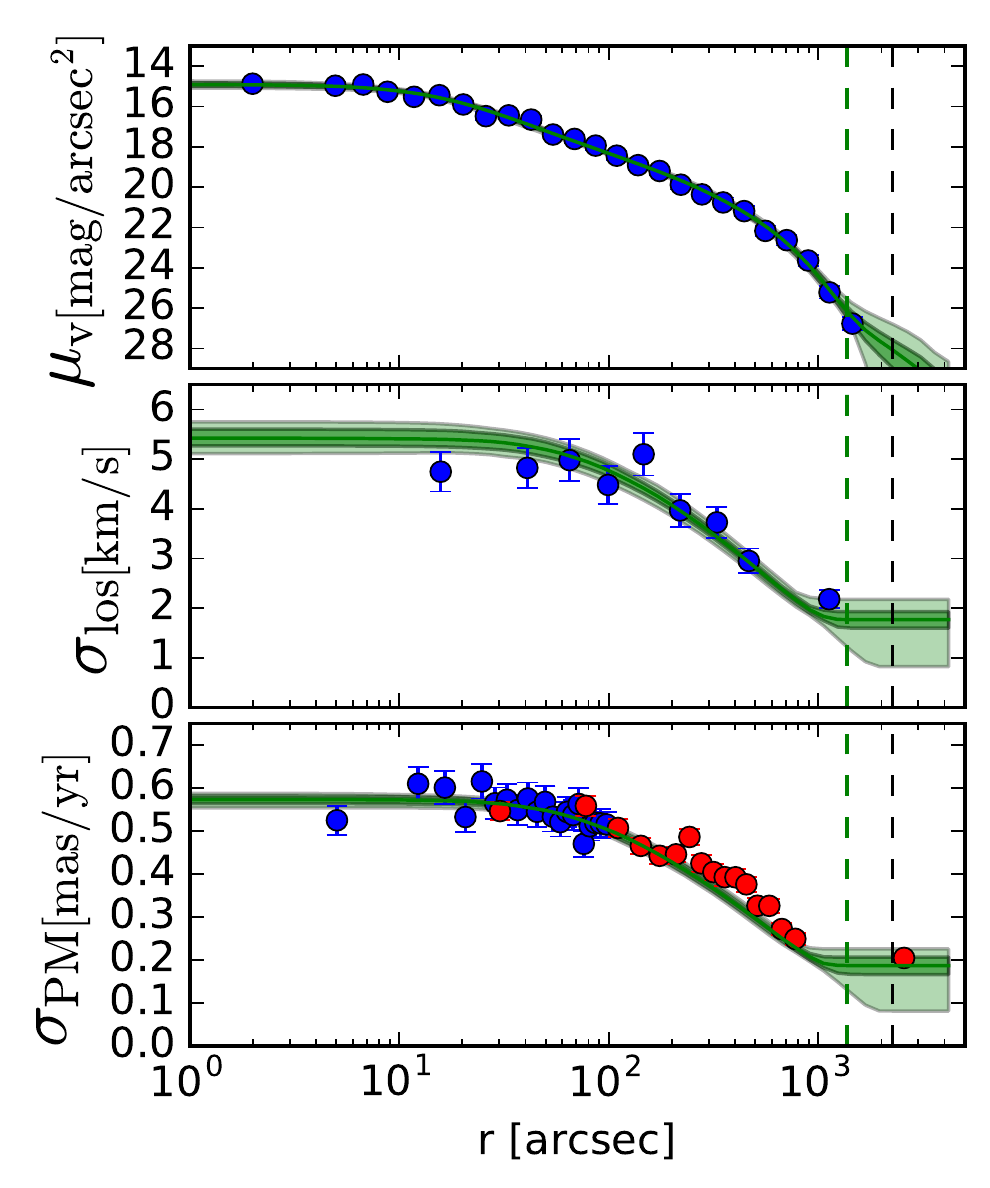}
    \caption{Fit of the model with PEs (\autoref{Sect:PEs}) to the dataset of case 3 (fitting to surface brightness, radial velocities, {\it HST} and {\it Gaia}-like PMs). The mock data for the surface brightness profile (top panel), LOS velocity dispersion profile (middle panel) and PM dispersion profile (bottom panel), are shown with blue filled circles, except the {\it Gaia} mock data shown in red. The green curve is the 50th percentile of the best-fit model with PEs and the shaded dark green region shows the 1$\sigma$ contours and the light green region shows the 2$\sigma$ contours. The dashed black line is the $r_{\rm J}$ and the dashed green line is $r_{\rm t}$.} 
    \label{fig:ian}
\end{figure}

\begin{table*} \renewcommand{\arraystretch}{1.2}\addtolength{\tabcolsep}{4pt}
\caption{Results of the \spes\ models (with PEs) fitted to different dataset combinations. The values in the last two columns should be compared to $r_{\rm J, true}=20.3\,$pc and $f_{\rm PE,true}=0.08$. The fitting parameters $M$, $\rh$, $\Upsilon_V$ and $D$ are given in \autoref{results_table}.}
\label{tab:pes}
\centering
\begin{tabular}{l|ccccc}
\hline
Data  & $W_0$  & $B$ &  $\eta $  & $r_{\rm t}$ [pc] & $f_{\rm PE}$ \\ 
\hline
(1) Only LOS  &  $9.32_{-0.36}^{+0.39}$ & $0.75_{-0.15}^{+0.09}$ & $0.339_{-0.035}^{+0.035}$  &  $13.8_{-3.05}^{+4.87}$ & $0.040_{-0.008}^{+0.100}$ \\ 
 (2) No Gaia  &  $9.17_{-0.26}^{+0.28}$ & $0.80_{-0.08}^{+0.07}$ & $0.321_{-0.036}^{+0.028}$  &  $13.0^{+1.62}_{-1.42}$ & $0.037^{+0.011}_{-0.100}$
\\
(3) All  &  $10.18_{-0.31}^{+0.37}$ & $0.52_{-0.15}^{+0.12}$ & $0.370_{-0.019}^{+0.019}$  &  $12.2^{+0.8}_{-0.6}$ & $ 0.043^{+0.007}_{-0.070}$ \\
\hline
 \end{tabular}
\end{table*}

We can also consider single-mass models which include the effects of energetically unbound stars, such as the Spherical Potential Escapers Stitched models \citep[\spes;][]{claydon18}. It has been shown that due to the interplay of the internal cluster potential and the Galactic tidal potential, there is a geometric condition for escape 
\citep{Fukushige2000, Baumgardt2001}, which causes stars with energies above the escape energy to remain associated with the cluster. These stars are known as potential escapers (hereafter PEs) and can stay trapped within the cluster for many crossing times, which causes an elevation in the velocity dispersion profile at large radii and an extension of the surface brightness/density profiles \citep{Kupper2010, 2017MNRAS.466.3937C, 2017MNRAS.468.1453D}. The fraction of PEs in a cluster depends on its mass, how dynamically evolved the GC is, its orbital properties and the properties of the host galaxy \citet{2017MNRAS.466.3937C}. In the context of the present study, their inclusion in a DF-based model provides a way of dealing with the last bin of the mock {\it Gaia} PM dispersion profiles for which the {\sc limepy} models do not have a prescription. 

Similarly to {\sc limepy}, \spes\ models have a parameter for the central potential $W_{0}$ and 
a truncation parameter $B$, which varies the sharpness of the truncation but with a non-zero density at the critical energy (for $B>0$), allowing the unbound population to be included continuously. \spes\ models are isotropic and therefore have no anisotropy parameter, but there is a third parameter $\eta$, which sets the width of the energy distribution of the unbound stars and therefore the ratio of the velocity dispersion at the edge of the model to (approximately) the dispersion in the centre. We also fit on a mass scale $M$ and a radial scale. Because the model can be solved beyond $r_{\rm t}$, we use $r_{\rm t}$ as the radial scale and as a fitting parameter and solve the model until the last data point, which for this $N$-body snapshot corresponds to $2r_{\rm J}$. 

We project the model in the same way as detailed in \citet{2015MNRAS.454..576G}, and when fitting on the surface brightness profile we also fit on the mass-to-light ratio $\Upsilon_V =M/L_V$ and a nuisance parameter that captures the uncertainty in the surface brightness $\sigma_\mu$. 
The fitting procedure is performed using the binned data and using \textsc{emcee} \citep{ForemanMackey2013}, as well as uniform priors are adopted for all parameters.

Results for the best-fit model parameters and the properties of the PEs are given in \autoref{tab:pes}, including $r_{\rm t}$ and  the inferred fraction of mass in PEs, $f_{\rm PE}$. 
The fact that the recovered $f_{\rm PE}$ is about half the true value is most likely a systematic problem in the model rather than an observational bias: $f_{\rm PE}$ is also a factor 2 or 3 too low when the model is compared to $N$-body data without observational uncertainties \citep{claydon18}. This is likely due to the assumption of spherical symmetry and the absence of a detailed treatment of the Galactic tidal potential in the model  \citep{claydon18}. 
\autoref{results_table} shows the best-fit values of $M$, $r_{\rm h}$, $\Upsilon_V$ and $D$ (when applicable) for the three different dataset combinations considered. \autoref{fig:ian} shows the best-fit model compared to the mock data profiles for case 3. The recovered values for $M$ and $\rh$ for the different data sets are similar to the results of the single-mass {\sc limepy} models discussed in \autoref{section:single-mass_limepy}. The accuracy of the inferred cluster mass $M$ improves when adding PMs, and the best-fit value is consistent with the true one within $1\sigma$ only when including {\it Gaia} PMs. The recovered value of $r_{\rm h}$ is also improved when adding PMs, but it is still underestimated, even with {\it Gaia} PMs, which is probably a systematic bias as the result of the fact that mass segregation is not included in the {\sc spes} models. The $\Upsilon_V$ and $D$ are better fitted when including {\it Gaia} PMs as opposed to just {\it HST}, but $\Upsilon_V$ is underestimated and $D$ is overestimated by $2\sigma$ (case 2) or $1\sigma$ (case 3). These deviations from the real values may be due to a fundamental limitation of isotropic, single-mass models.

Note that if we were to fix $r_{\rm t}=r_{\rm J}$ the model would struggle to simultaneously fit the low density and flat velocity dispersion near $r_{\rm J}$, whereas with the smaller $r_{\rm t}\simeq0.6r_{\rm J}$ value reached by the best-fit model a low density can be obtained while keeping the velocity dispersion profile flat in the outer regions. This may be further improved with a multimass version of this model, in which the giant stars would have a higher central density (relative to the density at $r_{\rm t}$) and a lower central velocity dispersion (relative to the dispersion at $r_{\rm t}$) because of mass segregation. That said, the model tested in this section remains the only DF-based model able to successfully match the outermost {\it Gaia} data. 
However, an equally good representation of the kinematics in the cluster outskirts is also provided by Jeans models (see \autoref{Jeans_Watkins}).

\section{Multimass models}
\label{Sect_multimass}

\subsection{(C) 3-component {\sc limepy} DF}
\label{Sect_3-comp}

In this section, we explore simple DF-based three-component models. The DF is that of an isotropic, lowered isothermal model, as implemented in \limepy\ \citep{2015MNRAS.454..576G}, and we approximate the cluster by three components: (1) (dark) low-mass stars; (2) (visible) stars and (3) (dark) remnants. The $j$-th component (with $j$ = 1, 2, or 3) is specified by a mean mass $m_j$ and a total mass $M_j$. The masses $m_j$  are internally expressed in units of the global mean mass of the cluster\footnote{The way the mean mass is defined affects some of the model parameters. See the discussion in Section~2 of \citet{peuten17}. } and $M_j$ is internally converted to a normalised mass fraction. The velocity scale of each component $s_j$ relates to $m_j$ as  $s_j \propto m_j^{-1/2}$ to include the effect of (partial) equipartition \citep[for details see][]{1976ApJ...206..128D, 2015MNRAS.454..576G}. Although more mass components are required to accurately describe the mass function of GCs, the general dynamical behaviour of a mass segregated cluster should be captured by a simple three-component model. In these models, there are five free parameters associated with the mass function, the three $m_j$ values, and any two of the $M_j$ values. (Note that without the total mass $M$  as a fitting parameter, there are six parameters needed to fully specify the mass function). We exclude the most distant data point of the binned {\it Gaia} mock velocity dispersion profile from the fit, because we encountered a similar issue as with the single-mass models discussed in \autoref{section:single-mass_limepy}. Although this issue is not as severe as for the single-mass models, we found that including this outermost data point led to best-fit models that significantly overestimate the radial extent of the cluster and provide a poor fit to the outer surface brightness profile. As in \autoref{section:single-mass}, the fitting procedure is performed using the binned data and using the \textsc{emcee} MCMC code \citep{ForemanMackey2013}, with a nuisance parameter $\sigma_\mu$ for the unknown uncertainty in the surface brightness values, and adopting uniform priors for all parameters.

\subsubsection{Model~A from Gunn \& Griffin (1979)}
 \label{GG79_model}
 As a first attempt, we use model A from \citet{1979AJ.....84..752G}. Informed by a stellar IMF, they propose a three-component model of the form $m_j = [0.5, 1, 1.5]$ and $M_j=[5, 1, 0.1]$. With the mass function fully defined, and using \citet{1966AJ.....71...64K} models (i.e. $g=1$ in \limepy), the model has one free parameter, namely  the dimensionless central potential $W_0$. In addition to $W_0$, we fit on two physical scales (the total mass $M$ and the half-mass radius $\rh$) and on the distance $D$. We fit the surface density of the second mass component to the surface brightness profile, where we use the mass-to-light ratio $\Upsilon_V=M/L_V$ as an additional fitting parameter, and a nuisance parameters $\sigma_\mu$ for the uncertainty in the surface brightness profile, bringing the total of fitting parameters to 6. 
 We use the LOS velocities, {\it HST}-like and {\it Gaia}-like PMs.  The resulting physical scales are $M\simeq (8.6\pm0.7)\times10^4\,\msun$ (and $\Upsilon_V \simeq 2.3\pm0.3$) and $\rh\simeq 2.9\pm0.1\,$pc. Hence, despite giving precise results (approximately 8\% error in $M$),  $M$ is systematically overestimated ($\sim2.5\sigma$). We also find that the surface brightness profile is not well described by this model, reflected by a large $\sigma_\mu = 0.8\,$mag. This is likely because the mass function is not a good representation of the real mass function of this particular cluster. By having only a small amount of mass in the dark remnants, the visible stars in the model are too much segregated to the centre with respect to low-mass stars, leading to an inflated total mass and radius. Comparing these results to the single-mass model fits of the previous section, we see that using the wrong mass function can give systematic biases of similar magnitude to those found with the single component models. A multi-component model is therefore not necessarily a better model.

\subsubsection{Three-component model with two mass function parameters}

In this section we consider a three-component model with freedom to adjust the mass function, guided by the data. We fix $m_1 = 0.3$ (low-mass stars) and $m_2 = 0.8$ (visible stars), appropriate values for an old GC, and we assume that the visible stars are tracers in the potential, and set their mass fraction to 1\% (i.e. $f_2=M_2/M = 0.01$), which is a reasonable assumption because the tracer stars dominating the light (evolved stars and upper main sequence stars) indeed make a negligible fraction of the total cluster mass. The motivation to treat them as tracers is also to reduce the number of model parameters. This model has two remaining free parameters with which the mass function is fully defined: the mass of the remnants  (i.e. $m_3$) and the mass fraction in the remnants ($f_3$). Apart from $W_0$, we also leave the truncation parameter $g$ as a free parameter and again fit on $\Upsilon_V$ by comparing the projected mass density profile of the second component to the surface brightness profile. 

\begin{figure}
    \centering
        \includegraphics[width=\linewidth]{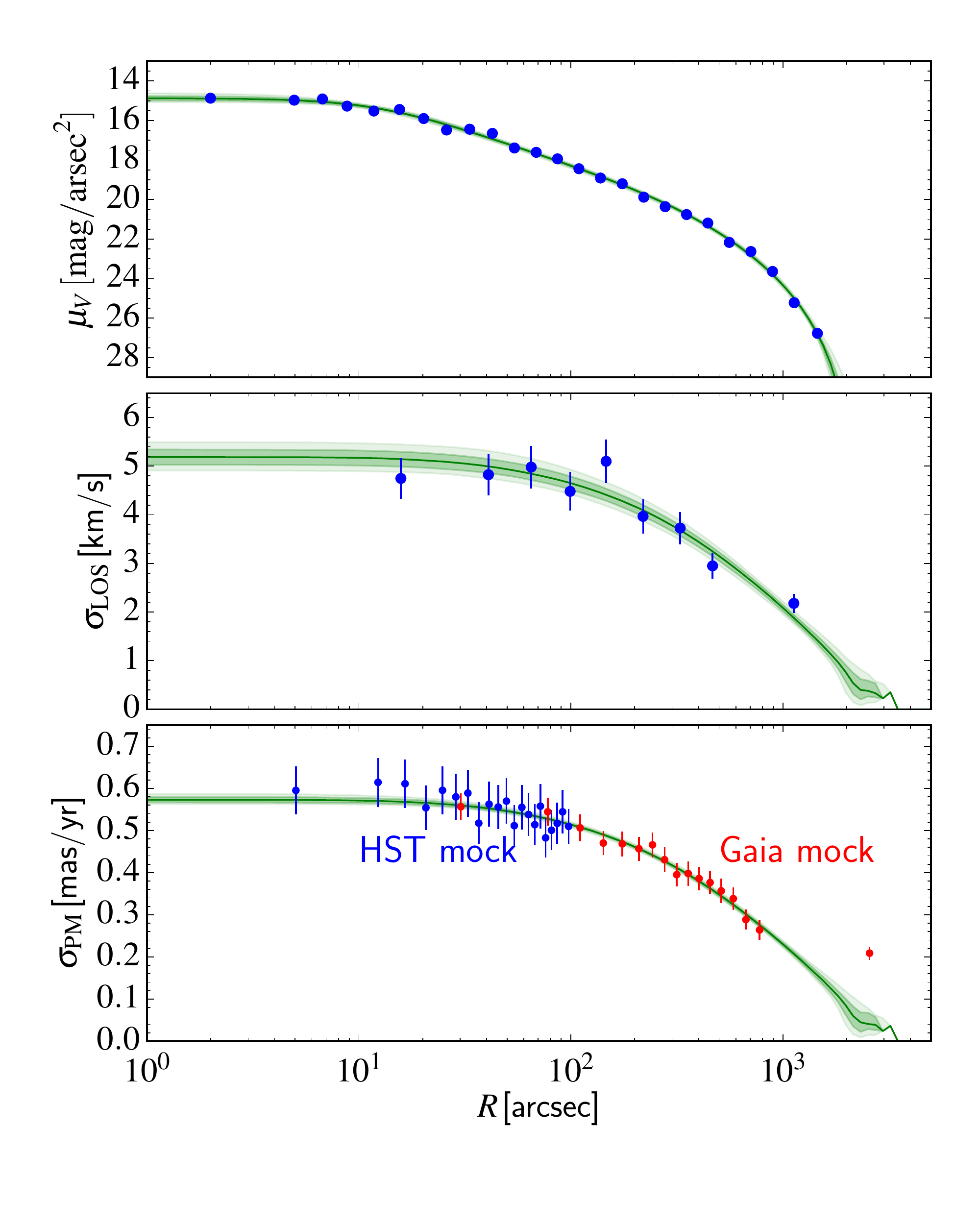}
    \caption{Fit of the three-component model from \autoref{Sect_3-comp} to the dataset of case 3 (fitting to surface brightness, LOS velocities, {\it HST} and {\it Gaia}-like PMs). The mock data for the surface brightness profile (top panel), LOS velocity dispersion profile (middle panel) and PM dispersion profile (bottom panel), are shown with blue filled circles, except the {\it Gaia} mock data shown in red. The most distant {\it Gaia} PM data point at $\sim3000$ arcsec was not included in the fit. Median values at each $R$ are shown with green lines, and the $1\sigma$ and $2\sigma$ contours are shown with dark and light green shaded regions, respectively. 
    }
    \label{fig:three_comp_rv_pm_gaia}
\end{figure}

The results of this model fitted to the full data set (case 3) are shown in \autoref{fig:three_comp_rv_pm_gaia} (the results are very similar for cases 1 and 2, so we do not show them here). Despite its relative simplicity, the model does a good job at describing the various mock profiles.  The true $M$ and $\rh$ are retrieved within $1\sigma$ (see \autoref{results_table}) when PMs are included. When using only LOS kinematics, the median $M$ is $1.5\sigma$ away from the true mass of the cluster.  

Results for model shape parameters and the properties of the third (dark remnant) component are given in \autoref{tab:three_comp}. 
From this we see that in the best-fit model, about $30\%$ of the mass is in the dark remnant component, which have a mass slightly higher ($\sim0.9\,\msun$) than that of visible stars ($0.8\,\msun$). The mass in the third component is lower than the mass fraction in remnants in the $N$-body snapshot ($\sim39\%$), which is probably due to the fact that $m_3$ is higher than the typical remnant mass ($0.67\,\msun$; dominated by white dwarfs).

A smaller value of $W_0$ is obtained for the 3-component model compared, for example, to single-mass models of \autoref{section:single-mass} and the multimass models of \autoref{Sect_multi_vincent}. As we noted above, the global mean mass is chosen as a reference mass to compute the \limepy \ models. $W_0$ represents the value of the dimensionless central potential for the mean mass group. The mean mass in the 3-component model is $\sim0.38\,\msun$, smaller than in the $N$-body snapshot and multimass models of \autoref{Sect_multi_vincent} where it is $0.5\,\msun$. Since lower-mass stars are less centrally concentrated, a smaller value of $W_0$ for the 3-component model is thus expected. We note that the $W_0$ of the second component (i.e. the visible stars) is $\sim 10$ for these models, very similar to the single-mass models, again as expected.

\begin{table} \renewcommand{\arraystretch}{1.2}\addtolength{\tabcolsep}{2pt}
\caption{Results of three-component model fits to different dataset combinations.  The fitting parameters $M$, $\rh$, $\Upsilon_V$ and $D$ are given in \autoref{results_table}.}
\label{tab:three_comp}
\centering
\begin{tabular}{lcccc}
\hline
 Data                    & $W_0$                          & $g$                                &  $m_3 $ & $f_3$              \\ \hline
 (1) Only LOS      &  $4.95_{-0.19}^{+0.22}$ & $1.20_{-0.17}^{+0.17}$ & $0.92_{-0.04}^{+0.05}$ & $0.26_{-0.09}^{+0.12}$\\
 (2) No Gaia         &  $4.88_{-0.23}^{+0.24}$ & $1.32_{-0.15}^{+0.16}$ & $0.94_{-0.04}^{+0.05}$ & $0.32_{-0.08}^{+0.10}$  \\
 (3) All                  &  $4.85_{-0.20}^{+0.31}$ & $1.29_{-0.17}^{+0.16}$ & $0.94_{-0.05}^{+0.05}$ & $0.31_{-0.06}^{+0.08}$  \\

  \hline
 \end{tabular}
\end{table}

\subsection{(D) Multimass {\sc limepy} DF}
\label{Sect_multi_vincent}

In this subsection and the following, we consider multi-component lowered isothermal models, as introduced in \autoref{Sect_3-comp}, but now including many more mass components to accurately describe the stellar mass function within GCs. In this case, we use as an additional observational constraint the mock stellar mass function of main sequence stars around the half-mass radius of the cluster (see \autoref{section:mock_data}). We want to determine whether these models can provide an improvement over simpler three-component models, and also to what extent they can recover information about the dark mass (low-mass stars and dark remnants) even if the kinematic and structural observational constraints are dominated by the bright visible stars. We examine two different approaches: the first one relies on the multi-component models implemented in \limepy\ \citep{2015MNRAS.454..576G} assuming power-law functional forms for the mass functions of visible stars and remnants, while the second approach in the next subsection uses the \citet{1979AJ.....84..752G} multimass DF, a flexible mass function for the visible component but a fixed prescription for the mass function of remnants \citep[as used for example by][]{2012ApJ...755..156S}. 

Here we consider more realistic \limepy\ models including more numerous mass bins. In addition to the five free parameters specifying the structural properties, anisotropy, and scales of the cluster ($W_0$, $g$, $r_a$, $M$, $r_{\rm h}$, defined as in previous sections), we adopt additional free parameters associated with the mass function. The mass function of visible stars is defined by a power law ($dN/dm \propto m^{-\alpha}$) with index $\alpha=\alpha_1$ for stars with $m<0.5 \ \msun$ and index $\alpha=\alpha_2$ for stars with $m>0.5 \ \msun$, with the second power-law truncated at the turn-off mass (0.85 M$_{\odot}$).  The mass function of dark remnants is specified by a power law with index $\alpha=\alpha_{\rm rem}$ spanning the mass range between the mass of the lowest-mass white dwarf and the mass of the most massive black hole predicted by the single-star evolution (SSE) package \citep[][]{2000MNRAS.315..543H} at the age and metallicity of the cluster ($\sim12$~Gyr, [Fe/H]$\sim -1$), corresponding to a range of 0.52 to 20.85 M$_{\odot}$ for an IMF extending to 100~M$_{\odot}$. 
Combined with the total cluster mass $M$, the four additional free parameters (power-law exponents $\alpha_1$, $\alpha_2$, $\alpha_{\rm rem}$, and the fraction of mass in dark remnants $M_{\rm dark}/M$) fully define the mass function, thus specifying all values of $m_j$ and $M_j$ needed to build the multimass model. We considered ten linearly spaced mass bins\footnote{See \citet{peuten17} for a discussion of the choice and minimum number of mass bins to ensure fast but stable solutions.} for main-sequence and evolved stars covering the range from 0.1~$\msun$ to the main-sequence turn-off mass. For the remnants, we define the edges of the mass bins such that we have eight remnant mass bins, three linearly spaced bins corresponding to white dwarfs, one bin for neutron stars, and four logarithmically spaced bins for black holes of different masses. The edges of the mass bins between the different types of stellar remnants are adopted as the masses at which SSE predicts a transition between white dwarfs and neutron stars, and between neutron stars and black holes. The values of $M_j$ (total mass in each bin) are computed according to the power law adopted for the mass function, assuming a continuous mass spectrum. Note that these are all related through the adopted single power-law mass function for the remnants, so the remnant mass bins are not independent of each other.

Part of the motivation for leaving freedom in the total mass fraction and mass distribution of dark remnants is that these are not known a priori and very uncertain, which can lead to large systematic uncertainties in the inferred cluster mass and mass profile \citep[e.g.][]{2012ApJ...755..156S}. We also want to explore how, with limited assumptions, the mass fraction and distribution of remnants can itself be constrained by multimass models.

We compare the multimass models introduced in this section to our mock data in a similar way to what was done with the models considered previously. However, the conversion of density profiles for different mass species to surface brightness profiles that can be compared to observations requires some extra attention because several mass components (with different spatial distributions) can contribute to the observed light and we need to specify their relative mass-to-light ratios. Here we couple our multimass models to predictions from the flexible stellar population synthesis (FSPS) models \citep{2009ApJ...699..486C, 2010ApJ...712..833C} based on Padova stellar isochrones \citep{2000A&AS..141..371G, 2007A&A...469..239M, 2008A&A...482..883M} to obtain the mass-to-light ratios as a function of age, metallicity, stellar mass and evolutionary stage, which we would need to compare our models with real observations. We fit on the shape of the surface brightness profile only (not its absolute values) so that the inferred total mass is controlled by the kinematics and not influenced by systematic differences between the absolute values of the mass-to-light ratios in the FSPS models and the $N$-body simulation used to generate mock data. The global mass-to-light ratio $\Upsilon_V$ is thus another free parameter. The kinematic mock data are compared to the model velocity dispersion values for tracer stars with the same mean mass as those of a given kinematic sample.

As in previous sections, the fitting procedure is performed on the binned data using the \textsc{emcee} MCMC code \citep{ForemanMackey2013}, with a nuisance parameter for the unknown uncertainty in the surface brightness values, and adopting uniform priors for all parameters (uniform in $\log{r_{\rm a}}$ for the anisotropy). 
We found that including the outermost data point of the {\it Gaia} PM dispersion profiles led to similar biases as those discussed in \autoref{Sect_3-comp} and we thus ignored these data in the fit.

\begin{figure*}
    \centering
        \includegraphics[width=\linewidth]{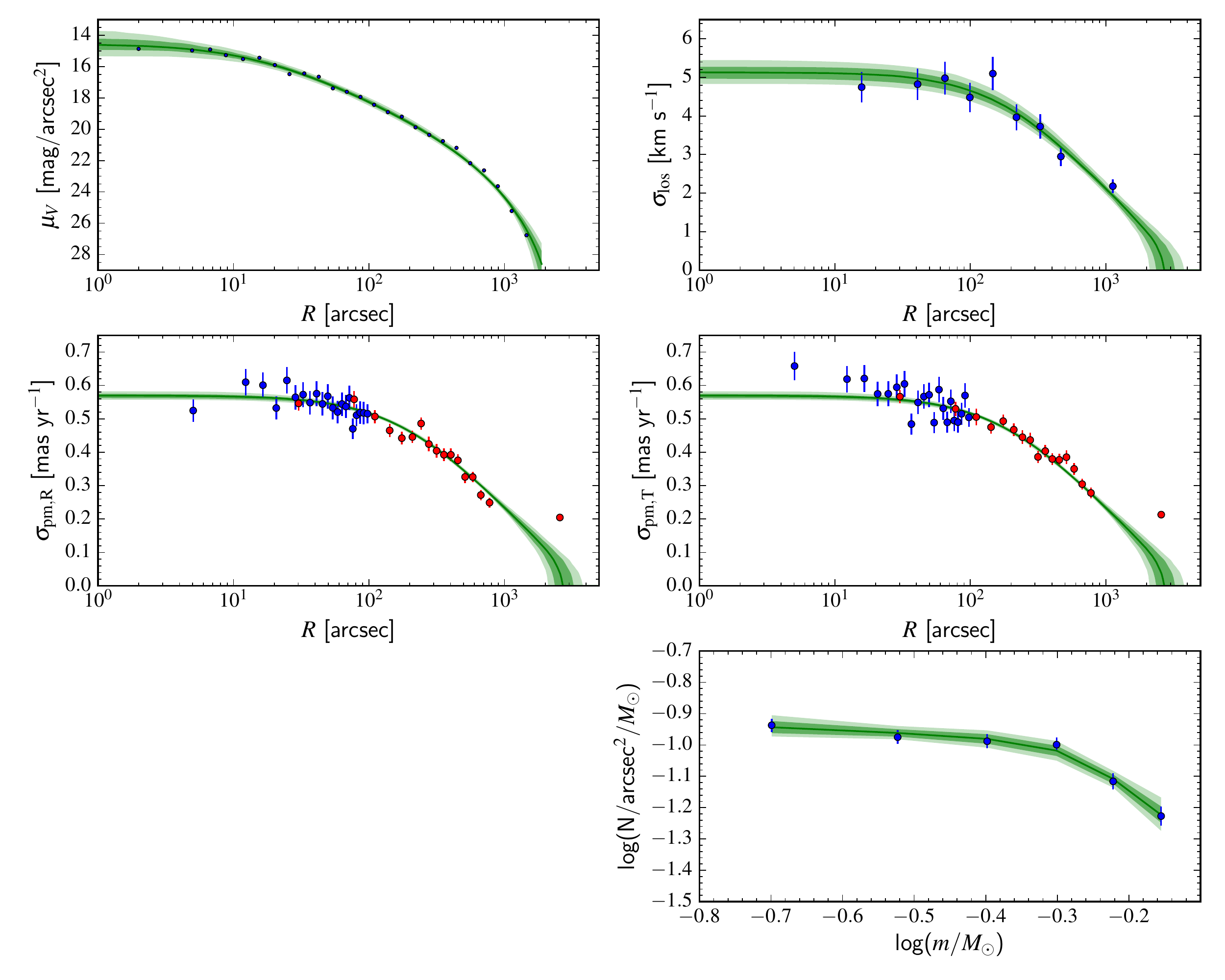}
    \caption{Fit of the multimass {\sc limepy} model from \autoref{Sect_multi_vincent} to the dataset of case 3 (fitting to surface brightness, LOS velocities, {\it HST} and {\it Gaia}-like PMs, and mass function in an annular region between 250 and 350$^{\prime\prime}$). The mock data for the surface brightness profile (upper left), LOS velocity dispersion profile (upper right), PM dispersion profiles (radial component in middle left, tangential component in middle right panel), and local stellar mass function near the half-mass radius (lower right panel) are shown with blue filled circles, except the {\it Gaia} mock data shown in red. The $1\sigma$ and $2\sigma$ contours on the best-fit model are shown with dark and light green shaded regions, respectively, and the median values are shown with green lines.}
    \label{fig:multi_limepy_rv_pm_gaia}
\end{figure*}

The results from fitting these multimass {\sc limepy} models to the full data set (case 3) are shown in \autoref{fig:multi_limepy_rv_pm_gaia} (the results are similar for cases 1 and 2, so we do not show the other fits). Again, the various mock profiles are well reproduced by the models, although the outermost data point on the {\it Gaia} PM dispersion profile cannot be matched by these DF-based models. 

The true $M$, $\Upsilon_V$, and $\rh$ are typically retrieved within $1-1.5\sigma$ (see \autoref{results_table}), but we notice some small systematic effects. 
These models tend to slightly overestimate the total mass (and thereby mass-to-light ratio), underestimate the half-mass radius, and overestimate the fraction of mass in dark remnants (although this last recovered quantity is always consistent with the true value within $1\sigma$). The systematic underestimation of the half-mass radius may be due to underestimating the mass in low-mass stars at large radius.

\begin{figure*}
    \centering
        \includegraphics[width=\linewidth]{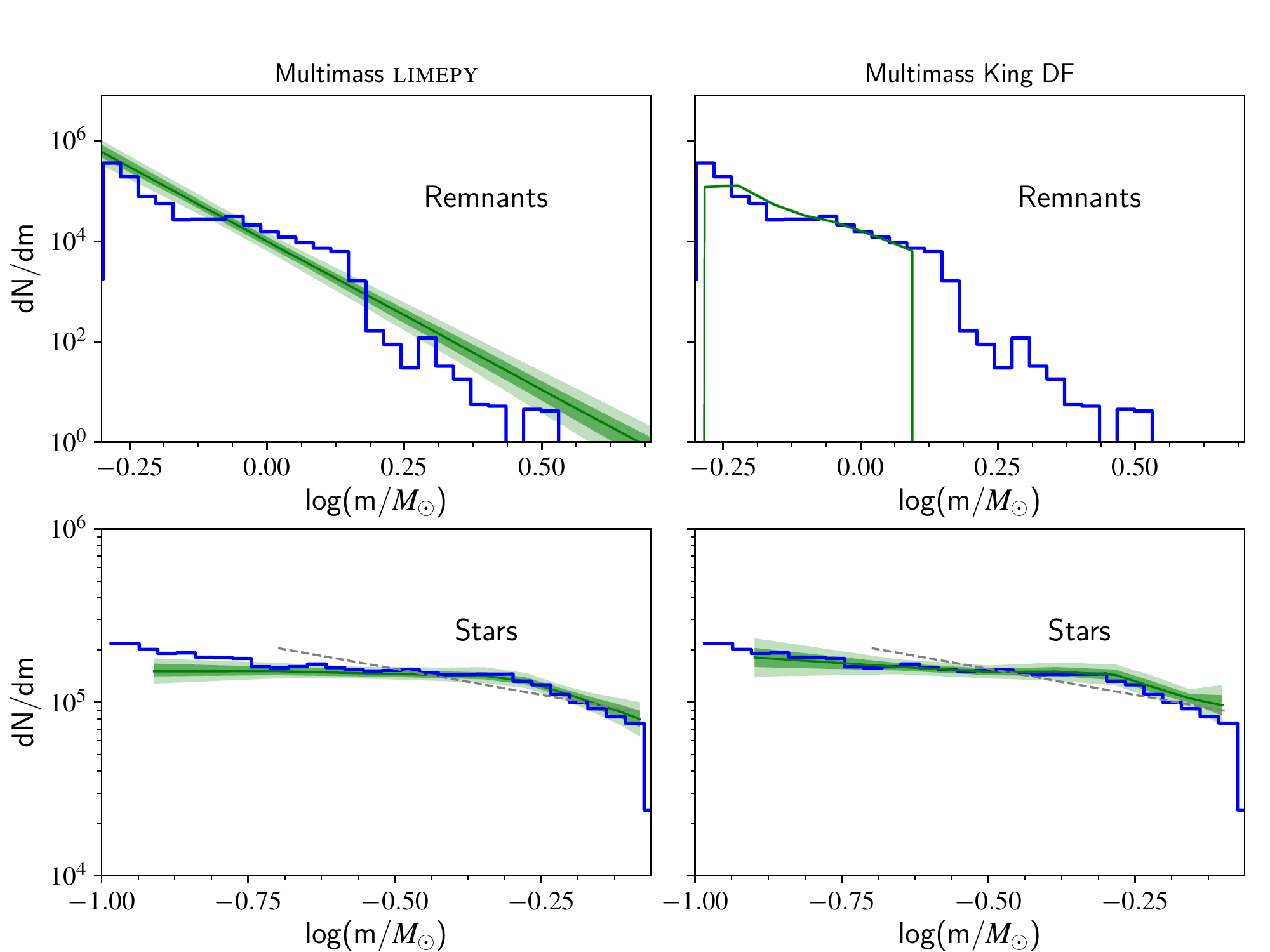}
    \caption{Inferred global mass function of dark remnants (upper panels) and visible stars (lower panels) from multimass model fits to the mock data from case~3. The results for the multimass {\sc limepy} models (\autoref{Sect_multi_vincent}) are shown in the left panels, while those from the multimass models of \autoref{Sect_multi_antonio} are shown in the right panels. The $1\sigma$ and $2\sigma$ contours are shown with dark and light green shaded regions, respectively, and the median values are shown with green lines. The true mass functions are shown with blue histograms. For comparison, we also show with a grey dashed line the best-fit global mass function (when describing the global mass function as single power-law in the range $0.2 < m < 0.8 \ \msun$) from the method of \autoref{Nbody_Holger} based on a grid of $N$-body models.}
    \label{fig:MF_rem}
\end{figure*}

Results for the model parameters, global mass function and mass distribution of dark remnants are given in \autoref{tab:multi_limepy}. We note that isotropic models are favoured (the best-fit anisotropy radius is much larger than the Jacobi radius of the model), 
as found for the single-mass models. It is interesting to see that our simple power-law parametrization for the mass function of dark remnants yields a successful recovery of the fraction of the total mass in remnants and a good approximation of their mass function, as can be seen from the top left panel of \autoref{fig:MF_rem}, where we overplot the inferred mass function on the true mass function from the $N$-body snapshot. In the bottom left panel of the same figure, we show that the global mass function of stars is also well recovered within uncertainties, except perhaps for a slight discrepancy at the very low-mass end (which comprises a very small fraction of the total cluster mass). The broken power-law we adopted is a good representation of the global mass function of the cluster.

\begin{table*} \renewcommand{\arraystretch}{1.4}\addtolength{\tabcolsep}{5.3pt} 
\caption{Results of multimass {\sc limepy} model fits to different dataset combinations, all including the mock local stellar mass function measurement. The fitting parameters $M$, $\rh$, $\Upsilon_V$ and $D$ are given in \autoref{results_table}. Note that when describing the global mass function of main-sequence stars in the $N$-body snapshot by a broken power law, it has values of $\alpha_1=0.19$ and $\alpha_2=1.17$.}
\label{tab:multi_limepy}
\centering
\begin{tabular}{lcccccc}
\hline
Data  & $W_0$                          & $g$ & log($r_a$/{\rm pc}) & $\alpha_1$ & $\alpha_2$ & $\alpha_{\rm rem}$   \\ \hline             
 (1) Only LOS (+ MF)     &  $7.00_{-0.61}^{+1.09}$ & $1.13_{-0.32}^{+0.30}$ & $3.63_{-1.17}^{+0.95}$ & $0.09_{-0.08}^{+0.08}$ & $1.15_{-0.25}^{+0.27}$ &  $6.35_{-0.54}^{+0.71}$ \\
 (2) No Gaia (+ MF)     &  $6.84_{-0.53}^{+0.77}$ & $1.43_{-0.28}^{+0.25}$ & $3.63_{-1.05}^{+0.94}$ & $0.07_{-0.08}^{+0.08}$ & $1.16_{-0.27}^{+0.26}$ &  $5.90_{-0.44}^{+0.50}$ \\
 (3) All (+ MF)     &  $6.91_{-0.49}^{+0.75}$ & $1.34_{-0.15}^{+0.16}$ & $3.86_{-0.95}^{+0.79}$ & $0.06_{-0.08}^{+0.08}$ & $1.12_{-0.25}^{+0.26}$ &  $5.93_{-0.31}^{+0.35}$ \\

  \hline
 \end{tabular}
\end{table*}

\subsection{(E) Multimass King DF}
\label{Sect_multi_antonio}

We now turn to the multimass modelling method used by \citet{2012ApJ...755..156S}. In the present case, the DF is given by the sum of the contributions of 12 different evenly-spaced mass groups from 0.1 to 1.3 $M_{\odot}$ each described by a Michie-King DF \citep{1963MNRAS.125..127M,1966AJ.....71...64K}:
\begin{equation}
f(E,L)=\sum_{j=1}^{12}f_{j}=\sum_{j=1}^{12}k_{j}\exp\left(\frac{A_{j}L^{2}}{\sigma_{\rm K}^{2}r_{\rm a}^{2}}\right)\exp\left(\frac{A_{j}E}{\sigma_{\rm K}^{2}}-1\right),
\label{King_DF}
\end{equation}
where $E$ and $L$ are the energy and angular momentum per unit mass, $r_{\rm a}$ is the radius at which orbits start to be radially biased, $\sigma_{\rm K}$ is a normalization term, the coefficients $A_{j}\propto m_{j}$ define the degree of mass segregation and the coefficients $k_{j}$ determine the MF. We refer to \autoref{appendix_King_DF} for the equations used to integrate the DF to obtain the density ($\rho_{j}$) and velocity dispersion profiles in both the radial and tangential components ($\sigma_{{\rm r},j}$, $\sigma_{{\rm t},j}$) of all mass species and projected on the plane of the sky ($\Sigma$, $\sigma_\mathrm{LOS}$, $\sigma_\mathrm{R}$, $\sigma_\mathrm{T}$).

The free parameters of this family of models are the central potential $W_{0}\equiv\Phi_{0}/\sigma_{K}^{2}$, the scale (core) radius $r_{c}\equiv \sqrt{9\sigma_{K}^{2}/4\pi  G \rho_{0}}$, and the coefficients $k_{j}$ which determine the MF \citep[see also]{1966AJ.....71...64K,1979AJ.....84..752G}.
Unlike the method described in \autoref{Sect_multi_vincent}, where the global mass function is parametrized by a broken power law, the shape of the mass function is free to vary in each of the 8 bins corresponding to visible stars, according to the values of the coefficients $k_j$. 
A population of remnants has been also simulated by passively evolving a \citet{2001MNRAS.322..231K} IMF defined between 0.1 M$_{\odot}$ and 8 M$_{\odot}$ using the prescriptions of \citet{2009A&A...507.1409K} and then retaining a number of remnants in each bin defined by
\begin{equation}
	N_{{\rm remn},j} = N_{{\rm remn},j}^\mathrm{ev} \frac{N_j^\mathrm{mod}} {N_j^\mathrm{IMF}}
\end{equation}
where $N_{j}^\mathrm{mod}$ and $N_{j}^\mathrm{IMF}$ are the number of visible stars adopted in the $j$-th mass bin by the model and that predicted by the \citet{2001MNRAS.322..231K} IMF, and $N_{\mathrm{remn},j}^\mathrm{ev}$ is the number of remnants after passive evolution in the same $j$-th mass bin. 
The upper limit of the adopted IMF assumes that the entire population of remnants is constituted by only white dwarfs. This simplification was initially motivated by the large natal kicks expected for neutron stars and the relatively rapid ejection of black holes expected from dynamical evolution. We note that there is recent theoretical and indirect observational evidence pointing to the possible retention of significant numbers of stellar-mass black holes to the present-day in GCs \citep[e.g.][]{2013MNRAS.432.2779B, 2015ApJ...800....9M, 2016MNRAS.462.2333P, 2018MNRAS.479.4652A, 2018MNRAS.478.1844A}, but an evolved cluster (given its present-day mass and mass function) like M\,4 modelled here is not expected to have retained a significant population of black holes. Indeed, only two black holes are still present in the snapshot studied, although the cluster has retained 130 neutron stars. The neutron stars make $\lesssim 0.3\%$ of the total cluster mass and their mean mass is only slightly larger than that of the most massive white dwarfs, so we do not expect that including them would result in a noticeable improvement of the fits.

The best-fit model has been calculated following the iterative procedure described in \citet{2012ApJ...755..156S}. Briefly, a first guess of the $k_j$ coefficients has been chosen and the values of $W_{0}$ and $r_\mathrm{c}$ providing the best fit to the number density profile have been searched. As a first guess, the coefficients corresponding to a \citet{2001MNRAS.322..231K} mass function have been adopted. The local mass function at the same projected radius as the mock observations has then been calculated and used to adjust the $k_{j}$ coefficients
\begin{equation}
k_{j}^{{\rm new}} = k_{j}^\mathrm{old} \left( \frac{N_{\mathrm{obs},j}}{N_{\mathrm{mod},j}} \right)^\eta
\end{equation}
where $N_{\mathrm{mod},j}$ and $N_{\mathrm{obs},j}$ are the predicted and observed number of particles (remnants excluded) in the $j$-th mass bin, and $\eta$ is a dampening factor set to 0.5 to avoid divergence. The above procedure is repeated until the parameters $W_{0}$, $r_\mathrm{c}$ and the coefficients $k_j$ start to fluctuate around equilibrium values.
The mass of the model is then normalized using a maximum-likelihood technique (see equation \ref{like_eq}). 
Note that since the shape of the mass profile (governed by the parameters $W_{0}$, $r_\mathrm{c}$ and by the mass function) is chosen using only the density profile and the mass function near the half-light radius, the derived half-mass radius does not depend on the adopted set of kinematical constraints with this method.
Different values of $r_\mathrm{a}$ have also been tested to account for different degrees of radial anisotropy. However, the isotropic model ($r_\mathrm{a}=\infty$) always provided the best fit to the data, as was found with the models of the previous sections.

Unlike what is done with {\sc limepy} models, where the distance is left as a free parameter throughout the fitting procedure, in the technique described here the distance was estimated (for cases 2 and 3) for any choice of $r_\mathrm{a}/r_{\rm c}$ at the end of each iteration cycle before normalization (the last step to derive the mass). 
For this purpose, the distance providing the proportion of dispersions in the three directions (LOS, radial and tangential; see eq. 6, 7 and 8) predicted by the considered anisotropy profile is chosen (e.g. in the case of the best-fit isotropic model, the distance providing the same dispersions in all projections was chosen). 

\begin{figure*}
\centering
\includegraphics[width=\linewidth]{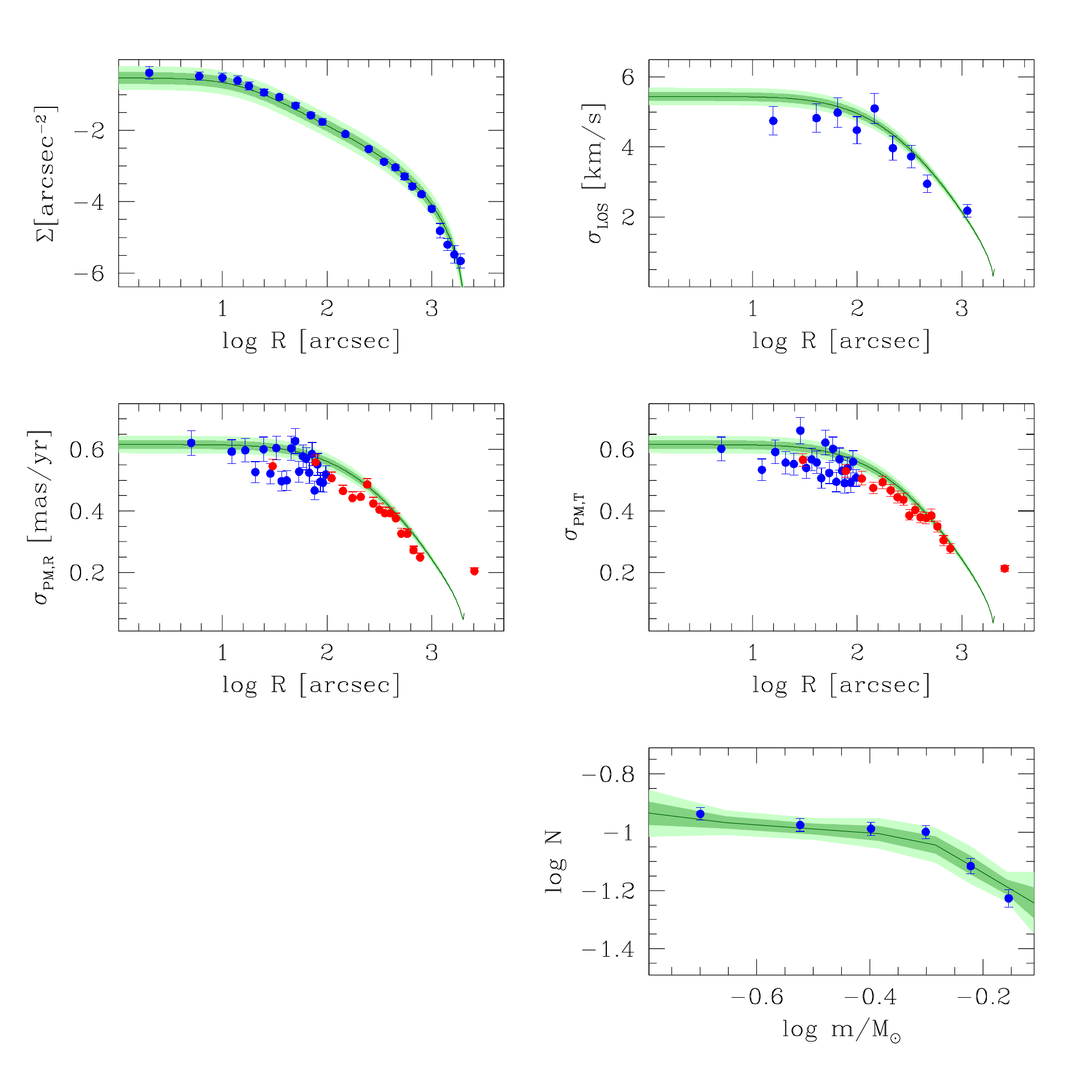}
\caption{
Fit of the multimass King DF model from \autoref{Sect_multi_antonio} to the dataset of case 3 (projected number density, LOS velocities, {\it HST} and {\it Gaia}-like PM dispersion profiles, and mass function in an annular region between 250 and 350$^{\prime\prime}$). The mock data for the projected number density profile (upper left), LOS velocity dispersion profile (upper right), PM dispersion profiles (radial component in middle left, tangential component in middle right panel), and local stellar mass function near the half-mass radius (lower right panel) are shown with blue filled circles, except the {\it Gaia} mock data shown in red. The $1\sigma$ and $2\sigma$ confidence regions for the best-fit model are indicated by the green shaded area in all panels, and the median values are shown with green lines.
}
\label{fig:mult_soll_prof}
\end{figure*}

The best-fit model to the mock data from case 3 (number density profile, LOS velocities, {\it HST} and {\it Gaia}-like PMs, with the addition of the mass function) is shown in \autoref{fig:mult_soll_prof}. The results are similar for the other cases considered and thus are not shown here. The various observables are once again well reproduced by the models, and as in the previous sections the outermost data point in the {\it Gaia} PM dispersion profile cannot be matched by these models.

As shown in \autoref{results_table}, the true $M$ and $\Upsilon_\mathrm{V}$ are recovered within $1\sigma$ in all cases, and $r_\mathrm{h}$ is systematically slightly underestimated (although within $\sim2\sigma$), as was found with the other flavour of multimass models in \autoref{Sect_multi_vincent}. The lower right panel of \autoref{fig:MF_rem} shows that the global mass function of stars is well recovered with this method. The upper right panel of that figure displays the adopted remnant mass function overplotted on the true remnant mass function. In the white dwarf regime, the adopted prescription for the mass function of remnants is a good match to the prescription used in the $N$-body model. The multimass models presented in this subsection however ignore the contribution of neutron stars and black holes, and they slightly underestimate the total mass in remnants.

\section{Jeans modelling}
\label{Sect_Jeans}

Another mass-modelling approach is the use of the Jeans equation to derive the global mass profile $M(<r)$ from the density $\nu$ and velocity dispersion
$\sigma_{r}$ profiles of a tracer population:
\begin{equation}
	\frac{1}{\nu} \frac{\mathrm{d} \nu \sigma_\mathrm{r}^2} {\mathrm{d} r} + 2 \beta \frac{\sigma_\mathrm{r}^2} {r} = -\frac{G M(<r)} {r^2},
	\label{jeans_eq}
\end{equation}
where
\begin{equation}
	\beta \equiv 1 - \frac{\sigma_\mathrm{t}^2} {2 \sigma_\mathrm{r}^2}
    \label{eqn:def_anisotropy}
\end{equation}
is the anisotropy profile.

The advantage of the above equation is that it links two observable quantities (density and velocity dispersion profiles) of an arbitrary sample population to a global quantity (the mass profile generating the gravitational potential) without any a priori assumption on the underlying distribution function. On the other hand, errors on the observational profiles are enhanced in the derivative calculation often leading to noisy and/or unphysical results. In the last decades this approach has been used by several authors on both GCs and dwarf galaxies \citep[e.g.][]{1995AJ....109..209G, 2013A&A...552A..49L, 2013MNRAS.428.3648I} using different techniques to minimize the noise in the derivatives involved in \autoref{jeans_eq}.

Here, we describe the results of two different Jeans-modelling approaches.

\subsection{(F) Variable M/L Jeans (Sollima)}
\label{Jeans_Sollima}

In this section we apply the technique described in \citet{2016MNRAS.462.1937S}. In particular, the stars brighter than the main-sequence turn-off (with $V<17$) have been used as the tracer population. For these stars both LOS velocities and PMs are obtainable with current observational facilities and with {\it Gaia}. The projected number density profile of this stellar population has been fitted with a sum of \textbf{27} Gaussian functions with different variances\footnote{Note that according to equation \ref{mgauss}, in this formalism the Gaussian functions are defined as a function of the 3D radius ($r$) and projected into the plane of the sky to be compared with the observed profile.}(see also \autoref{Jeans_Watkins}):
\begin{equation}
	\Sigma = \sqrt{2 \pi} \sum_{i} \nu_{i} s_{i} \exp \left( -\frac{r^2}{2 s_{i}^2} \right) .
\label{mgauss}
\end{equation}
The coefficients $\nu_{i}$ are used to compute the 3D density profile of the tracer population:
\begin{equation}
	\nu = \sum_{i} \nu_{i} \exp \left( -\frac{r^2} {2 s_{i}^2} \right) .
\end{equation}
The variances $s_{i}$ have been chosen to increase in logarithmic steps of 0.1 in the range $-1.6<\log(r/\mathrm{pc})<1$. A global density profile ($\rho_\mathrm{t}$) is iteratively updated 
by choosing appropriate coefficients $\nu_{i}$ in equation \ref{mgauss}, and equation \ref{jeans_eq} has been solved for $\sigma_\mathrm{r}^2$:
\begin{equation}
	\sigma_\mathrm{r}^2 = \frac{G (B_{\infty} - B_{r})} {\nu \exp \left( 2 \int \frac{\beta}{r'} \mathrm{d} r' \right)},
\end{equation}
where
\begin{equation}
	B_{r} = \int_0^r \frac{M \nu \exp\left( 2 \int \frac{\beta}{r''} \mathrm{d} r'' \right)}{r'^{2}} \mathrm{d} r' .
\end{equation}
For simplicity, the Osipkov-Merritt \citep{1979PAZh....5...77O, 1985AJ.....90.1027M} anisotropy profile
\begin{equation}
	\beta(r) = \frac{\left( {r}/{r_\mathrm{a}} \right)^2} {1 + \left( {r}/{r_\mathrm{a}} \right)^2}
\end{equation}
has been adopted, where $r_\mathrm{a}$ is the radius beyond which orbits become
significantly radially biased. The corresponding projections of the velocity moments onto the plane of the sky ($\sigma_\mathrm{LOS}$, $\sigma_\mathrm{R},$ and $\sigma_\mathrm{T}$) can then be computed (see \autoref{appendix_Jeans_Sollima}).

The best fit coefficients $\nu_\mathrm{i}$ have been chosen by maximizing the
log-likelihood
\begin{equation}
	\label{like_eq}
	\ln~\mathcal{L} = -\frac{1}{2} \sum_{j} \sum_{k=\mathrm{LOS,R,T}} \left( \frac{(v_{{k},j} - \overline{v_{k}})^2} {\sigma_{k}^2 + \delta v_{k,j}^2} + \ln \left( \sigma_{k}^2 + \delta
v_{k,j}^2 \right) \right)
\end{equation}
using a Markov Chain Monte Carlo algorithm. In the above equation the terms $\delta v_{k,j}^{2}$ are the uncertainties in the k-th velocity component of the j-th star. Only positive coefficients have been chosen to ensure a monotonically decreasing profile. The positivity of the distribution function across the energy domain has been tested using the \citet{1916MNRAS..76..572E} formula \citep[see][]{2016MNRAS.462.1937S}.

To convert the velocities in the plane of the sky from physical units into angular ones we adopted the true cluster distance in case 1, while for cases 2 and 3 we adopted the distance providing the same scaling factor between $\sigma_\mathrm{LOS}$, $\sigma_\mathrm{R}$ and $\sigma_\mathrm{T}$ in models and mock observations.

The surface brightness profile has been interpolated using equation \ref{mgauss} providing a 3D luminosity density profile. Errors have been calculated using a Monte Carlo algorithm in which, for each mock measurement (either LOS velocity or PM), a synthetic velocity has been simulated at the distance of the observed stars from the cluster centre with velocities randomly extracted from Gaussian functions with dispersions given by the convolution of the local intrinsic dispersion and the individual velocity error of the selected star. The same algorithm applied to the observations has then been applied to the 100 synthetic samples thus estimating the uncertainties on all inferred quantities. For all three considered sets of constraints, the best fit has been obtained by assuming very large values of $r_\mathrm{a}$, corresponding to isotropic models. 

\begin{figure}
    \centering
        \includegraphics[width=\linewidth]{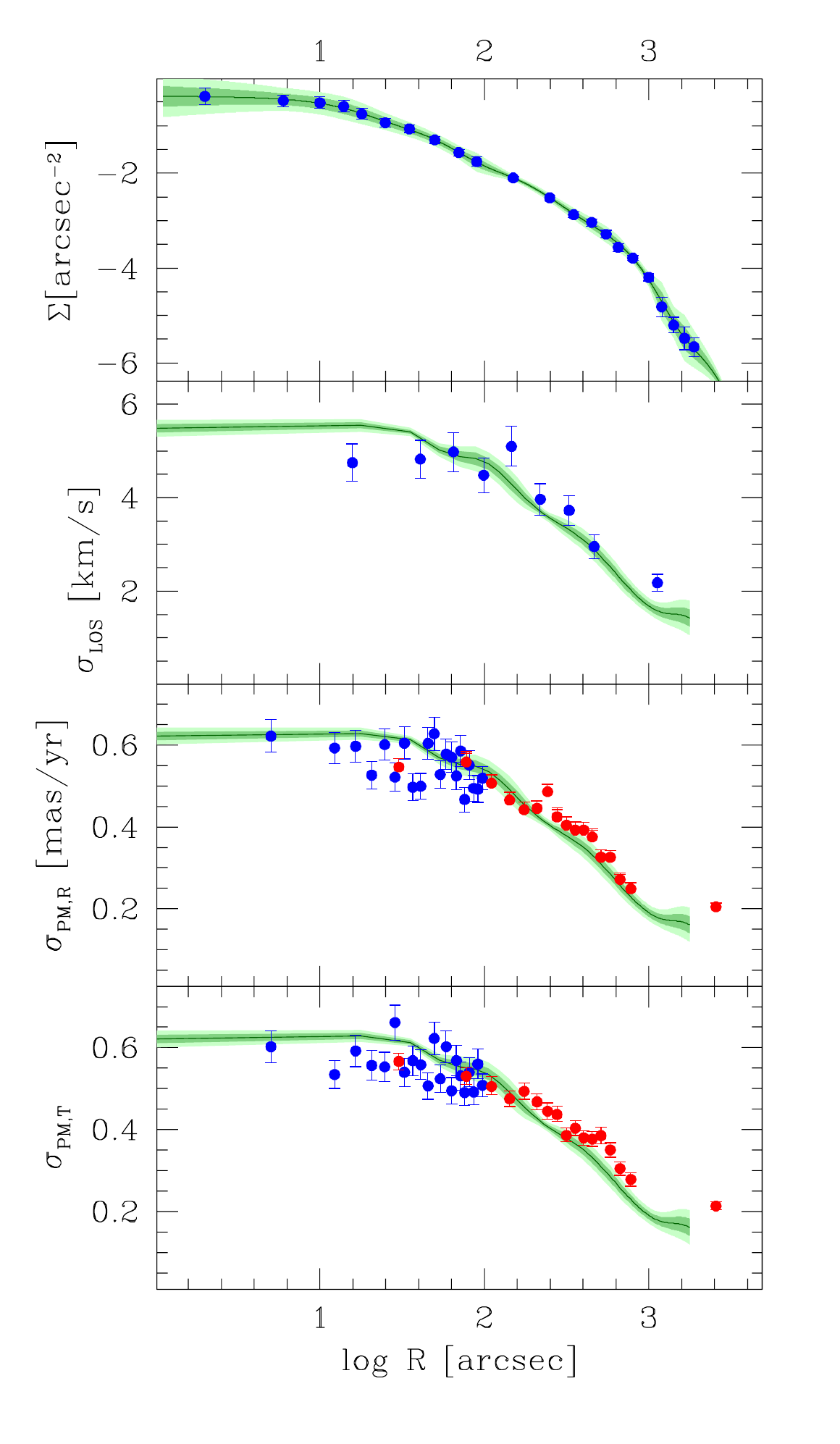}
    \caption{Best-fit Jeans model (using the technique by \citealt{2016MNRAS.462.1937S}; \autoref{Jeans_Sollima}) to the projected number density, LOS velocity, projected radial and tangential PM dispersion profiles (from top to bottom, respectively) to the dataset from case 3. The 1$\sigma$ confidence level is indicated by the green shaded area in all panels.}
    \label{fig:jeans_soll_prof}
\end{figure}

The best fit model to the full dataset (case 3) is shown in \autoref{fig:jeans_soll_prof}. Again, all the mock profiles are well reproduced by the model. The Jeans model, which unlike the truncated DF-based models presented in the previous sections does not impose a priori restrictions on the shape of the outer profiles, is better capturing the flattening of the velocity dispersion profile seen in the {\it Gaia}-like PMs. The $M$, $\Upsilon_\mathrm{V}$, and $\rh$ of the cluster (\autoref{results_table}) are all recovered within $1\sigma$. Note however that the uncertainties on these quantities are large and much more conservative than was found with DF-based models. This can be attributed to the flexibility of the Jeans model and to the scarcity of data tracing the mass in the outermost parts of the cluster. With the mass distribution in the outskirts poorly constrained by the data and not constrained or restricted by an imposed distribution function, global quantities like $M$ and $\rh$ become less precisely constrained. In that sense, the recovered cumulative mass or mass-to-light ratio profiles (\autoref{comparison}) are perhaps a fairer way to assess the performance of Jeans models than these global quantities.

\subsection{(G) Variable M/L Jeans (JAM)}
\label{Jeans_Watkins}


In this section, we use the spherical Jeans Anisotropy Multi-Gaussian Expansion (JAM, MGE) models developed in \citet{Cappellari2008} and extended in \citet{Cappellari2015}. These models assume that the cluster is spherically symmetric and not rotating\footnote{For cylindrically-symmetric models that can optionally include rotation, we would turn to the axisymmetric JAM models \citep{Cappellari2008, Watkins2013, 2012arXiv1211.7009C}.}, which preliminary analysis of the available mock data suggests is reasonable for this cluster, but do allow for velocity anisotropy. 

Many aspects of the method presented in this section are similar to the Jeans modelling approach presented in \autoref{Jeans_Sollima}. We thus refer to \autoref{appendix_JAM} for details of how the JAM models are computed and fitted to data, and simply summarize here some of the key differences and assumptions with this method.

The tracer density profile $\nu(r)$ and mass density $\rho(r)$ of the cluster are again both provided in the form of a Multi-Gaussian Expansion (MGE), but unlike the models of \autoref{Jeans_Sollima}, we use only 6 Gaussians for the JAM models. Instead of assuming an Osipkov-Merritt anisotropy profile, anisotropy is provided here per Gaussian component of the tracer MGE and allowed to vary non-parametrically throughout the cluster. To capture the effect of the Galactic tidal field, we further include a contaminant population that is dominant at large radii from the cluster centre (see \autoref{appendix_JAM}). The models are also fitted to the data discretely, not using the binned profiles.



\begin{figure}
    \centering
	\includegraphics[width=\linewidth]{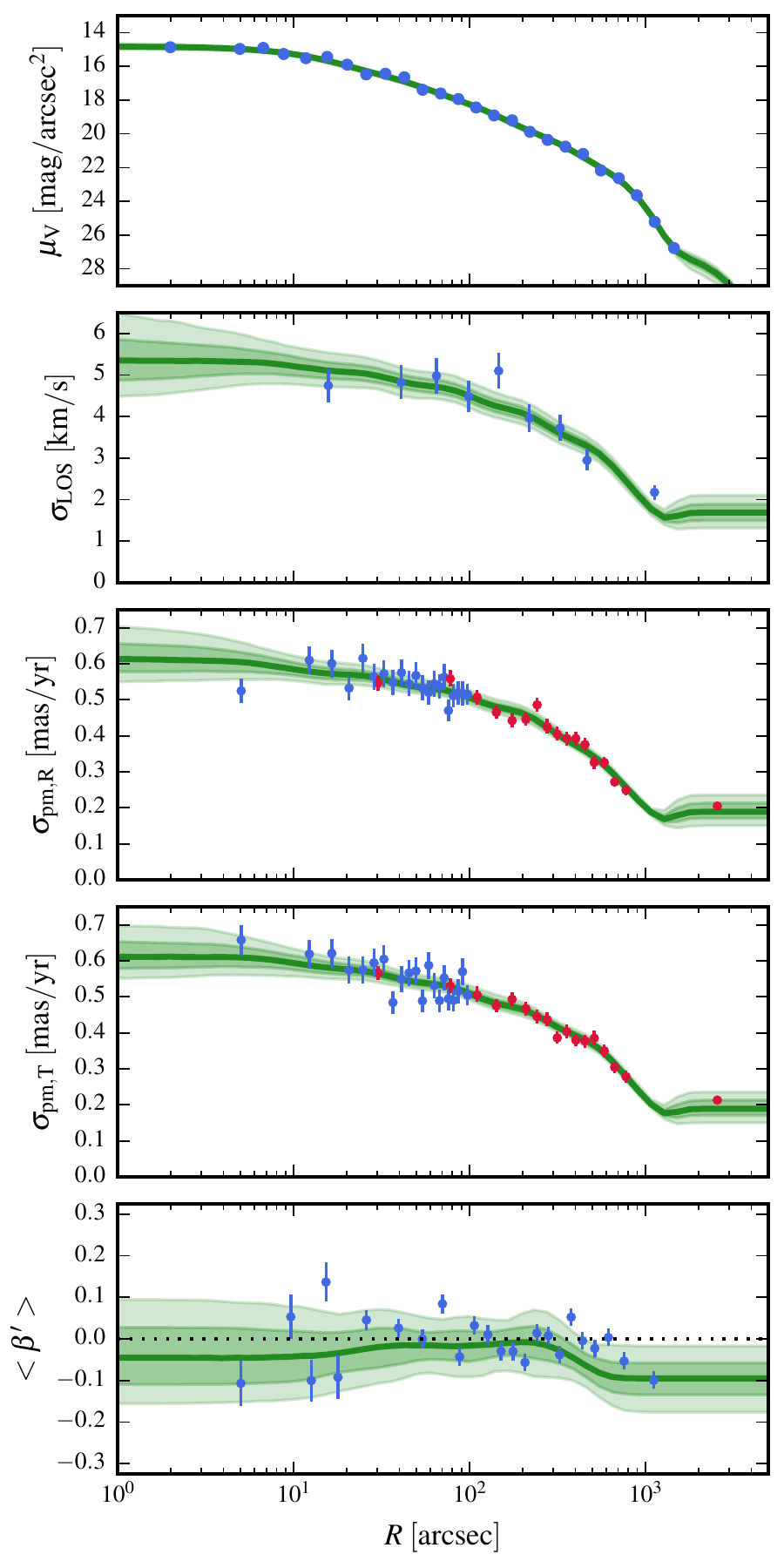}
    \caption{From top to bottom: model fits to the surface brightness, projected LOS velocity dispersion, projected radial PM dispersion, projected tangential PM dispersion, and anisotropy profiles for case 3 using the Discrete JAM method (\autoref{Jeans_Watkins}). The dark green lines show the medians of the profiles, the darker shaded regions encompass the 1$\sigma$ confidence limits and the lighter shaded regions encompass the 2$\sigma$ confidence limits. The blue and red points show the profiles calculated from the simulated data. The inclusion of a contaminant population is seen as the outer bump in the surface brightness profile and the flattening of the velocity dispersion profiles at large radii.}
    \label{figure:laura_fits}
\end{figure}

The surface brightness, velocity dispersion, and anisotropy fits for case 3 are shown in \autoref{figure:laura_fits}. We draw 1000 samples from the final posterior distribution of the fit parameters and then use these parameters to calculate the corresponding profiles. The dark green lines show the medians of the profiles, the darker shaded regions encompass the 15.9 through 84.1 percentiles (effectively the 1$\sigma$ confidence regions) and the lighter shaded regions encompass the 2.3 through 97.7 percentiles (effectively the 2$\sigma$ confidence regions). In the upper panel, the blue points show the surface brightness profile calculated from the simulated data, to which we fit directly. In the middle three panels, the blue and red points show the binned dispersion profiles calculated from the simulated data, however, note that we fit to the measurements for the individual stars and not to these binned dispersion profiles. The binned dispersion profiles are shown here for comparison purposes only.

The utility in the discrete fits is apparent with our inclusion here of a contaminant population. Most other methods were forced to remove the outermost point in the binned PM dispersion profile as the models were not able to fit this flattening in the profile. Our discrete approach allows us to include a contaminant population to capture these outer stars instead of removing them entirely, and furthermore, allows the models themselves to determine which stars are more likely to be cluster members and which are more likely to be contaminants rather than us having to decide at the outset which stars to keep or exclude.

The contaminant population is seen as the flattening of the velocity dispersion profiles at large radii, which nicely fits the outermost data point. It also gives the outer bump in the surface brightness profile.\footnote{This outer shape is somewhat arbitrary; the contaminant density enters primarily via the probability of cluster membership calculated in equation \ref{equation:Pcluster} and so for large radii $P_\mathrm{cluster} = 0$ (i.e. $P_\mathrm{contam} = 1$ for constant background density $\nu_\mathrm{contam}$ and any Gaussian width $s_\mathrm{contam}$). This motivated our choice to fix the width of the contaminant Gaussian.}

Another feature of these models is the anisotropy profile that we were able to recover from the fits, as shown in the lower panel of \autoref{figure:laura_fits}, where isotropy $\beta' = 0$ (see equation \ref{equation:modified_beta}) is marked as the dotted line. To calculate the `true' anisotropy profile for this comparison, we selected all stars in the snapshot brighter than 17.5~mag (the adopted faint limit for our \textit{HST} mock proper motion data) but with no radial limit; these were then binned in radius and the spherical anisotropy was calculated using equation \ref{eqn:def_anisotropy}, where $\sigma_\mathrm{t}^2 = \frac{1}{2} \left( \sigma_\theta^2 + \sigma_\phi^2 \right)$. Again, we did not fit directly to these binned data points, they are shown for comparison only.

We see that the cluster is close to isotropic, so in this case assuming isotropy would not be disastrous, as can be seen by the successful fits from other methods that assume isotropy. Nevertheless, we do see some mild deviations from isotropy, particularly in the outer regions of the cluster that our models are able to capture. Tangential anisotropy is not captured by any of the models of the previous sections, which all allow only for radially-biased anisotropy or isotropy.

The inferred $M$, $\Upsilon_\mathrm{V}$, and $\rh$ (\autoref{results_table}) are all recovered well within $1\sigma$ with this method. The uncertainties on these values are rather conservative, especially when compared to the DF-based models, for the same reasons as mentioned in \autoref{Jeans_Sollima} for the other flavour of Jeans models.

\section{Grid of $N$-body models}
\label{Nbody_Holger}

In the approach presented in this section, the cluster properties and best-fit profiles were determined by comparing the mock data (dispersion profiles, surface brightness profile, stellar mass function) to a grid of $\sim$1000  scaled $N$-body simulations, as described in \citet{2017MNRAS.464.2174B} and \citet{2018MNRAS.478.1520B}. These are $N$-body simulations of isolated star clusters containing $N=10^5$ stars initially and ran with the GPU-enabled version of the collisional $N$-body code NBODY6 \citep{1999PASP..111.1333A, 2012MNRAS.424..545N}.

The clusters initially follow a \citet{1962AJ.....67..471K} density profile, with concentrations spanning the range $0.2 \leq c \leq 2.5$ and initial half-mass radii in the range $2 \leq r_{\rm h} \leq 35$~pc. The simulations were run up to an age of 13.5 Gyr, but the cluster models considered in this work were calculated from snapshots centered around the age of the mock M\,4 ($\simeq 12$ Gyr). To account for the preferential removal of low-mass stars, simulations were run with a Kroupa IMF as well as IMFs depleted in low-mass stars in order to be able to match the mock mass function and more accurately estimate the cluster parameters. The adopted IMFs are defined as a combination of five connected power-laws between 0.1 and 15 $\msun$ with breaks at 0.2, 0.5, 0.8, and 1.0 $\msun$. We considered 6 different sets of parameters defining the power-law slopes in different mass regimes and their mass limits \citep[for details see Table 1 of][]{2018MNRAS.478.1520B}. The parameters are chosen to describe a time sequence of the mass function of a star cluster starting with a Kroupa IMF in different stages of its dissolution. Due to the fact that the simulated clusters are isolated, the chosen IMFs remain more or less unchanged throughout the evolution, apart from the high-mass end where stellar evolution has turned stars into remnants. In the simulations, a 10\% retention fraction 
was assumed for the black holes and neutron stars. The remaining 90\% received kick velocities large enough that they immediately left the clusters. The clusters did not contain primordial binaries.

For each model, 10 snapshots in time spaced by 50 Myr were combined. The combined $N$-body snapshots were then scaled in radius to match the half-light radius of the mock cluster. During the scaling the mass of the cluster was increased such that the relaxation time of the cluster was constant. The velocities are also scaled following equation 2 from \citet{2017MNRAS.464.2174B}. Best-fitting models are obtained by linearly interpolating the grid of $N$-body simulations with varying initial cluster concentration, half-mass radius, and IMF to find the set of parameters that simultaneously provides the best match to all the mock data considered. To compare with the mock stellar mass function, the model clusters were projected onto the plane of the sky and stars were selected in the same area as the mock data.

\begin{figure*}
\centering
\includegraphics[width=\linewidth]{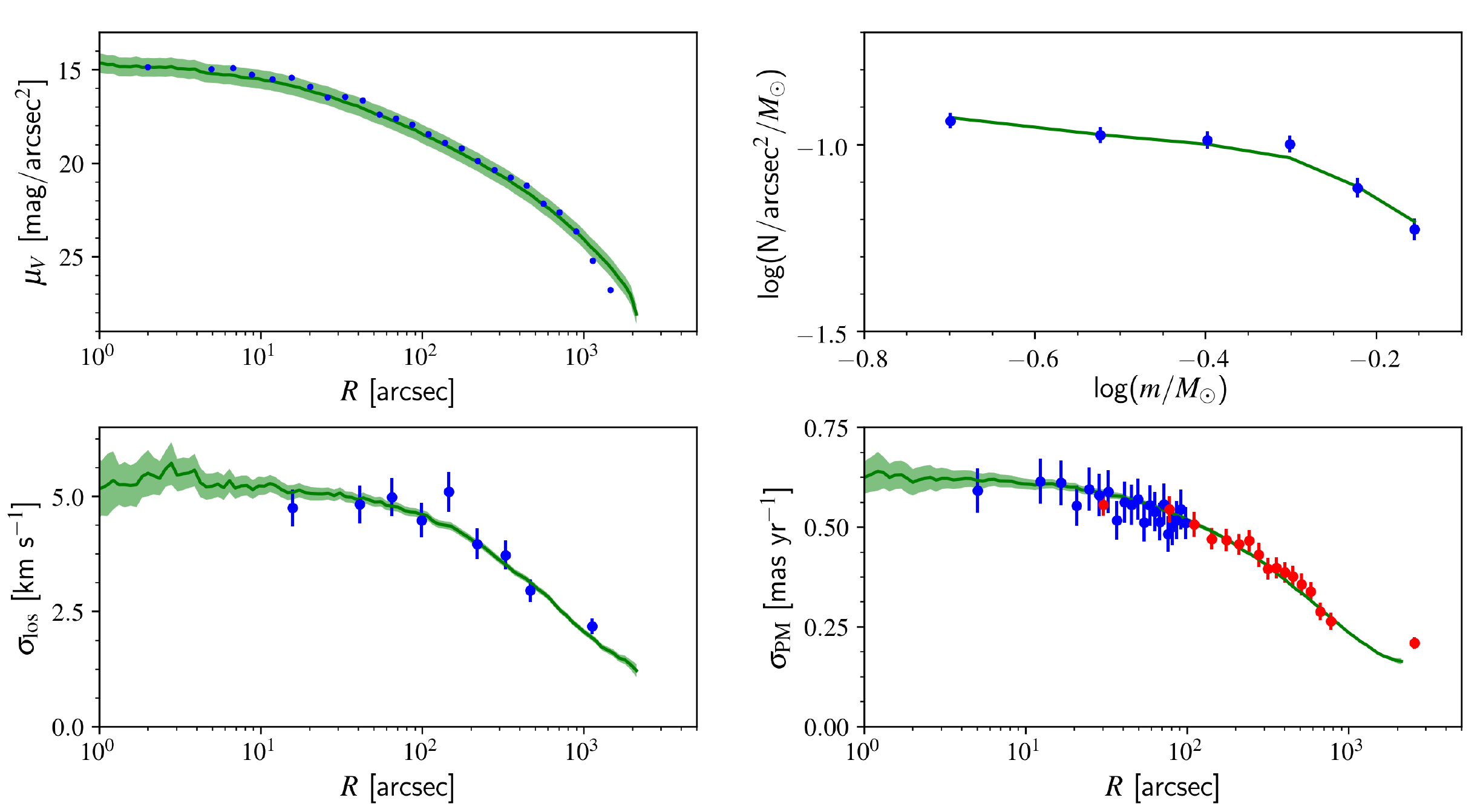}
\caption{
Fit of the grid of $N$-body models from \autoref{Nbody_Holger} to the dataset of case 3 (surface brightness, LOS velocities, {\it HST} and {\it Gaia}-like PM dispersion profiles, and mass function in an annular region between 250 and 350$^{\prime\prime}$). 
The mock data for the surface brightness profile (upper left), local stellar mass function (upper right panel), LOS velocity dispersion profile (lower left), and PM dispersion profile (lower right; $\sigma_{\rm PM} = (\sigma_{\rm PM,R}+\sigma_{\rm PM,T})/2$) are shown with blue filled circles, except the {\it Gaia} mock data shown in red. The $1\sigma$ confidence regions for the best-fit model are indicated (when available) by the green shaded area, and the median values are shown with green lines.
}
\label{fig:nbody_case3}
\end{figure*}

The results from fitting the grid of $N$-body models to the mock data (case 3) are shown in \autoref{fig:nbody_case3} (similar results were obtained for cases 1 and 2, so we do not show the other fits). The mock profiles are generally well reproduced by the models. The best-fit local mass function is not a perfect match to the mock data, which can be attributed to the limited flexibility of the current grid of $N$-body models regarding the choice of IMF, which is only varied along a single dissolution sequence. 

From the values reported in \autoref{results_table}, we see that the true $M$ is recovered within $1\sigma$ for cases 1 and 2, and $\sim 1.5\sigma$ away from the true value for case 3 ($\sim 1500\ \msun$ off). The method tends to systematically underestimate $\rh$ (with a $2-3 \sigma$ difference between the recovered and true value). $\Upsilon_{\rm V}$ is also underestimated, although only in case 2 is the bias in the recovered $\Upsilon_{\rm V}$ significantly larger than the statistical uncertainties. The dark remnant fraction is also recovered well, within uncertainties for all cases. Note that mass function uncertainties are not accounted for in the quoted uncertainties, but as mentioned above the current grid can vary the mass function only along a single sequence. This may explain some of the systematic differences noted above.

We find best-fit global mass functions with a power-law index $\alpha=-0.54$, $\alpha=-0.58$, and $\alpha=-0.65$ (when describing the global mass function as single power-law in the range $0.2 < m < 0.8 \ \msun$) respectively for cases 1, 2, and 3. Comparing this with the observed global mass function of stars and the best-fit broken-power law for the multimass models (see \autoref{fig:MF_rem}), we conclude that the $N$-body method also recovers the global mass function reasonably well.

We note that the fits in this section were performed by considering only the stars (in the models and observations) within $2000^{\prime\prime}$ in projection from the cluster centre, i.e. roughly within the Jacobi radius of the cluster. Because the method currently relies on a grid of isolated $N$-body models, it does not allow to capture the effect of tides on the dynamics of stars near the Jacobi radius and beyond. We indeed see from \autoref{fig:nbody_case3} that the best-fit model does not match the outermost PM data point. In practice, with real data, one could adopt a similar approach by estimating the Jacobi radius with an initial estimate of the cluster orbit (thanks to {\it Gaia}) and cluster mass.

\section{Comparison of the enclosed mass and mass-to-light ratio profiles}
\label{comparison}
\vspace{0.3cm}

The strengths, limitations and general performance of the different methods considered in this work are already becoming apparent from the previous sections and \autoref{results_table}, but here we push the comparison further by directly comparing the inferred enclosed mass and mass-to-light ratio profiles. 

\begin{figure*}
    \centering
	\includegraphics[width=0.48\linewidth]{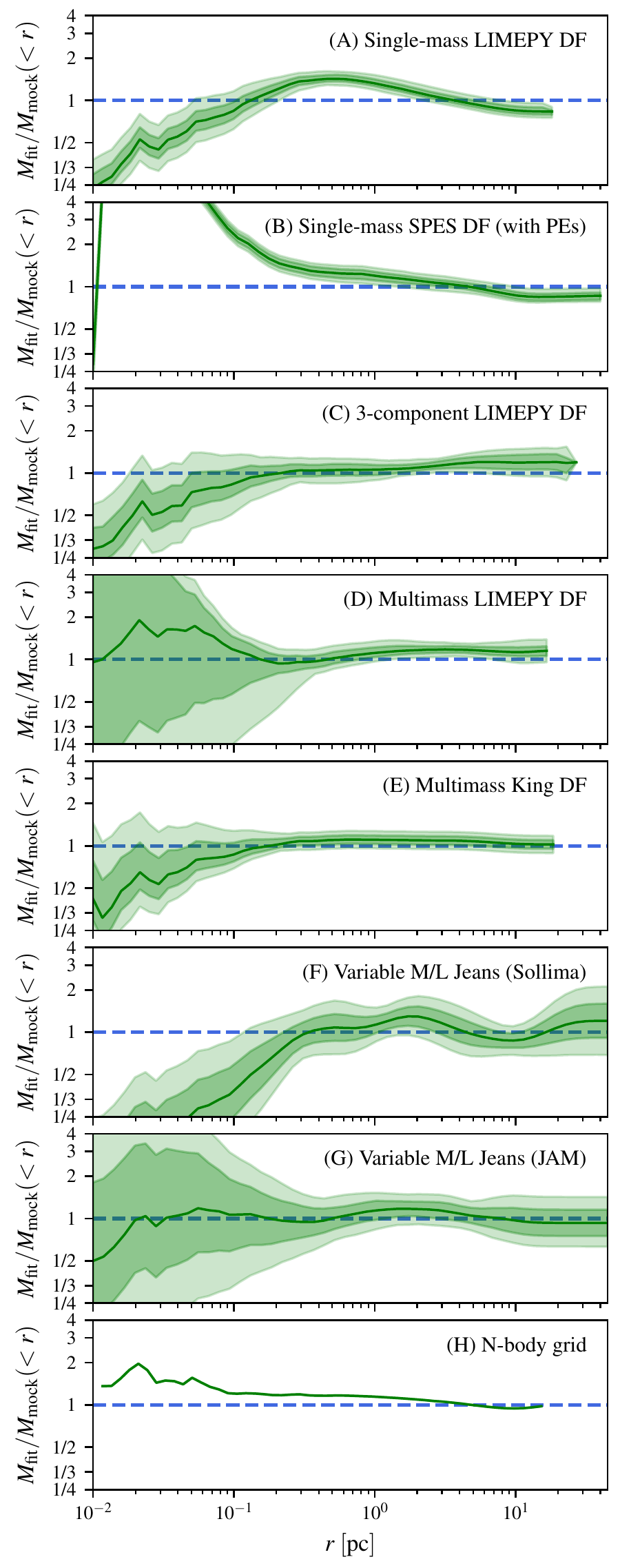}
    \includegraphics[width=0.48\linewidth]{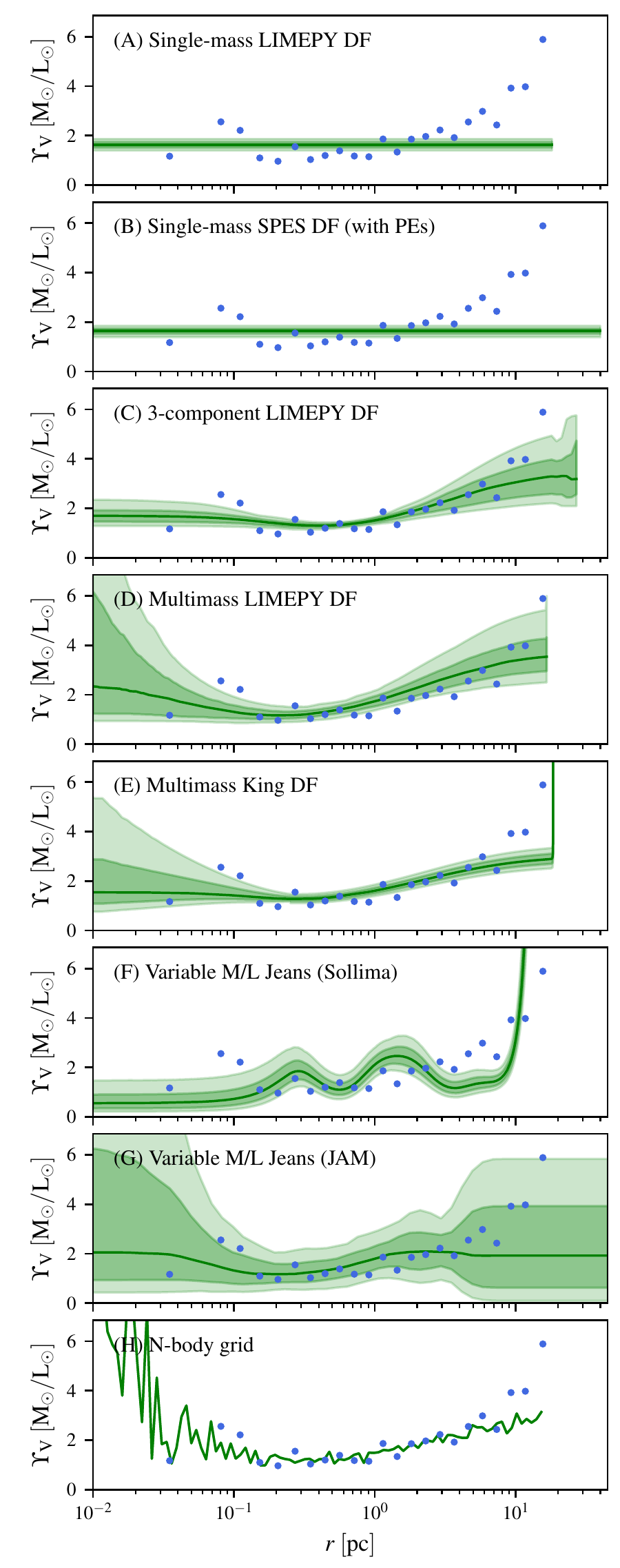}
    \caption{{\it Left:} Comparison of the inferred enclosed mass profiles in 3D (divided by the true enclosed mass profile) from the different methods for the mock dataset of case 1. {\it Right:} Comparison of the inferred 3D $V$-band mass-to-light profiles from the different methods for the mock dataset of case 1. The true mass-to-light ratio profile computed from the $N$-body model snapshot is represented by blue filled circles. Solid lines indicate the median, dark shaded green regions the $1\sigma$ contours, and light shaded green regions the $2\sigma$ contours (when available). 
    }
    \label{fig:comp_encmass_rv}
\end{figure*}

\begin{figure*}
    \centering
	\includegraphics[width=0.48\linewidth]{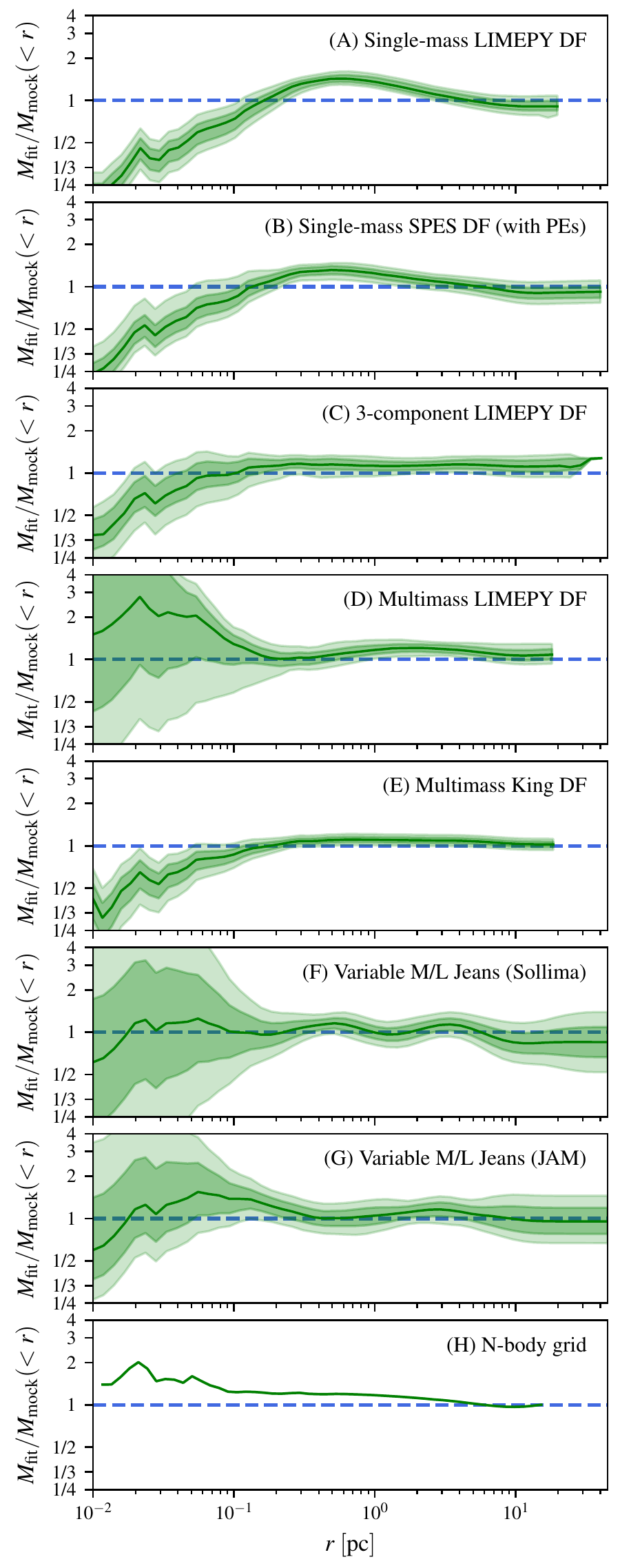}
    	\includegraphics[width=0.48\linewidth]{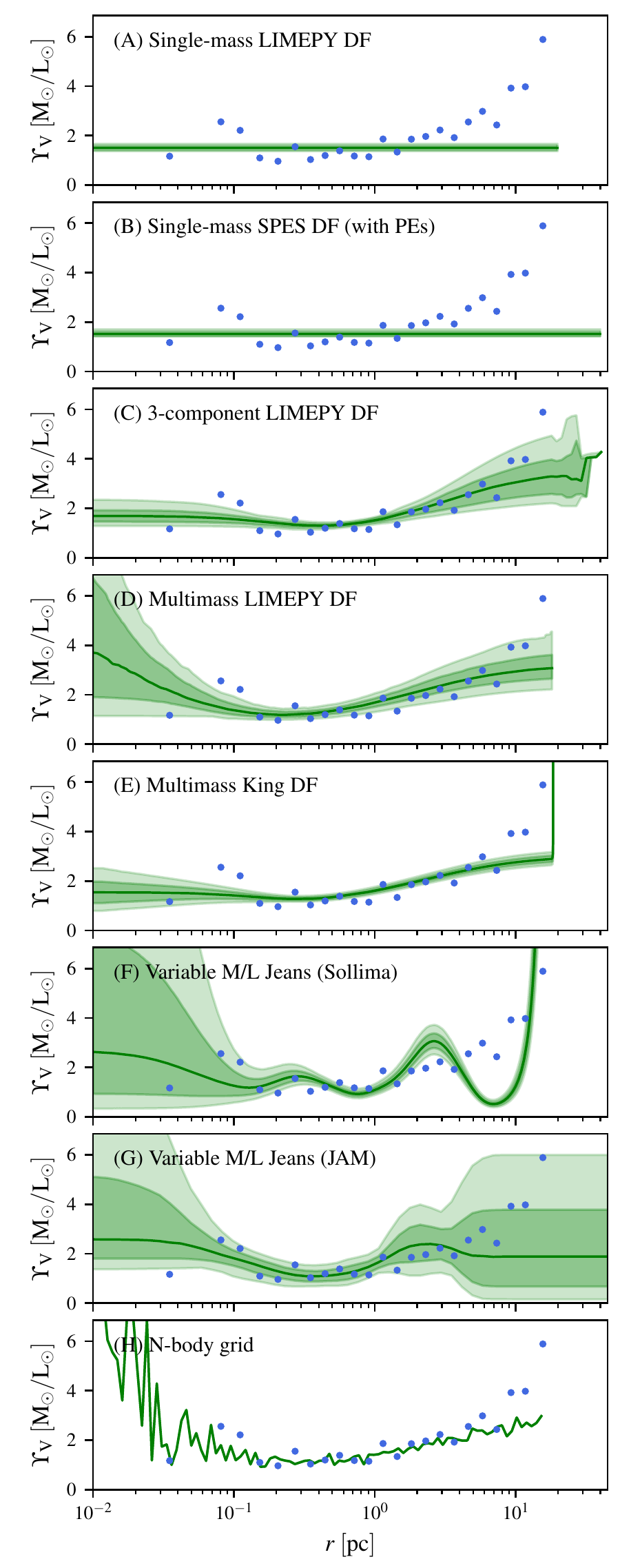}
    \caption{Same as Figure \ref{fig:comp_encmass_rv} but for the mock dataset of case 2. }
    \label{fig:comp_encmass_rv_hst}
\end{figure*}

\begin{figure*}
    \centering
	\includegraphics[width=0.48\linewidth]{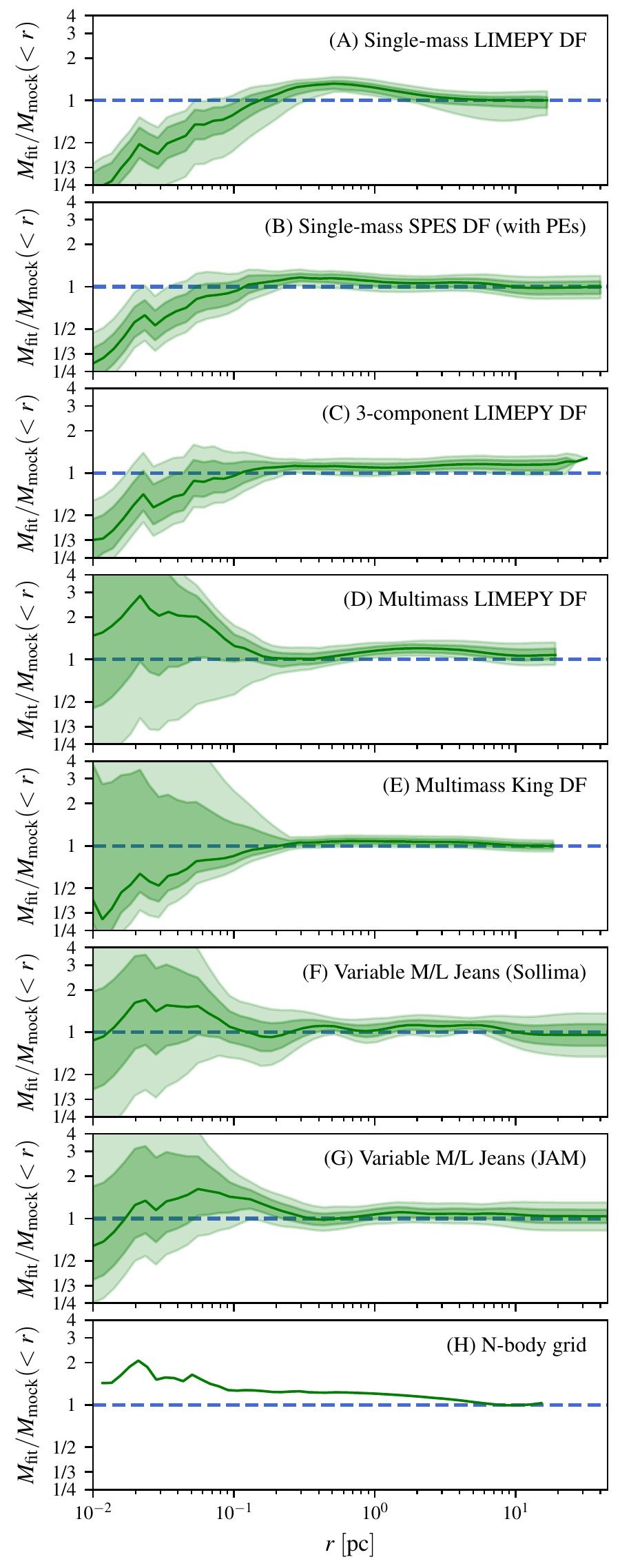}
	\includegraphics[width=0.48\linewidth]{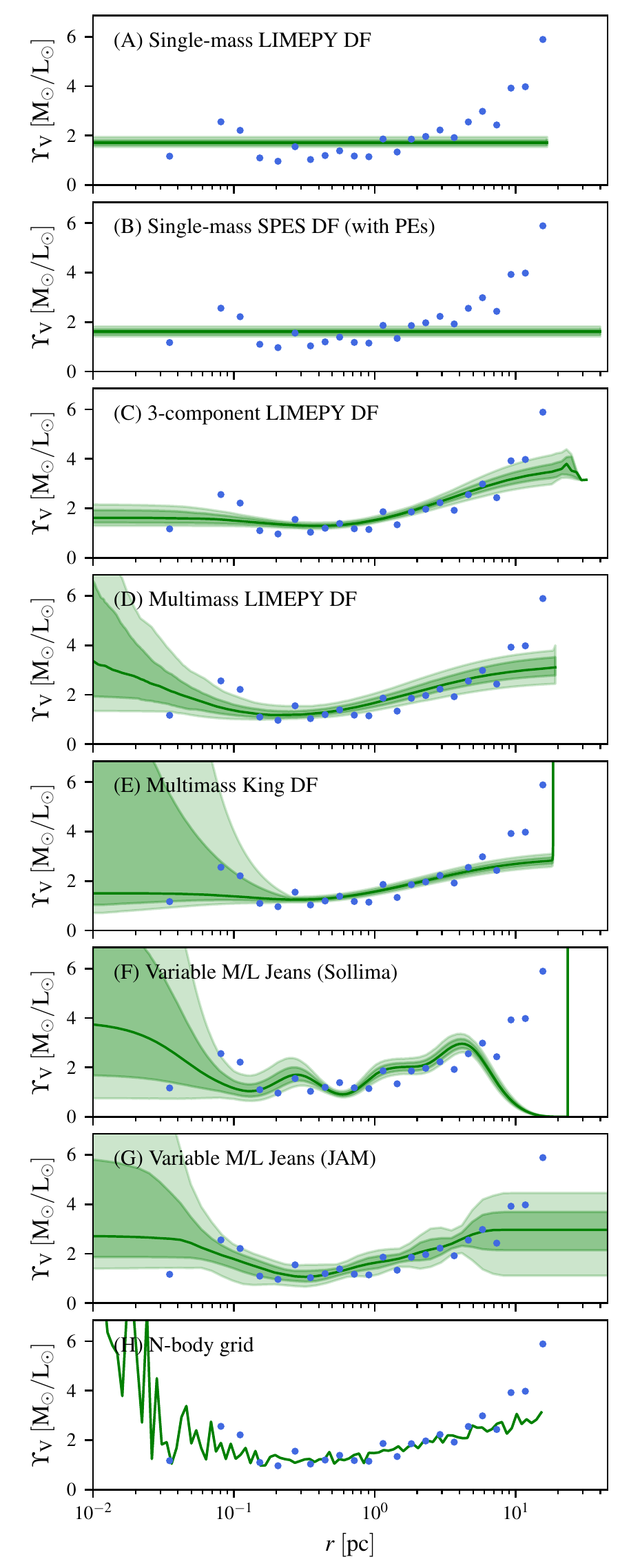}
\caption{Same as Figure \ref{fig:comp_encmass_rv} but for the mock dataset of case 3. }
    \label{fig:comp_encmass_rv_hst_gaia}
\end{figure*}

For all methods, \autoref{fig:comp_encmass_rv} displays for the case 1 dataset the ratio of the inferred cumulative mass profile to the true cumulative mass profile (in 3D; left panel), and the inferred 3D mass-to-light ratio profile (right panel). \autoref{fig:comp_encmass_rv_hst} and \autoref{fig:comp_encmass_rv_hst_gaia} display the same quantities but for the case 2 and case 3 datasets, respectively. Unsurprisingly, including the \textit{HST}-like PMs usually improves the recovered profiles in the central regions compared to the fits with LOS velocities alone, and including the \textit{Gaia}-like PMs then improves the outer profiles. As expected, the error bars on the recovered mass and mass-to-light ratio profiles generally shrink in the radial range of the additional kinematic data considered.

The obvious limitations of single-mass DF-based are readily apparent from looking at these figures, regardless of the richness of the kinematic data available. These models always systematically underestimate the mass in the very central region, then overestimate the cumulative mass at intermediate distances from the centre but slightly underestimate the total mass (unless kinematic data covers the external regions of the cluster). The cumulative mass profile comparison shows that mass is clearly missing from these models in the inner and outer regions, most likely due to the inability of the single-mass DF-based models to account for the presence and different spatial distribution of heavy dark remnants (concentrated in the inner regions) and faint low-mass stars (preferentially located in the external parts of the cluster). The way for the model to mitigate these limitations while still providing a satisfying fit to the mock data appears to be to overpredict the amount of mass in the intermediate regions (within and around the half-mass radius). So not only can single-mass models underestimate the total mass \citep[e.g.][]{2015MNRAS.448L..94S, 2015MNRAS.451.2185S}, they may also lead to an overestimate of the gravitational acceleration in the most commonly studied regions of clusters. This has potentially important consequences for the interpretation of pulsar timing data and inferred accelerations (used to probe the gravitational potential of GCs) when based on simple dynamical models in which mass follows light \citep[e.g.][]{2017ApJ...845..148P, 2017MNRAS.471..857F}. We note that single-mass DF-based models with potential escapers yield similar results, but they allow to trace the mass profile further out by including the population of potential escapers at large radii.

Multimass DF-based models provide a better description of the cumulative mass profile. Both the three-component model (\autoref{Sect_3-comp}) and the multimass model from \autoref{Sect_multi_antonio} (based on \citealt{1979AJ.....84..752G}) however underestimate the mass in the core (although only within a region corresponding to $\sim3\%$ of the half-mass radius), possibly because they slightly underestimate the total mass fraction in remnants and ignore the contribution of heavier remnants such as neutron stars. The multimass {\sc limepy} models from \autoref{Sect_multi_vincent}, which allowed more freedom in the total mass and mass function of remnants, recover the inner mass profile within the uncertainties even though these uncertainties are large in the innermost region. 
All multimass models very slightly overestimate the mass at intermediate radii.

The Jeans models (both the flavours from \autoref{Jeans_Sollima} and \autoref{Jeans_Watkins}) are able to reproduce the enclosed mass profiles well overall, typically as well as the best of the DF-based models; this is because they are able (indirectly) to account for the presence of different mass populations and account for their different spatial distributions, although they do not produce a mass function or directly quantify the degree of mass segregation as the multimass models are able to do. Global quantities, like $M$ and $\rh$, 
are accurately but not as precisely determined by Jeans models (larger error bars than other methods), however the models do provide the most conservative and realistic error bars 
on the recovered cumulative mass profile (the true profile is generally recovered within the uncertainties), with no significant bias in a particular direction. This is where the strength associated with the flexibility of these models becomes apparent.

For the method based on a grid of $N$-body models (\autoref{Nbody_Holger}), for which the effect of mass segregation is built in from first principles, the cumulative mass profile is generally well recovered. The enclosed mass is only slightly overestimated (but within $\sim20\%$ or better from the true value) for most of the radial extent of the cluster. Only in the inner 0.1 pc is the enclosed mass more significantly overestimated (a factor $\sim1.5$ to $2$ larger than the true value). This discrepancy in the centre could be related to the fact that the $N$-body models have no or too few binaries to heat the cluster core. Note that in its current implementation, the method does not yield uncertainties on the recovered mass and mass-to-light ratio profiles.

Apart from the single-mass DF-based models which have the obvious limitation of assuming a constant mass-to-light ratio profile, all other models considered satisfyingly reproduce the mass-to-light ratio profile, regardless of the combination of kinematic datasets used. Again, Jeans models provide the most conservative and realistic error bars on the mass-to-light ratio profiles, especially in the outskirts of the cluster. Multimass DF-based models however still have the advantage that the variation of the mass-to-light ratio with radius can easily be connected with mass segregation and the relative contribution of different mass species at different distances from the centre. This can be useful, for example, to establish if white dwarfs, neutron stars, or stellar-mass black holes are dominating the inferred increase in the mass-to-light ratio towards the cluster centre \citep[e.g.][]{1977ApJ...218L.109I, 2018MNRAS.473.4832G}. In the case of the M\,4 $N$-body snapshot studied here, the increase in the mass-to-light ratio towards the centre is caused by white dwarfs dominating the mass density in the inner regions, a piece of information that is straightforward to extract from the best-fit multimass models.

The inferred enclosed mass and mass-to-light ratio profiles for the Jeans models display fluctuations (especially in the outer regions) that have no physical basis. 
Both types of Jeans models treat the densities as a sum of Gaussians, and there is a delicate balance to choosing the number of Gaussians for a fit: when using too few of them, the model does not accurately represent the cluster\footnote{The models in \autoref{Jeans_Watkins} were also run with 5 Gaussians but the quality of the fits was significantly worse.}, but when using too many there is the risk of overfitting the data. The models in \autoref{Jeans_Sollima} use 27 Gaussians, with widths evenly spaced in $\log$ radius at fixed values, chosen so as to sample the density everywhere; whereas the models in \autoref{Jeans_Watkins} use just 6 Gaussians and allow the widths to vary so as to position themselves where they need to be to get the best fit. Clearly both approaches are successful and are able to provide a good fit to the cluster overall. The larger fluctuations in the models from \autoref{Jeans_Sollima} are likely because of the large number of Gaussians used that are able to fit any fluctuations in the data at the location where that Gaussian dominates, physical or not. In principle, one could try to improve this aspect of the Jeans modelling by tuning the number of Gaussians used to fit the profiles and/or by adopting specific priors, but with real data for which the underlying mass profile is not know this is not necessarily practical.

\section{Discussion}
\label{Sect_discussion}

We compared different mass-modelling methods against mock data from an $N$-body simulation of M\,4 and found that all the models considered generally provide a satisfying fit to the data. However, depending on one's scientific goal and available data, differences between the approaches and models mean that different techniques can be more efficient, suitable, or biased. That dynamical models are able to fit the data well is perhaps not so surprising given the number of free parameters in some of these models. The fact that a mass model provides a good fit to structural and kinematic data of a GC however does not necessarily imply that this model is a faithful representation of the mass distribution within the cluster. This is why we also compared other quantities such as global properties ($M$, $r_{\rm h}$, $\Upsilon_{\rm V}$, etc.) and cumulative mass or mass-to-light ratio profiles that are predicted by the best-fit mass models but not directly involved in the fitting procedure (and not available when fitting real data).

Single-mass DF-based models are still a possible way to obtain a first-order estimate of the global properties of a stellar cluster, such as its total mass and half-mass radius. Note however that if the kinematic tracers (generally bright stars more massive than the mean cluster star) are concentrated in the inner regions of the cluster and do not cover the full radial extent, these models will tend to significantly underestimate the total mass and half-mass radius, while also overestimating the amount of mass at intermediate radii. When the kinematic data cover a significant extent of the cluster (e.g. including {\it Gaia} proper motions in the outer parts), the single-mass models perform better as the kinematics of the tracer stars in the outer parts is much more similar to the kinematics of the average cluster star.

However, as data further and further out is included, and the velocity dispersion profile starts to flatten due to potential escapers, the inability of lowered isothermal models to account for the outermost kinematic data can drive the fits towards models with largely overestimated masses and half-mass radii, and poor fits to the surface brightness profile (this caveat also applies to multimass lowered isothermal models). In this case the outermost data points would have to be ignored in a fitting procedure to provide a satisfying description of the cluster closer in. Since the region where potential escapers affect the observed kinematics is generally poorly constrained, 
unknown biases may creep in the analysis when blindly fitting truncated models to data covering the outskirts of clusters. Similar care is recommended when fitting a grid of isolated $N$-body models (as in \autoref{Nbody_Holger}) because these will not capture the flattening of the velocity dispersion profile due to tides. We have shown that distribution function-based models including potential escapers are a promising way to avoid possible biases caused by the outermost data points. At the moment, however, only a single-mass version of these models has been developed, and so these models are still subject to biases due to ignoring the effect of mass segregation. We have also shown that by including a simple population of `contaminants' that is dominant at large radii, Jeans models can in principle account for the outermost data points without fine-tuning. The extremely simple contaminant model used here does not offer the same physical insight about tidal effects that is provided by DF-based models with potential escapers, although a more sophisticated model that encompasses the nature of the contaminants could be included.

We have considered three-component DF-based models as the simplest way to improve on single-mass models by approximately including the effect of a mass spectrum and mass segregation without a measurement of the stellar mass function inside the cluster. We note that multimass models are not necessarily better (see \autoref{GG79_model} for details) in that using a wrong or ill-informed mass function that is not representative of the mass function of the cluster can give systematic biases of similar magnitude as the single-mass models (an overestimate of the total mass and half-mass radius in the case studied here). 
However, when leaving some freedom to adjust the stellar mass function of the models guided by the kinematic and structural data, we found that three-component models can perform very well and properly capture the effect of mass segregation. This was done by fitting on the mean mass of dark remnants and the fraction of mass in remnants, adopting standard assumptions for the mean mass of the other mass bins (low-mass stars and bright visible stars) and assuming that visible stars are tracers with negligible contribution to the total cluster mass. These models would be particularly well-suited to apply to partially resolved clusters or resolved clusters with limited data available (e.g. no stellar mass function measurement), for example to infer dynamical masses and avoid biases due to neglecting mass segregation.

When a measurement of the local stellar mass function within the cluster is available, multimass DF-based models with a full mass spectrum (i.e. a large number of mass bins) and fitting a grid of $N$-body models are reliable ways to infer the global mass function and global parameters such as the total mass, half-mass radius, mass-to-light ratio, and fraction of cluster mass in remnants. They also satisfyingly recover the cumulative mass and mass-to-light ratio profiles. However, given the limited variety of initial mass functions that can be considered in the grid of $N$-body models, this method has less flexibility and may not recover the global/stellar mass functions as precisely as the multimass models. It also makes it more challenging to quantify uncertainties on the mass function. 

In addition to testing a previously used version of multimass DF-based models with a prescription for the mass function of remnants, we introduced and tested a variant 
with the total mass in remnants and remnant mass function free to vary. The success of this type of model in recovering the total mass in remnants and estimating their mass function (while only using information from visible stars) opens up exciting possibilities to weigh populations of heavy remnants in GCs with fast and flexible DF-based models. More testing will be needed on mock data and models with different remnant mass functions (e.g. with a larger population of black holes), but ultimately it will be interesting to compare such an approach with other methods trying to infer populations of black holes in GCs based on their present-day structural and internal kinematic properties, usually based on a sparse grid and limited number of evolutionary models \citep{2018MNRAS.478.1844A, 2018MNRAS.479.4652A, 2016MNRAS.462.2333P, 2018ApJ...864...13W, 2018ApJ...855L..15K}. The size of present-day black hole populations in Milky Way GCs is an important ingredient to understand the formation of binary black holes generating gravitational wave signals.

If one is not concerned with inferring the global mass function 
and directly modelling the dynamics of different mass components, then Jeans modelling is another efficient way to constrain the non-constant mass-to-light ratio profile caused by mass segregation. We found that the flexibility of Jeans models tends to lead to more conservative and realistic errors bars (free of systematic biases) on the global cluster properties and on the recovered cumulative mass and mass-to-light ratio profiles, even though global quantities like the total mass and half-mass radius are not precisely measured due to the lack of constraints in the outer parts of the cluster (both from scarcity of data and the absence of restrictions in the model). One could envisage situations where Jeans modelling is first used to constrain the underlying mass profile, which is then used as a prior for a different mass-modelling technique, for example a multimass DF-based model with which further insight is obtained on the mass function and distribution of different mass components. Jeans modelling also offers another way to infer the fraction of cluster mass in dark remnants by determining the mass profile (using kinematics) and subtracting the stellar mass from star counts (if deep enough photometry is available) to obtain the fraction of remnants as a function of distance from the cluster centre \citep{2016MNRAS.462.1937S}.

\section{Summary and conclusions}
\label{summary}

\begin{table*} \renewcommand{\arraystretch}{1.2}
\caption{Summary of the mass-modelling methods compared in this work.}
\label{summary_table}
\centering
\begin{tabular}{p{3.cm}|p{4.4cm}|p{4.4cm}|p{4.4cm}}
\hline
\bf{Modelling method}  & \bf{Pros}  & \bf{Cons} & \bf{Minimal recommended data}  \\ \hline     
\hline
A. Single-mass \limepy \ DF \newline (\autoref{section:single-mass_limepy}) & - Fast and simple way to estimate global cluster properties & - Prone to biases: underestimates the total mass and half-mass radius \newline- Effect of PEs not included, kinematics in cluster outskirts should be ignored for good fit \newline - Assumes constant M/L profile \newline - Systematically underestimates the mass in the very central region and overestimates the mass at intermediate radii & - Velocity dispersion profile \newline - Surface density or brightness profile \\
\hline
B. Single-mass SPES DF (with PEs) \newline (\autoref{Sect:PEs})& - Only DF model to capture the effect of PEs and the kinematics in cluster outskirts & - Effect of mass segregation not captured, so prone to same biases as method A \newline - Systematically underestimates the mass in the very central region and overestimates the mass at intermediate radii \newline - Assumes constant M/L profile \newline - Two additional free parameters  & - Velocity dispersion profile \newline - Surface density or brightness profile \newline - Good kinematics and star counts needed in the outer parts to constrain the two additional parameters \\
\hline
C. 3-component \limepy \ DF \newline (\autoref{Sect_3-comp}) & - Simplest DF model to capture the effect of mass segregation, the variation of M/L with radius, and to estimate cluster ``dark" mass \newline - No mass local mass function measurement required & - Effect of PEs not included, kinematics in cluster outskirts should be ignored for good fit & - Velocity dispersion profile \newline - Surface density or brightness profile \\
\hline
D. Multimass \limepy \ DF \newline(\autoref{Sect_multi_vincent}) & - Allows to infer present-day global stellar mass function, and fraction of mass in dark remnants \newline - Captures variation of M/L with radius \newline - Freedom in remnants can be used to constrain remnant mass function & - Effect of PEs not included, kinematics in cluster outskirts should be ignored for good fit & - Velocity dispersion \newline - Surface density or brightness profile \newline - Local stellar mass function(s) or density profile for several mass components  \\
\hline
E. Multimass King DF \newline(\autoref{Sect_multi_antonio}) & - Allows to infer present-day global stellar mass function, and fraction of mass in dark remnants \newline - Captures variation of M/L with radius & - Effect of PEs not included \newline - No freedom in remnant mass function (fixed prescription used)& - Velocity dispersion \newline - Surface density or brightness profile \newline - Local stellar mass function(s) or density profile for several mass components \\
\hline
F. Variable M/L Jeans (Sollima) \newline(\autoref{Jeans_Sollima}) & - Captures variation of M/L with radius \newline - More conservative and realistic uncertainties on enclosed mass profile (no bias)  & - Flexibility (number of Gaussians to fit the tracer density profile) can lead to unrealistic/unphysical fluctuations in the inferred mass and mass-to-light ratio profiles \newline - Outer parts of models less constrained, so larger uncertainties on global quantities like total mass and half-mass radius & - Velocity dispersion \newline - Surface density profile \newline - Optional: deep photometry (to estimate stellar mass from star counts and subtract it from total mass to estimate fraction of remnants)  \\
\hline
G. Variable M/L Jeans (JAM) \newline(\autoref{Jeans_Watkins}) & - Captures variation of M/L with radius \newline  - More conservative and realistic uncertainties on enclosed mass profile (no bias) \newline - Kinematics in the cluster outskirts can be captured by including a population of contaminants dominant at large radii \newline - More flexibility with anisotropy & - Outer parts of models less constrained, so larger uncertainties on global quantities like total mass and half-mass radius & - Velocity dispersion \newline - Surface density or brightness profile \newline - Optional: deep photometry (to estimate stellar mass from star counts and subtract it from total mass to estimate fraction of remnants)  \\
\hline
H. $N$-body grid \newline(\autoref{Nbody_Holger})  & - Allows to infer present-day global stellar mass function, and fraction of mass in dark remnants \newline  - Captures variation of M/L with radius& - Effect of PEs and flattening of the velocity dispersion profile due to tides not captured by grid of isolated $N$-body models \newline - Less flexibility, limited IMF variety in current grid of models, global mass function may not be recovered as precisely as multimass models and uncertainties on global mass function more challenging to estimate& - Velocity dispersion \newline - Surface density or brightness profile \newline - Local stellar mass function(s) or density profile for several mass components \\
  \hline
 \end{tabular}
\end{table*}

We summarize our findings with the few take-home messages below, as well as a list of pros, cons, and minimal recommended data for the different methods in \autoref{summary_table}. These can serve as a starting point for choosing the most appropriate mass-modelling method for a given problem.

\begin{itemize}
\item{Single-mass DF-based models (\autoref{section:single-mass}) can be used to estimate the global properties of a cluster but they are prone to biases because they do not capture the effect of mass segregation. They tend to significantly underestimate the total mass and half-mass radius, although this can be mitigated by including kinematic data covering a wide extent of the cluster.}
\item{Three-component DF-based models (\autoref{Sect_3-comp}), leaving freedom to adjust the mass function, represent a simple way to overcome the biases of single-mass models. 
They would be well-suited to cases where data is limited (e.g. no stellar mass function) but a more accurate estimate of the dynamical mass or half-mass radius is needed (compared to single-mass models).}
\item{The use of multimass DF-based models with a more realistic mass spectrum (\autoref{Sect_multi_vincent} and \autoref{Sect_multi_antonio}) or a grid of $N$-body models are indicated to infer the present-day global stellar mass function of a cluster from knowledge of its local stellar mass function, kinematics, and structural properties.}
\item{Multimass DF-based models (including the three-component flavour), a grid of $N$-body models and Jeans models are all suitable options to recover the radially-varying mass-to-light ratio profile of GCs, with the first two allowing to estimate the contribution of different types of stars to this profile. On the other hand, Jeans models, because of their flexibility can lead to unrealistic fluctuations.}
\item{Methods based on Jeans models (in combination with deep enough photometry to estimate the mass in stars), a grid of $N$-body models and multimass models can all be used to infer the fraction of cluster mass in dark remnants. The latter have the advantage that the effect of different remnant mass functions on the cluster dynamics can be explored, and by leaving it free to vary this remnant mass function can potentially be constrained (see \autoref{Sect_multi_vincent}).} 
\item{Multimass DF-based models, Jeans models, and fitting a grid of $N$-body models can all reliably recover the mass profile of a mass segregated cluster, as opposed to single-mass DF-based models which systematically underestimate the mass in the very central region and overestimate the mass at intermediate radii (\autoref{comparison}).}
\item{Jeans models tend to provide the most conservative and reasonable uncertainties on the recovered mass profile, with no significant bias (\autoref{comparison}). However, because the outer parts of these models are less constrained, global quantities such as the total mass and half-mass radius tend to be less precisely determined (although with good accuracy, i.e. within the errors).}
\item{An important caveat to the points above is that when including kinematic data in the very outskirts of a cluster (as {\it Gaia} will enable), the use of lowered isothermal DF-based models (single-mass and multimass) or of a grid of isolated $N$-body model reach a fundamental limitation. These models cannot reproduce the flattening in the outer velocity dispersion profile due to potential escapers. Unless these kinematic data in the cluster outskirts are ignored when fitting the models, this can lead to overestimating the cluster mass and radial extent.}
\item{DF-based models including potential escapers (\autoref{Sect:PEs}) are necessary to capture the kinematics in the cluster outskirts and overcome the limitations of lowered isothermal models. This can also be achieved with Jeans models (\autoref{Sect_Jeans}) by including a population of contaminants that is dominant at large distances from the cluster centre.}
\end{itemize}

Our study is limited to one snapshot from a specific $N$-body simulation, 
so we have to be careful not to overgeneralize our results, and we caution that some of the remarks above may not apply to certain situations. We also note that we analyzed mock data corresponding to the most nearby GC and optimistic conditions. With lower-quality or limited kinematic or mass function data, some of the applications discussed above (like constraining the mass function of remnants) may prove challenging and would warrant further testing. Nevertheless, our comparison provides initial guidelines for the choice and applicability of different mass-modelling approaches. We also hope that it serves as a template and example to follow for future tests of dynamical modelling methods on mock data before these are applied to real data to draw conclusions.

\section*{Acknowledgements}
We thank the anonymous referee for their useful comments and constructive feedback. We thank Douglas Heggie for making the output of his $N$-body simulation of M\,4 publicly available. This work was done as part of the {\it Gaia Challenge}\footnote{\url{http://astrowiki.ph.surrey.ac.uk/dokuwiki/doku.php?id=start}}, and we would like to thank all the other participants of the Collisional Systems working group for insightful discussions, in particular Adriano Agnello, Emanuele Dalessandro, Miklos Peuten, and Anna Lisa Varri. We thank the organisers of the {\it Gaia Challenge} Workshop in Stockholm (2016) during which this specific study was initiated. We also thank the International Space Science Institute (ISSI, Bern, CH) for welcoming the activities of the Team 407 ``Globular Clusters in the {\it Gaia} Era" (team leaders VHB \& MG), during which part of this work was conducted. VHB acknowledges support from the NRC-Canada Plaskett Fellowship and from the Radboud Excellence Initiative. MG acknowledges financial support from the Royal Society (University Research Fellowship) and MG and IC thank the European Research Council (ERC StG-335936, CLUSTERS) for financial support. LLW acknowledges support provided by NASA through \textit{HST} grants AR-14322 (PI: Watkins) and AR-15055 (PI: Watkins), from the Space Telescope Science Institute, which is operated by the Association of Universities for Research in Astronomy, Inc., under NASA contract NAS 5-26555. AZ is supported by the ESA Research Fellowship, and acknowledges the Royal Society for financial support through the Newton International Fellowship Follow-up Funding.

This research has made use of Astropy\footnote{\url{http://www.astropy.org}}, a community-developed core Python package for Astronomy \citep{astropy2013}. This research made use of the R programming language (\url{https://www.R-project.org/}), its data.table package, and the Rstudio environment (\url{https://www.rstudio.com/}). This research has made use of NASA's Astrophysics Data System.

\appendix

\bibliographystyle{mn2e} 
\bibliography{MM_GC}

\begin{thebibliography}{100}
\expandafter\ifx\csname natexlab\endcsname\relax\def\natexlab#1{#1}\fi

\bibitem[{{Aarseth}(1999)}]{1999PASP..111.1333A}
{Aarseth} S.~J., 1999, \pasp, 111, 1333

\bibitem[{{Abbott} {et~al}\mbox{.}(2016{\natexlab{a}}){Abbott}, {Abbott},
  {Abbott}, {Abernathy}, {Acernese}, {Ackley}, {Adams}, {Adams}, {Addesso},
  {Adhikari}, \& et~al.}]{2016ApJ...818L..22A}
{Abbott} B.~P. {et~al.}, 2016{\natexlab{a}}, \apjl, 818, L22

\bibitem[{{Abbott} {et~al}\mbox{.}(2016{\natexlab{b}}){Abbott}, {Abbott},
  {Abbott}, {Abernathy}, {Acernese}, {Ackley}, {Adams}, {Adams}, {Addesso},
  {Adhikari}, \& et~al.}]{2016PhRvL.116f1102A}
{Abbott} B.~P. {et~al.}, 2016{\natexlab{b}}, Physical Review Letters, 116,
  061102

\bibitem[{{Antonini} \& {Rasio}(2016)}]{2016ApJ...831..187A}
{Antonini} F., {Rasio} F.~A., 2016, \apj, 831, 187

\bibitem[{{Arca Sedda} {et~al}\mbox{.}(2018){Arca Sedda}, {Askar}, \&
  {Giersz}}]{2018MNRAS.479.4652A}
{Arca Sedda} M., {Askar} A., {Giersz} M., 2018, \mnras, 479, 4652

\bibitem[{{Askar} {et~al}\mbox{.}(2018){Askar}, {Arca Sedda}, \&
  {Giersz}}]{2018MNRAS.478.1844A}
{Askar} A., {Arca Sedda} M., {Giersz} M., 2018, \mnras, 478, 1844

\bibitem[{{Askar} {et~al}\mbox{.}(2017){Askar}, {Szkudlarek},
  {Gondek-Rosi{\'n}ska}, {Giersz}, \& {Bulik}}]{2017MNRAS.464L..36A}
{Askar} A., {Szkudlarek} M., {Gondek-Rosi{\'n}ska} D., {Giersz} M., {Bulik} T.,
  2017, \mnras, 464, L36

\bibitem[{{Astropy Collaboration} {et~al}\mbox{.}(2013){Astropy Collaboration},
  {Robitaille}, {Tollerud}, {Greenfield}, {Droettboom}, {Bray}, {Aldcroft},
  {Davis}, {Ginsburg}, {Price-Whelan}, {Kerzendorf}, {Conley}, {Crighton},
  {Barbary}, {Muna}, {Ferguson}, {Grollier}, {Parikh}, {Nair}, {Unther},
  {Deil}, {Woillez}, {Conseil}, {Kramer}, {Turner}, {Singer}, {Fox}, {Weaver},
  {Zabalza}, {Edwards}, {Azalee Bostroem}, {Burke}, {Casey}, {Crawford},
  {Dencheva}, {Ely}, {Jenness}, {Labrie}, {Lim}, {Pierfederici}, {Pontzen},
  {Ptak}, {Refsdal}, {Servillat}, \& {Streicher}}]{astropy2013}
{Astropy Collaboration} {et~al.}, 2013, \aap, 558, A33

\bibitem[{{Baumgardt}(2001)}]{Baumgardt2001}
{Baumgardt} H., 2001, MNRAS, 325, 1323

\bibitem[{{Baumgardt}(2017)}]{2017MNRAS.464.2174B}
{Baumgardt} H., 2017, \mnras, 464, 2174

\bibitem[{{Baumgardt} \& {Hilker}(2018)}]{2018MNRAS.478.1520B}
{Baumgardt} H., {Hilker} M., 2018, \mnras, 478, 1520

\bibitem[{{Baumgardt} \& {Makino}(2003)}]{2003MNRAS.340..227B}
{Baumgardt} H., {Makino} J., 2003, \mnras, 340, 227

\bibitem[{{Baumgardt} \& {Sollima}(2017)}]{2017MNRAS.472..744B}
{Baumgardt} H., {Sollima} S., 2017, \mnras, 472, 744

\bibitem[{{Bedin} {et~al}\mbox{.}(2003){Bedin}, {Piotto}, {King}, \&
  {Anderson}}]{2003AJ....126..247B}
{Bedin} L.~R., {Piotto} G., {King} I.~R., {Anderson} J., 2003, \aj, 126, 247

\bibitem[{{Bellini} {et~al}\mbox{.}(2014){Bellini}, {Anderson}, {van der
  Marel}, {Watkins}, {King}, {Bianchini}, {Chanam{\'e}}, {Chandar}, {Cool},
  {Ferraro}, {Ford}, \& {Massari}}]{Bellini2014}
{Bellini} A. {et~al.}, 2014, \apj, 797, 115

\bibitem[{{Bianchini} {et~al}\mbox{.}(2018){Bianchini}, {van der Marel}, {del
  Pino}, {Watkins}, {Bellini}, {Fardal}, {Libralato}, \&
  {Sills}}]{Bianchini2018}
{Bianchini} P., {van der Marel} R.~P., {del Pino} A., {Watkins} L.~L.,
  {Bellini} A., {Fardal} M.~A., {Libralato} M., {Sills} A., 2018, ArXiv
  e-prints, arXiv:1806.02580

\bibitem[{{Braga} {et~al}\mbox{.}(2015){Braga}, {Dall'Ora}, {Bono}, {Stetson},
  {Ferraro}, {Iannicola}, {Marengo}, {Neeley}, {Persson}, {Buonanno},
  {Coppola}, {Freedman}, {Madore}, {Marconi}, {Matsunaga}, {Monson}, {Rich},
  {Scowcroft}, \& {Seibert}}]{2015ApJ...799..165B}
{Braga} V.~F. {et~al.}, 2015, \apj, 799, 165

\bibitem[{{Breen} \& {Heggie}(2013)}]{2013MNRAS.432.2779B}
{Breen} P.~G., {Heggie} D.~C., 2013, \mnras, 432, 2779

\bibitem[{{B{\"u}denbender} {et~al}\mbox{.}(2015){B{\"u}denbender}, {van de
  Ven}, \& {Watkins}}]{Buedenbender2015}
{B{\"u}denbender} A., {van de Ven} G., {Watkins} L.~L., 2015, \mnras, 452, 956

\bibitem[{{Cappellari}(2008)}]{Cappellari2008}
{Cappellari} M., 2008, \mnras, 390, 71

\bibitem[{{Cappellari}(2012)}]{2012arXiv1211.7009C}
{Cappellari} M., 2012, ArXiv e-prints, arXiv:1211.7009

\bibitem[{{Cappellari}(2015)}]{Cappellari2015}
{Cappellari} M., 2015, ArXiv e-prints, arXiv:1504.05533

\bibitem[{{Claydon} {et~al}\mbox{.}(2018){Claydon}, {Gieles}, {Varri},
  {Heggie}, \& {Zocchi}}]{claydon18}
{Claydon} I., {Gieles} M., {Varri} A.~L., {Heggie} D.~C., {Zocchi} A., 2018,
  \mnras, to be submitted

\bibitem[{{Claydon} {et~al}\mbox{.}(2017){Claydon}, {Gieles}, \&
  {Zocchi}}]{2017MNRAS.466.3937C}
{Claydon} I., {Gieles} M., {Zocchi} A., 2017, \mnras, 466, 3937

\bibitem[{{Conroy} \& {Gunn}(2010)}]{2010ApJ...712..833C}
{Conroy} C., {Gunn} J.~E., 2010, \apj, 712, 833

\bibitem[{{Conroy} {et~al}\mbox{.}(2009){Conroy}, {Gunn}, \&
  {White}}]{2009ApJ...699..486C}
{Conroy} C., {Gunn} J.~E., {White} M., 2009, \apj, 699, 486

\bibitem[{{Da Costa} \& {Freeman}(1976)}]{1976ApJ...206..128D}
{Da Costa} G.~S., {Freeman} K.~C., 1976, \apj, 206, 128

\bibitem[{{Daniel} {et~al}\mbox{.}(2017){Daniel}, {Heggie}, \&
  {Varri}}]{2017MNRAS.468.1453D}
{Daniel} K.~J., {Heggie} D.~C., {Varri} A.~L., 2017, \mnras, 468, 1453

\bibitem[{{Eddington}(1916)}]{1916MNRAS..76..572E}
{Eddington} A.~S., 1916, \mnras, 76, 572

\bibitem[{{Foreman-Mackey} {et~al}\mbox{.}(2013){Foreman-Mackey}, {Hogg},
  {Lang}, \& {Goodman}}]{ForemanMackey2013}
{Foreman-Mackey} D., {Hogg} D.~W., {Lang} D., {Goodman} J., 2013, \pasp, 125,
  306

\bibitem[{{Freire} {et~al}\mbox{.}(2017){Freire}, {Ridolfi}, {Kramer},
  {Jordan}, {Manchester}, {Torne}, {Sarkissian}, {Heinke}, {D'Amico}, {Camilo},
  {Lorimer}, \& {Lyne}}]{2017MNRAS.471..857F}
{Freire} P.~C.~C. {et~al.}, 2017, \mnras, 471, 857

\bibitem[{{Fukushige} \& {Heggie}(2000)}]{Fukushige2000}
{Fukushige} T., {Heggie} D.~C., 2000, MNRAS, 318, 753

\bibitem[{{Gaia Collaboration} {et~al}\mbox{.}(2018{\natexlab{a}}){Gaia
  Collaboration}, {Brown}, {Vallenari}, {Prusti}, {de Bruijne}, {Babusiaux},
  {Bailer-Jones}, {Biermann}, {Evans}, {Eyer}, \& et~al.}]{Gaia2018}
{Gaia Collaboration} {et~al.}, 2018{\natexlab{a}}, \aap, 616, A1

\bibitem[{{Gaia Collaboration} {et~al}\mbox{.}(2018{\natexlab{b}}){Gaia
  Collaboration}, {Helmi}, {van Leeuwen}, {McMillan}, {Massari}, {Antoja},
  {Robin}, {Lindegren}, {Bastian}, {Arenou}, \& et~al.}]{Helmi2018}
{Gaia Collaboration} {et~al.}, 2018{\natexlab{b}}, \aap, 616, A12

\bibitem[{{Gaia Collaboration} {et~al}\mbox{.}(2016){Gaia Collaboration},
  {Prusti}, {de Bruijne}, {Brown}, {Vallenari}, {Babusiaux}, {Bailer-Jones},
  {Bastian}, {Biermann}, {Evans}, \& et~al.}]{Gaia2016}
{Gaia Collaboration} {et~al.}, 2016, \aap, 595, A1

\bibitem[{{Gebhardt} \& {Fischer}(1995)}]{1995AJ....109..209G}
{Gebhardt} K., {Fischer} P., 1995, \aj, 109, 209

\bibitem[{{Gieles} {et~al}\mbox{.}(2018){Gieles}, {Balbinot}, {Yaaqib},
  {H{\'e}nault-Brunet}, {Zocchi}, {Peuten}, \& {Jonker}}]{2018MNRAS.473.4832G}
{Gieles} M., {Balbinot} E., {Yaaqib} R.~I.~S.~M., {H{\'e}nault-Brunet} V.,
  {Zocchi} A., {Peuten} M., {Jonker} P.~G., 2018, \mnras, 473, 4832

\bibitem[{{Gieles} \& {Zocchi}(2015)}]{2015MNRAS.454..576G}
{Gieles} M., {Zocchi} A., 2015, \mnras, 454, 576

\bibitem[{{Giersz} \& {Heggie}(2011)}]{2011MNRAS.410.2698G}
{Giersz} M., {Heggie} D.~C., 2011, \mnras, 410, 2698

\bibitem[{{Girardi} {et~al}\mbox{.}(2000){Girardi}, {Bressan}, {Bertelli}, \&
  {Chiosi}}]{2000A&AS..141..371G}
{Girardi} L., {Bressan} A., {Bertelli} G., {Chiosi} C., 2000, \aaps, 141, 371

\bibitem[{{Gunn} \& {Griffin}(1979)}]{1979AJ.....84..752G}
{Gunn} J.~E., {Griffin} R.~F., 1979, \aj, 84, 752

\bibitem[{{Heggie}(2014)}]{Heggie2014}
{Heggie} D.~C., 2014, \mnras, 445, 3435

\bibitem[{{H{\'e}non}(1961)}]{1961AnAp...24..369H}
{H{\'e}non} M., 1961, Annales d'Astrophysique, 24, 369

\bibitem[{{Hurley} {et~al}\mbox{.}(2000){Hurley}, {Pols}, \&
  {Tout}}]{2000MNRAS.315..543H}
{Hurley} J.~R., {Pols} O.~R., {Tout} C.~A., 2000, \mnras, 315, 543

\bibitem[{{Ibata} {et~al}\mbox{.}(2013){Ibata}, {Nipoti}, {Sollima},
  {Bellazzini}, {Chapman}, \& {Dalessandro}}]{2013MNRAS.428.3648I}
{Ibata} R., {Nipoti} C., {Sollima} A., {Bellazzini} M., {Chapman} S.~C.,
  {Dalessandro} E., 2013, \mnras, 428, 3648

\bibitem[{{Illingworth} \& {King}(1977)}]{1977ApJ...218L.109I}
{Illingworth} G., {King} I.~R., 1977, \apjl, 218, L109

\bibitem[{{Jordi} {et~al}\mbox{.}(2010){Jordi}, {Gebran}, {Carrasco}, {de
  Bruijne}, {Voss}, {Fabricius}, {Knude}, {Vallenari}, {Kohley}, \&
  {Mora}}]{Jordi2010}
{Jordi} C. {et~al.}, 2010, \aap, 523, A48

\bibitem[{{Kamann} {et~al}\mbox{.}(2018){Kamann}, {Husser}, {Dreizler},
  {Emsellem}, {Weilbacher}, {Martens}, {Bacon}, {den Brok}, {Giesers},
  {Krajnovi{\'c}}, {Roth}, {Wendt}, \& {Wisotzki}}]{2018MNRAS.473.5591K}
{Kamann} S. {et~al.}, 2018, \mnras, 473, 5591

\bibitem[{{Kimmig} {et~al}\mbox{.}(2015){Kimmig}, {Seth}, {Ivans}, {Strader},
  {Caldwell}, {Anderton}, \& {Gregersen}}]{2015AJ....149...53K}
{Kimmig} B., {Seth} A., {Ivans} I.~I., {Strader} J., {Caldwell} N., {Anderton}
  T., {Gregersen} D., 2015, \aj, 149, 53

\bibitem[{{King}(1962)}]{1962AJ.....67..471K}
{King} I., 1962, \aj, 67, 471

\bibitem[{{King}(1966)}]{1966AJ.....71...64K}
{King} I.~R., 1966, \aj, 71, 64

\bibitem[{{Kremer} {et~al}\mbox{.}(2018){Kremer}, {Ye}, {Chatterjee},
  {Rodriguez}, \& {Rasio}}]{2018ApJ...855L..15K}
{Kremer} K., {Ye} C.~S., {Chatterjee} S., {Rodriguez} C.~L., {Rasio} F.~A.,
  2018, \apjl, 855, L15

\bibitem[{{Kroupa}(2001)}]{2001MNRAS.322..231K}
{Kroupa} P., 2001, \mnras, 322, 231

\bibitem[{{Kruijssen}(2009)}]{2009A&A...507.1409K}
{Kruijssen} J.~M.~D., 2009, \aap, 507, 1409

\bibitem[{{K{\"u}pper} {et~al}\mbox{.}(2010){K{\"u}pper}, {Kroupa},
  {Baumgardt}, \& {Heggie}}]{Kupper2010}
{K{\"u}pper} A.~H.~W., {Kroupa} P., {Baumgardt} H., {Heggie} D.~C., 2010,
  MNRAS, 407, 2241

\bibitem[{{Leonard} {et~al}\mbox{.}(1992){Leonard}, {Richer}, \&
  {Fahlman}}]{1992AJ....104.2104L}
{Leonard} P.~J.~T., {Richer} H.~B., {Fahlman} G.~G., 1992, \aj, 104, 2104

\bibitem[{{Libralato} {et~al}\mbox{.}(2018){Libralato}, {Bellini}, {van der
  Marel}, {Anderson}, {Watkins}, {Piotto}, {Ferraro}, {Nardiello}, \&
  {Vesperini}}]{Libralato2018}
{Libralato} M. {et~al.}, 2018, \apj, 861, 99

\bibitem[{{L{\"u}tzgendorf} {et~al}\mbox{.}(2013){L{\"u}tzgendorf},
  {Kissler-Patig}, {Gebhardt}, {Baumgardt}, {Noyola}, {de Zeeuw}, {Neumayer},
  {Jalali}, \& {Feldmeier}}]{2013A&A...552A..49L}
{L{\"u}tzgendorf} N. {et~al.}, 2013, \aap, 552, A49

\bibitem[{{Marigo} \& {Girardi}(2007)}]{2007A&A...469..239M}
{Marigo} P., {Girardi} L., 2007, \aap, 469, 239

\bibitem[{{Marigo} {et~al}\mbox{.}(2008){Marigo}, {Girardi}, {Bressan},
  {Groenewegen}, {Silva}, \& {Granato}}]{2008A&A...482..883M}
{Marigo} P., {Girardi} L., {Bressan} A., {Groenewegen} M.~A.~T., {Silva} L.,
  {Granato} G.~L., 2008, \aap, 482, 883

\bibitem[{{Merritt}(1985)}]{1985AJ.....90.1027M}
{Merritt} D., 1985, \aj, 90, 1027

\bibitem[{{Meylan} \& {Heggie}(1997)}]{MeylanHeggie1997}
{Meylan} G., {Heggie} D.~C., 1997, \aapr, 8, 1

\bibitem[{{Meylan} \& {Mayor}(1986)}]{1986A&A...166..122M}
{Meylan} G., {Mayor} M., 1986, \aap, 166, 122

\bibitem[{{Michie}(1963)}]{1963MNRAS.125..127M}
{Michie} R.~W., 1963, \mnras, 125, 127

\bibitem[{{Milone} {et~al}\mbox{.}(2018){Milone}, {Marino},
  {Mastrobuono-Battisti}, \& {Lagioia}}]{2018MNRAS.479.5005M}
{Milone} A.~P., {Marino} A.~F., {Mastrobuono-Battisti} A., {Lagioia} E.~P.,
  2018, \mnras, 479, 5005

\bibitem[{{Morscher} {et~al}\mbox{.}(2015){Morscher}, {Pattabiraman},
  {Rodriguez}, {Rasio}, \& {Umbreit}}]{2015ApJ...800....9M}
{Morscher} M., {Pattabiraman} B., {Rodriguez} C., {Rasio} F.~A., {Umbreit} S.,
  2015, \apj, 800, 9

\bibitem[{{Nitadori} \& {Aarseth}(2012)}]{2012MNRAS.424..545N}
{Nitadori} K., {Aarseth} S.~J., 2012, \mnras, 424, 545

\bibitem[{{Osipkov}(1979)}]{1979PAZh....5...77O}
{Osipkov} L.~P., 1979, Pisma v Astronomicheskii Zhurnal, 5, 77

\bibitem[{{Pancino} {et~al}\mbox{.}(2017){Pancino}, {Bellazzini}, {Giuffrida},
  \& {Marinoni}}]{Pancino2017}
{Pancino} E., {Bellazzini} M., {Giuffrida} G., {Marinoni} S., 2017, \mnras,
  467, 412

\bibitem[{{Paresce} \& {De Marchi}(2000)}]{2000ApJ...534..870P}
{Paresce} F., {De Marchi} G., 2000, \apj, 534, 870

\bibitem[{{Pasquali} {et~al}\mbox{.}(2004){Pasquali}, {De Marchi}, {Pulone}, \&
  {Brigas}}]{2004A&A...428..469P}
{Pasquali} A., {De Marchi} G., {Pulone} L., {Brigas} M.~S., 2004, \aap, 428,
  469

\bibitem[{{Peuten} {et~al}\mbox{.}(2016){Peuten}, {Zocchi}, {Gieles},
  {Gualandris}, \& {H{\'e}nault-Brunet}}]{2016MNRAS.462.2333P}
{Peuten} M., {Zocchi} A., {Gieles} M., {Gualandris} A., {H{\'e}nault-Brunet}
  V., 2016, \mnras, 462, 2333

\bibitem[{{Peuten} {et~al}\mbox{.}(2017){Peuten}, {Zocchi}, {Gieles}, \&
  {H{\'e}nault-Brunet}}]{peuten17}
{Peuten} M., {Zocchi} A., {Gieles} M., {H{\'e}nault-Brunet} V., 2017, \mnras,
  470, 2736

\bibitem[{{Phinney}(1993)}]{1993ASPC...50..141P}
{Phinney} E.~S., 1993, in Astronomical Society of the Pacific Conference
  Series, Vol.~50, Structure and Dynamics of Globular Clusters, {Djorgovski}
  S.~G., {Meylan} G., eds., p. 141

\bibitem[{{Portegies Zwart} {et~al}\mbox{.}(2010){Portegies Zwart}, {McMillan},
  \& {Gieles}}]{2010ARA&A..48..431P}
{Portegies Zwart} S.~F., {McMillan} S.~L.~W., {Gieles} M., 2010, \araa, 48, 431

\bibitem[{{Posti} \& {Helmi}(2018)}]{Posti2018}
{Posti} L., {Helmi} A., 2018, ArXiv e-prints, arXiv:1805.01408

\bibitem[{{Prager} {et~al}\mbox{.}(2017){Prager}, {Ransom}, {Freire},
  {Hessels}, {Stairs}, {Arras}, \& {Cadelano}}]{2017ApJ...845..148P}
{Prager} B.~J., {Ransom} S.~M., {Freire} P.~C.~C., {Hessels} J.~W.~T., {Stairs}
  I.~H., {Arras} P., {Cadelano} M., 2017, \apj, 845, 148

\bibitem[{{Pryor} {et~al}\mbox{.}(1986){Pryor}, {Smith}, \&
  {McClure}}]{1986AJ.....92.1358P}
{Pryor} C., {Smith} G.~H., {McClure} R.~D., 1986, \aj, 92, 1358

\bibitem[{{Read} {et~al}\mbox{.}(2006){Read}, {Wilkinson}, {Evans}, {Gilmore},
  \& {Kleyna}}]{Read2006}
{Read} J.~I., {Wilkinson} M.~I., {Evans} N.~W., {Gilmore} G., {Kleyna} J.~T.,
  2006, \mnras, 367, 387

\bibitem[{{Robin} {et~al}\mbox{.}(2003){Robin}, {Reyl{\'e}}, {Derri{\`e}re}, \&
  {Picaud}}]{Robin2003}
{Robin} A.~C., {Reyl{\'e}} C., {Derri{\`e}re} S., {Picaud} S., 2003, \aap, 409,
  523

\bibitem[{{Rodriguez} {et~al}\mbox{.}(2016){Rodriguez}, {Chatterjee}, \&
  {Rasio}}]{2016PhRvD.93h4029R}
{Rodriguez} C.~L., {Chatterjee} S., {Rasio} F.~A., 2016, \prd, 93, 084029

\bibitem[{{Sarajedini} {et~al}\mbox{.}(2007){Sarajedini}, {Bedin}, {Chaboyer},
  {Dotter}, {Siegel}, {Anderson}, {Aparicio}, {King}, {Majewski},
  {Mar{\'{\i}}n-Franch}, {Piotto}, {Reid}, \&
  {Rosenberg}}]{2007AJ....133.1658S}
{Sarajedini} A. {et~al.}, 2007, \aj, 133, 1658

\bibitem[{{Shanahan} \& {Gieles}(2015)}]{2015MNRAS.448L..94S}
{Shanahan} R.~L., {Gieles} M., 2015, \mnras, 448, L94

\bibitem[{{Sollima} \& {Baumgardt}(2017)}]{2017MNRAS.471.3668S}
{Sollima} A., {Baumgardt} H., 2017, \mnras, 471, 3668

\bibitem[{{Sollima} {et~al}\mbox{.}(2015){Sollima}, {Baumgardt}, {Zocchi},
  {Balbinot}, {Gieles}, {H{\'e}nault-Brunet}, \& {Varri}}]{2015MNRAS.451.2185S}
{Sollima} A., {Baumgardt} H., {Zocchi} A., {Balbinot} E., {Gieles} M.,
  {H{\'e}nault-Brunet} V., {Varri} A.~L., 2015, \mnras, 451, 2185

\bibitem[{{Sollima} {et~al}\mbox{.}(2012){Sollima}, {Bellazzini}, \&
  {Lee}}]{2012ApJ...755..156S}
{Sollima} A., {Bellazzini} M., {Lee} J.-W., 2012, \apj, 755, 156

\bibitem[{{Sollima} {et~al}\mbox{.}(2016){Sollima}, {Ferraro}, {Lovisi},
  {Contenta}, {Vesperini}, {Origlia}, {Lapenna}, {Lanzoni}, {Mucciarelli},
  {Dalessandro}, \& {Pallanca}}]{2016MNRAS.462.1937S}
{Sollima} A. {et~al.}, 2016, \mnras, 462, 1937

\bibitem[{{Trager} {et~al}\mbox{.}(1995){Trager}, {King}, \&
  {Djorgovski}}]{1995AJ....109..218T}
{Trager} S.~C., {King} I.~R., {Djorgovski} S., 1995, \aj, 109, 218

\bibitem[{{van de Ven} {et~al}\mbox{.}(2006){van de Ven}, {van den Bosch},
  {Verolme}, \& {de Zeeuw}}]{2006A&A...445..513V}
{van de Ven} G., {van den Bosch} R.~C.~E., {Verolme} E.~K., {de Zeeuw} P.~T.,
  2006, \aap, 445, 513

\bibitem[{{Vasiliev}(2018)}]{2018arXiv180709775V}
{Vasiliev} E., 2018, ArXiv:1807.09775

\bibitem[{{Vesperini} \& {Heggie}(1997)}]{1997MNRAS.289..898V}
{Vesperini} E., {Heggie} D.~C., 1997, \mnras, 289, 898

\bibitem[{{Wang} {et~al}\mbox{.}(2016){Wang}, {Spurzem}, {Aarseth}, {Giersz},
  {Askar}, {Berczik}, {Naab}, {Schadow}, \&
  {Kouwenhoven}}]{2016MNRAS.458.1450W}
{Wang} L. {et~al.}, 2016, \mnras, 458, 1450

\bibitem[{{Watkins} {et~al}\mbox{.}(2013){Watkins}, {van de Ven}, {den Brok},
  \& {van den Bosch}}]{Watkins2013}
{Watkins} L.~L., {van de Ven} G., {den Brok} M., {van den Bosch} R.~C.~E.,
  2013, \mnras, 436, 2598

\bibitem[{{Watkins} {et~al}\mbox{.}(2015{\natexlab{a}}){Watkins}, {van der
  Marel}, {Bellini}, \& {Anderson}}]{2015ApJ...803...29W}
{Watkins} L.~L., {van der Marel} R.~P., {Bellini} A., {Anderson} J.,
  2015{\natexlab{a}}, \apj, 803, 29

\bibitem[{{Watkins} {et~al}\mbox{.}(2015{\natexlab{b}}){Watkins}, {van der
  Marel}, {Bellini}, \& {Anderson}}]{Watkins2015b}
{Watkins} L.~L., {van der Marel} R.~P., {Bellini} A., {Anderson} J.,
  2015{\natexlab{b}}, \apj, 812, 149

\bibitem[{{Watkins} {et~al}\mbox{.}(2018){Watkins}, {van der Marel}, {Sohn}, \&
  {Evans}}]{Watkins2018}
{Watkins} L.~L., {van der Marel} R.~P., {Sohn} S.~T., {Evans} N.~W., 2018,
  ArXiv e-prints, arXiv:1804.11348

\bibitem[{{Weatherford} {et~al}\mbox{.}(2018){Weatherford}, {Chatterjee},
  {Rodriguez}, \& {Rasio}}]{2018ApJ...864...13W}
{Weatherford} N.~C., {Chatterjee} S., {Rodriguez} C.~L., {Rasio} F.~A., 2018,
  \apj, 864, 13

\bibitem[{{Zhu} {et~al}\mbox{.}(2016{\natexlab{a}}){Zhu}, {Romanowsky}, {van de
  Ven}, {Long}, {Watkins}, {Pota}, {Napolitano}, {Forbes}, {Brodie}, \&
  {Foster}}]{Zhu2016a}
{Zhu} L. {et~al.}, 2016{\natexlab{a}}, \mnras, 462, 4001

\bibitem[{{Zhu} {et~al}\mbox{.}(2016{\natexlab{b}}){Zhu}, {van de Ven},
  {Watkins}, \& {Posti}}]{Zhu2016b}
{Zhu} L., {van de Ven} G., {Watkins} L.~L., {Posti} L., 2016{\natexlab{b}},
  \mnras, 463, 1117

\bibitem[{{Zocchi} {et~al}\mbox{.}(2017){Zocchi}, {Gieles}, \&
  {H{\'e}nault-Brunet}}]{2017MNRAS.468.4429Z}
{Zocchi} A., {Gieles} M., {H{\'e}nault-Brunet} V., 2017, \mnras, 468, 4429

\end{thebibliography}

\section{Multimass King DF}
\label{appendix_King_DF}

The DF defined in equation \ref{King_DF} is integrated in the velocity domain at each radius to obtain the density ($\rho_{j}$) and velocity dispersion profiles in both the radial and tangential components ($\sigma_{{\rm r},j}$, $\sigma_{{\rm t},j}$) of all mass species and projected on the plane of the sky ($\Sigma$, $\sigma_\mathrm{LOS}$, $\sigma_\mathrm{R}$, $\sigma_\mathrm{T}$):

\begin{eqnarray}
	\rho_{j}(r) & = & 4 \pi \int_0^{\sqrt{-2 \phi}} \int_0^{\sqrt{-2 \phi - v_\mathrm{r}^2}} v_\mathrm{t} f_{j} \left( r, v_\mathrm{r}, v_\mathrm{t} \right) \mathrm{d} v_\mathrm{t} \mathrm{d} v_\mathrm{r} \\
	\sigma_{{\rm r},j}^2 (r) & = & \frac{4 \pi}{\rho_j(r)} \int_0^{\sqrt{-2 \phi}}  v_\mathrm{r}^{2} \int_0^{\sqrt{-2 \phi - v_\mathrm{r}^{2}}} v_\mathrm{t} f_{j} \left( r, v_\mathrm{r}, v_\mathrm{t} \right) \mathrm{d} v_\mathrm{t} \mathrm{d} v_\mathrm{r} \\
	\sigma_{{\rm t},j}^2 (r) & = & \frac{4 \pi}{\rho_j(r)} \int_0^{\sqrt{-2 \phi}} \int_0^{\sqrt{-2\phi - v_\mathrm{r}^{2}}} v_\mathrm{t}^{3} f_{j} \left( r, v_\mathrm{r}, v_\mathrm{t} \right) \mathrm{d} v_\mathrm{t} \mathrm{d} v_\mathrm{r} .
\end{eqnarray}

\begin{eqnarray}
	\Sigma_{j}(R) & = & 2 \int_0^{r_\mathrm{t}} \frac{r \rho_{j}}{\sqrt{r^2 - R^2}} \mathrm{d} r \\
	\sigma_{{\rm LOS},j}^2 (R) & = & \frac{1}{\Sigma_{j}(R)} \int_0^{r_\mathrm{t}} \frac{\rho_{j}[2 \sigma_{{\rm r},j}^2 (r^2 - R^2) + \sigma_{{\rm t},j}^2 R^2]}{r \sqrt{r^2 - R^2}} \mathrm{d} r \\
	\sigma_{{\rm R},j}^2 (R) & = & \frac{1}{\Sigma_{j} (R)} \int_0^{r_\mathrm{t}} \frac{\rho_{j}[2 \sigma_{{\rm r},j}^2 R^2 + \sigma_{{\rm t},j}^2 (r^2 - R^2)]} {r \sqrt{r^2 - R^2}} \mathrm{d} r \\
	\sigma_{{\rm T},j}^2 (R) & = & \frac{1}{\Sigma_{j}(R)} \int_0^{r_\mathrm{t}} \frac{r \rho_{j} \sigma_{{\rm t},j}^2} {\sqrt{r^2 - R^2}} \mathrm{d} r
\end{eqnarray}

\section{Variable M/L Jeans models (Sollima)}
\label{appendix_Jeans_Sollima}

With the tracer density profile and and the velocity dispersion profile determined as indicated in \autoref{Jeans_Sollima}, the projections of the velocity moments onto the plane of the sky ($\sigma_\mathrm{LOS}$, $\sigma_\mathrm{R},$ and $\sigma_\mathrm{T}$) as a function of projected radius $R$ for the tracer population are derived through the following relations:
\begin{eqnarray}
\sigma_{\rm LOS}^{2}(R)&=&\frac{1}{\Sigma}\int_{R}^{+\infty}\frac{\nu r
\sigma_{\rm r}^{2} \left(1-\beta\frac{R^{2}}{r^{2}}\right)}{\sqrt{r^{2}-R^{2}}} dr\nonumber\\
\sigma_{\rm R}^{2}(R)&=&\frac{1}{\Sigma}\int_{R}^{+\infty}\frac{\nu r
\sigma_{\rm r}^{2} \left(1-\beta+\beta\frac{R^{2}}{r^{2}}\right)}{\sqrt{r^{2}-R^{2}}} dr\nonumber\\
\sigma_{\rm T}^{2}(R)&=&\frac{1}{\Sigma}\int_{R}^{+\infty}\frac{\nu r
\sigma_{\rm r}^{2} \left(1-\beta\right)}{\sqrt{r^{2}-R^{2}}} dr\nonumber
\end{eqnarray}

\section{Variable M/L Jeans (JAM) models}
\label{appendix_JAM}

\subsection{JAM models}

A key feature of the JAM models is that the tracer density profile $\nu(r)$ and mass density $\rho(r)$ of the cluster are both provided in the form of a Multi-Gaussian Expansion (MGE). Under the assumption of spherical symmetry, each Gaussian component $j$ of the tracer MGE and $k$ of the mass MGE is defined by two parameters: a central density ($\nu_{j}$ or $\rho_{k}$), and a width ($s_{j}$ or $s_{k}$). Then the tracer density profile is
\begin{equation}
    \nu \left( r \right) = \sum_{j=1}^{N_\mathrm{TD}} \nu_{j}
    	\exp \left( -\frac{r^2}{2 s_{j}} \right),
	\label{eqn:tracer_mge_equation}
\end{equation}
and the mass density profile is
\begin{equation}
    \rho \left( r \right) = \sum_{k=1}^{N_\mathrm{MD}} \rho_\mathrm{k}
    	\exp \left( -\frac{r^2}{2 s_\mathrm{k}} \right),
	\label{eqn:mass_mge_equation}
\end{equation}
where $N_\mathrm{TD}$ is the total number of tracer density Gaussian components and $N_\mathrm{MD}$ is the total number of mass density Gaussian components.\footnote{This approach is similar to that in \autoref{Jeans_Sollima}, though it is not standard for all flavours of Jeans models.} Parameterising the profiles in this form makes them extremely easy to project onto the plane of the sky (or, conversely, to deproject surface density profiles) and the projected profiles are themselves MGEs. This is particularly useful as, in practice, we can typically only measure the surface tracer density profile from observations, so we fit to the projected profile and then deproject it to get the 3D profile.

It is typical for analyses of real clusters to use the luminosity density profile $\mu(r)$ as a proxy for the tracer density profile -- that is, $\nu(r) = C \mu(r)$ for some constant $C$ -- so we adopt this approach here\footnote{The two profiles can, in fact, be very different owing to the internal processes that have shaped the cluster. However, in this case, the assumption is fortunately quite reasonable.}. In fact, for the JAM models, we need only the shape of the tracer density profile and not its normalisation, so it is more accurate to say that we use the shape of the luminosity density profile as a proxy for the shape of the tracer density profile; we do not need to know or even assume anything about the relative normalisations of the two profiles and can assume that $C = 1$.

Using the mass density and luminosity density profiles we can also calculate the mass-to-light profile $\Upsilon(r)$ of the cluster as
\begin{equation}
	\Upsilon \left( r \right) = \frac{\rho \left( r \right)}
    	{\mu \left( r \right)} .
\end{equation}

Here, we assume that the MGEs have the same number of Gaussian components ($N_\mathrm{TD} = N_\mathrm{MD}$), and that the Gaussian widths are the same $\{ s_{j} \} = \{ s_{k} \}$; the widths themselves are free parameters. The central densities $\rho_{k}$ and $\nu_{k}$ are also left free and are independent of each other, which allows us to fit the light and mass density profiles non-parametrically, and allows the mass-to-light ratio to vary throughout the cluster. In this way, the models account for the different spatial distributions of different mass populations, albeit without explicitly identifying the underlying mass function.

Anisotropy is provided per Gaussian component of the tracer MGE. We wish to allow the anisotropy to vary non-parametrically through the cluster, so assign each component $k$ an anisotropy $\beta_{k}$. As defined, anisotropy (equation~\ref{eqn:def_anisotropy}) lies between $1$ (radial anisotropy) and $-\infty$ (tangential anisotropy), with isotropy at $\beta=0$. The high anti-symmetry in $\beta$ and the infinite lower limit on $\beta$ both make it a hard parameter to fit for, so instead we follow \citet{Read2006} and use a modified anisotropy parameter
\begin{equation}
	\beta' = \frac{\beta}{2 - \beta}
    \label{equation:modified_beta}
\end{equation}
for which $\beta' = 0$ still indicates isotropy, positive $\beta'$ values still indicate radial anisotropy, and negative $\beta'$ values still indicate tangential anisotropy, but which has the pleasing properties of being symmetric about isotropy and finite, restricted to the range $[-1,1]$.

Overall, the cluster models have $4N$ free parameters: $ \{ \nu_{k}, \rho_{k}, s_{k}, \beta_{k} \}$. We choose to use $N=6$ here, so we have 24 free parameters for the cluster\footnote{On the one hand, we want as few Gaussians as possible to limit the number of free parameters, and on the other hand we need a sufficient number of Gaussians to adequately describe the cluster. We ran models with $N=5$ and found that these did not perform well at the centre. Using $N=6$ improved the central fits. Using $N=7$ did not make a significant difference compared to the $N=6$ models.}.

The simulated cluster was evolved in a tidal field and so was affected by the interaction with the Galactic tidal field. Some tidally stripped stars or potential escapers remain in the simulated data set, so we further include a contaminant population that is dominant at large radii to correctly fit these stars. We model the contaminant density as an extra Gaussian with surface brightness $\nu_\mathrm{contam}$ and width $s_\mathrm{contam}$. In practice, $\nu_\mathrm{contam}$ and $s_\mathrm{contam}$ are degenerate, so we choose to allow $\nu_\mathrm{contam}$ to be free and fix $s_\mathrm{contam}$ to be the value of the outermost data point in the surface brightness profile. We assume that the velocity dispersion of the contaminant population is isotropic and equal to $\sigma_\mathrm{contam}$, which is also left free. This introduces a further two free parameters: $\{ \nu_\mathrm{contam}$, $\sigma_\mathrm{contam} \}$.

Both the JAM models for the cluster and our model for the contaminant population make predictions for the velocity dispersions in physical units. To compare with the mocks, we convert the velocities ($v_\mathrm{R}$, $v_\mathrm{T}$) into PMs ($\mu_\mathrm{R}$, $\mu_\mathrm{T}$) using the distance $D$ to the cluster. This adds another free parameter (apart from case 1 where the distance is fixed to the true distance because only LOS velocities are used). In total, we thus have 27 (26 for case 1) free parameters for our models.

\subsection{Model fitting}

We fit the JAM models to the surface brightness profile and the velocity dispersion profiles simultaneously. To recover the surface brightness profile, we fit to the binned profile calculated from the simulated data. We compute the likelihood of each data point $i$ at location $r_{i}$ with value $\mu_{\mathrm{V},i} \pm \delta_{\mathrm{V},i}$ given the model prediction $\mu_\mathrm{model} (r_{i})$,
\begin{equation}
	\mathcal{L}_\mathrm{SB} = \prod_{i}^{N_\mathrm{SB}} \frac{1}{\sqrt{2 \pi} \delta_{\mathrm{V},i}} \exp \left( \frac{-\left( \mu_\mathrm{model} \left( r_{i} \right) - \mu_{\mathrm{V},i} \right)^2}{2 \delta_{\mathrm{V},i}^2} \right) ,
\end{equation}
where $N_\mathrm{SB}$ is the number of points in the surface brightness profile, and the uncertainties $\delta_{\mathrm{V},i}$ are assumed to be 10\% of the surface brightness values $\mu_{\mathrm{V},i}$\footnote{We adopted $\sim10\%$ uncertainties on the surface brightness profile as rough error estimates based on bootstrapping experiments with mock cluster data, but in any case the uncertainties on the overall fits are dominated by the uncertainties on the kinematics, so the models are insensitive to this choice.}.

The models make predictions about the velocity dispersion profiles for the cluster and the contaminant population. To assess how well a model dispersion profile fits a given data set, we use the discrete approach described in \citet{Watkins2013} whereby we compute the likelihood $\mathcal{L}_{i}$ of observing a star $i$ at projected distance $R_{i}$ from the cluster centre with velocity $(v_{\mathrm{R},i} \pm \delta_{\mathrm{R},i}, v_{\mathrm{T},i} \pm \delta_{\mathrm{T},i}, v_{\mathrm{LOS},i} \pm \delta_{\mathrm{LOS},i})$ given the model prediction for the velocity distribution at $R_{i}$, which we assume to be Gaussian with velocity dispersions $(\sigma_{\mathrm{R},i}, \sigma_{\mathrm{T},i}, \sigma_{\mathrm{LOS},i})$. As the models are non-rotating, the mean velocity is everywhere 0, and the dispersions are uncorrelated, that is the cross-terms in the velocity ellipsoid tensor are 0. In this case, the likelihood of each velocity component can be considered independently. The projected radial velocity likelihood $\mathcal{L}_{R,i}$ for star $i$ is then
\begin{equation}
	\mathcal{L}_{{\rm R},i} = \frac{1}{\sqrt{2 \pi \left( \delta_{{\rm R},i}^2 + \sigma_{{\rm R},i}^2 \right)}} \exp \left( \frac{-v_{{\rm R},i}^2}{2 \left( \delta_{{\rm R},i}^2 + \sigma_{{\rm R},i}^2 \right)} \right),
\end{equation} 
and similarly for the projected tangential velocity likelihood $\mathcal{L}_{\mathrm{T},i}$ and the LOS velocity likelihood $\mathcal{L}_{\mathrm{LOS},i}$\footnote{This is the same approach adopted in \autoref{Jeans_Sollima}, see equation \ref{like_eq}.}. The likelihood of the model given all the line-of-sight velocity measurements is then
\begin{equation}
	\mathcal{L}_\mathrm{RV} = \prod_{i=1}^{N_\mathrm{RV}} \mathcal{L}_{\mathrm{LOS},i},
\end{equation}
where $N_\mathrm{RV}$ is the size of the LOS velocity dataset, and the likelihood of the model given all PM measurements is
\begin{equation}
	\mathcal{L}_\mathrm{PM} = \prod_{i=1}^{N_\mathrm{PM}} \mathcal{L}_{\mathrm{R},i} \mathcal{L}_{\mathrm{T},i},
\end{equation}
where $N_\mathrm{PM}$ is the size of the PM dataset. Note that in all of these calculations we have assumed that PMs ($\mu_\mathrm{R}$, $\mu_\mathrm{T}$) have been converted to velocities ($v_\mathrm{R}$, $v_\mathrm{T}$) using the distance $D$ to the cluster. When we have only LOS velocities (case 1), the total likelihood from the kinematics for a given population is $\mathcal{L}_\mathrm{kin,pop} = \mathcal{L}_\mathrm{RV}$. When we have both LOS velocity and PM datasets (cases 2 and 3), the total likelihood from the kinematics is $\mathcal{L}_\mathrm{kin,pop} = \mathcal{L}_\mathrm{RV} \mathcal{L}_\mathrm{PM}$.

For each star, we assess the likelihood of the observed kinematics given the cluster model $\mathcal{L}_\mathrm{kin,cluster}$ and the likelihood of the observed kinematics given the contaminant model $\mathcal{L}_\mathrm{kin,contam}$. To calculate the total likelihood from the kinematics, we combine the likelihoods from the two populations via
\begin{equation}
	\mathcal{L}_\mathrm{kin} = P_\mathrm{cluster} \mathcal{L}_\mathrm{kin,cluster}
    	+ (1-P_\mathrm{cluster}) \mathcal{L}_\mathrm{kin,contam} ,
\end{equation}
where $P_\mathrm{cluster}$ is the probability of the star being a cluster member and not a contaminant, which is derived from the densities of the two populations via
\begin{equation}
	P_\mathrm{cluster} = \frac{\nu_\mathrm{cluster} \left( r \right)}
    	{\nu_\mathrm{cluster} \left( r \right)
        + \nu_\mathrm{contam} \left( r \right)} .
	\label{equation:Pcluster}
\end{equation}
Finally, the total likelihood is given by
\begin{equation}
	\mathcal{L} = \mathcal{L}_\mathrm{SB} \mathcal{L}_\mathrm{kin} .
\end{equation}

We restrict the range of certain parameters using priors. To avoid having components with negative light or negative mass, we force the $\nu_{k}$ and $\rho_{k}$ parameters to be positive, but otherwise use a flat prior for these values. Similarly, we require the distance $D$ to be positive, but otherwise use a flat prior on the distance. For the Gaussian widths $s_{k}$, we require that $s_{k} < s_{k+1}$ and restrict the extent of the innermost and outermost Gaussians by insisting that $s_1 > R_\mathrm{min}$ and $s_6 < R_\mathrm{max} / \sqrt{3}$, where $R_\mathrm{min}$ and $R_\mathrm{max}$ are the minimum and maximum radial points of the mock surface brightness profile. The factor $\sqrt{3}$ in the latter restriction forces the profiles to have an outer slope of 3 or larger, and thus to be finite.

As discussed, $\beta'$ ranges between [-1,1], so in theory all we need is a flat prior in this range. However, anisotropy characterisation requires at least two orthogonal velocity components, so with LOS velocities only we are unable to fit for the anisotropy and, thus, break the degeneracy between anisotropy and mass. This is a problem for case 1 where we have LOS velocities only, but also for the outer regions of the cluster in case 2 as the \textit{HST}-like PMs only cover the central regions. We could further assume that the cluster is isotropic (i.e. $\beta'_{k} = 0$) in the regions where we lack PM data, as is usual for studies of real clusters in the absence of prior anisotropy information, however as the Gaussian widths are left free, it is not possible to set a priori which Gaussians cover the range of the PM data (where the anisotropy can be left free) and which are outside (where we may wish to assume a value). Instead, here we choose to leave the anisotropy free everywhere but to provide a strong prior on the $\beta'_\mathrm{k}$ values:
\begin{equation}
	P \left( \beta'_{k} \right) =
\begin{cases}
    	 1 ,& \beta'_{k} \le \mu_{\beta'} \nonumber \\
 \displaystyle    \frac{1}{\sqrt{2 \pi} \sigma_{\beta'}} \exp \left( -\frac{\left(\beta'_{k} - \mu_{\beta'} \right)^2} {2 \sigma_{\beta'}^2}
        \right), & \beta'_{k} > \mu_{\beta'} .
\end{cases}
\end{equation}
This imposes a flat prior for $\beta'$ in $[-\mu_{\beta'}, \mu_{\beta'}]$ and then uses a Gaussian fall off for $\beta'_{k}$ values outside of this range. As we do not expect the anisotropy to be extreme in either direction, we choose $\mu_{\beta'} = 0.2$ and $\sigma_{\beta'} = 0.02$. Essentially, we are assuming that the system is near isotropic but acknowledge that it may not be exactly isotropic and so we let it explore some mildly-anisotropic options. Further, the anisotropy for each Gaussian is affected by the anisotropy of the neighbouring Gaussians, so by leaving them all some freedom, they can respond to the neighbouring values where there may be better constraints. For the RV-only case, we could simply assume isotropy, but preferred to treat all cases the same.

Finally, we calculate the posterior probability by multiplying together the likelihood and the priors. To determine the best-fitting family of models, we wish to find the region of parameter space for which the posterior is maximised. To efficiently explore our parameter space and determine the region where the best-fitting models are located, we use the affine-invariant Markov chain Monte Carlo (MCMC) ensemble sampler \textsc{emcee} \citep{ForemanMackey2013}. Our MCMC runs each use 1000 walkers and we run for 10\,000 steps.

\label{lastpage}
\end{document}